\documentclass[longauth]{aa_mod} 

\usepackage{upquote} 
\usepackage{graphicx}
\usepackage{txfonts}
\usepackage[]{xcolor}
\usepackage{url}
\usepackage{hyperref}
\usepackage[]{datetime}
\usepackage{rotating}
\usepackage{multirow}
\usepackage{diagbox}
\usepackage{amsmath,amssymb}

\newcommand{\orcit}[1]{\protect\href{https://orcid.org/#1}{\protect\includegraphics[width=8pt]{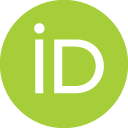}}}

\makeatletter
\renewcommand*\maketitle{%
  \thispagestyle{firstpage}
\begingroup
    \if@wideboxfn
    \setlength\bibindent{1.4\parindent}
    \else
    \setlength\bibindent{\parindent}
    \fi
    \renewcommand*\thefootnote{\@fnsymbol\c@footnote}%
    \renewcommand\@makefntext[1]{%
    \ifaa@longfn\hsize\textwidth\fi
    \noindent
    \hb@xt@\bibindent{\hss\@makefnmark\enspace}##1}
  \ifaa@twocolumn
  \begingroup
    \begin{aa@strip}
          \aa@maketitle
    \end{aa@strip}
    \@thanks	  	
  \endgroup
  \else
    \begingroup
      \let\thanks\footnote
      \aa@maketitle
    \endgroup
  \fi
\endgroup
  \setcounter{footnote}{0}%
}
\makeatother

\providecommand{\kms}{\ensuremath{\textrm{km\,s}^{-1}}}

\providecommand{\degree}{\ensuremath{^\circ}}

\newcommand{\matfont}[1]{\ensuremath\boldsymbol{\mathsf{#1}}}
\newcommand{\mat}[1]{\matfont{#1}}

\providecommand{\gaia}{\textit{Gaia}\xspace}
\providecommand{\gdr}[1]{\textit{Gaia}~DR{#1}\xspace}
\providecommand{\gedr}[1]{\textit{Gaia}~EDR{#1}\xspace}

\providecommand{\gmag}{\ensuremath{G}\xspace}

\providecommand{\gbp}{\ensuremath{G_{\rm BP}}\xspace}
\providecommand{\grp}{\ensuremath{G_{\rm RP}}\xspace}

\providecommand{\gminrp}{\ensuremath{G\!-\!G_{\rm RP}}\xspace}
\providecommand{\bpming}{\ensuremath{G_{\rm BP}\!-\!G}\xspace}
\providecommand{\bpminrp}{\ensuremath{G_{\rm BP}\!-\!G_{\rm RP}}\xspace}
\providecommand{\bprp}{\ensuremath{{\rm BP/RP}}\xspace}

\providecommand{\new}{}
\providecommand{\newer}{}

\providecommand{\linktodoc}{https://gea.esac.esa.int/archive/documentation/GDR3/}
\providecommand{\linktosec}[1]{\href{\linktodoc/#1}{online documentation}}
\providecommand{\linktotable}[2]{\href{\linktodoc/Gaia_archive/chap_datamodel/#1}{\tt{#2}\xspace}}
\providecommand{\linktoparam}[2]{\href{\linktodoc/Gaia_archive/chap_datamodel/#1}{\tt{#2}\xspace}}

\begin{document} 

\title{\gaia Data Release 3: The extragalactic content\thanks{Table~\ref{tab:qso_composite_fluxes}
    is only available in electronic form at the CDS at
    http://cdsweb.u-strasbg.fr/cgi-bin/qcat?J/A+A/}}
\titlerunning{\gaia Data Release 3: The extragalactic content}
\authorrunning{Gaia Collaboration}
\author{
{\it Gaia} Collaboration
\and     C.A.L.~                  Bailer-Jones\inst{\ref{inst:0001}}
\and         D.~                      Teyssier\orcit{0000-0002-6261-5292}\inst{\ref{inst:0002}}
\and         L.~                    Delchambre\orcit{0000-0003-2559-408X}\inst{\ref{inst:0003}}
\and         C.~                     Ducourant\orcit{0000-0003-4843-8979}\inst{\ref{inst:0004}}
\and         D.~                      Garabato\orcit{0000-0002-7133-6623}\inst{\ref{inst:0005}}
\and         D.~                Hatzidimitriou\orcit{0000-0002-5415-0464}\inst{\ref{inst:0006},\ref{inst:0007}}
\and       S.A.~                       Klioner\orcit{0000-0003-4682-7831}\inst{\ref{inst:0008}}
\and         L.~                     Rimoldini\orcit{0000-0002-0306-585X}\inst{\ref{inst:0009}}
\and         I.~                Bellas-Velidis\inst{\ref{inst:0007}}
\and         R.~                      Carballo\orcit{0000-0001-7412-2498}\inst{\ref{inst:0011}}
\and       M.I.~                     Carnerero\orcit{0000-0001-5843-5515}\inst{\ref{inst:0012}}
\and         C.~                        Diener\inst{\ref{inst:0013}}
\and         M.~                     Fouesneau\orcit{0000-0001-9256-5516}\inst{\ref{inst:0001}}
\and         L.~                     Galluccio\orcit{0000-0002-8541-0476}\inst{\ref{inst:0015}}
\and         P.~                        Gavras\orcit{0000-0002-4383-4836}\inst{\ref{inst:0016}}
\and         A.~                 Krone-Martins\orcit{0000-0002-2308-6623}\inst{\ref{inst:0017},\ref{inst:0018}}
\and       C.M.~                       Raiteri\orcit{0000-0003-1784-2784}\inst{\ref{inst:0012}}
\and         R.~                      Teixeira\orcit{0000-0002-6806-6626}\inst{\ref{inst:0020}}
\and     A.G.A.~                         Brown\orcit{0000-0002-7419-9679}\inst{\ref{inst:0021}}
\and         A.~                     Vallenari\orcit{0000-0003-0014-519X}\inst{\ref{inst:0022}}
\and         T.~                        Prusti\orcit{0000-0003-3120-7867}\inst{\ref{inst:0023}}
\and     J.H.J.~                    de Bruijne\orcit{0000-0001-6459-8599}\inst{\ref{inst:0023}}
\and         F.~                        Arenou\orcit{0000-0003-2837-3899}\inst{\ref{inst:0025}}
\and         C.~                     Babusiaux\orcit{0000-0002-7631-348X}\inst{\ref{inst:0026},\ref{inst:0025}}
\and         M.~                      Biermann\inst{\ref{inst:0028}}
\and       O.L.~                       Creevey\orcit{0000-0003-1853-6631}\inst{\ref{inst:0015}}
\and       D.W.~                         Evans\orcit{0000-0002-6685-5998}\inst{\ref{inst:0013}}
\and         L.~                          Eyer\orcit{0000-0002-0182-8040}\inst{\ref{inst:0031}}
\and         R.~                        Guerra\orcit{0000-0002-9850-8982}\inst{\ref{inst:0032}}
\and         A.~                        Hutton\inst{\ref{inst:0033}}
\and         C.~                         Jordi\orcit{0000-0001-5495-9602}\inst{\ref{inst:0034}}
\and       U.L.~                       Lammers\orcit{0000-0001-8309-3801}\inst{\ref{inst:0032}}
\and         L.~                     Lindegren\orcit{0000-0002-5443-3026}\inst{\ref{inst:0036}}
\and         X.~                          Luri\orcit{0000-0001-5428-9397}\inst{\ref{inst:0034}}
\and         F.~                       Mignard\inst{\ref{inst:0015}}
\and         C.~                         Panem\inst{\ref{inst:0039}}
\and         D.~            Pourbaix$^\dagger$\orcit{0000-0002-3020-1837}\inst{\ref{inst:0040},\ref{inst:0041}}
\and         S.~                       Randich\orcit{0000-0003-2438-0899}\inst{\ref{inst:0042}}
\and         P.~                    Sartoretti\inst{\ref{inst:0025}}
\and         C.~                      Soubiran\orcit{0000-0003-3304-8134}\inst{\ref{inst:0004}}
\and         P.~                         Tanga\orcit{0000-0002-2718-997X}\inst{\ref{inst:0015}}
\and       N.A.~                        Walton\orcit{0000-0003-3983-8778}\inst{\ref{inst:0013}}
\and         U.~                       Bastian\orcit{0000-0002-8667-1715}\inst{\ref{inst:0028}}
\and         R.~                       Drimmel\orcit{0000-0002-1777-5502}\inst{\ref{inst:0012}}
\and         F.~                        Jansen\inst{\ref{inst:0049}}
\and         D.~                          Katz\orcit{0000-0001-7986-3164}\inst{\ref{inst:0025}}
\and       M.G.~                      Lattanzi\orcit{0000-0003-0429-7748}\inst{\ref{inst:0012},\ref{inst:0052}}
\and         F.~                   van Leeuwen\inst{\ref{inst:0013}}
\and         J.~                        Bakker\inst{\ref{inst:0032}}
\and         C.~                      Cacciari\orcit{0000-0001-5174-3179}\inst{\ref{inst:0055}}
\and         J.~                 Casta\~{n}eda\orcit{0000-0001-7820-946X}\inst{\ref{inst:0056}}
\and         F.~                     De Angeli\orcit{0000-0003-1879-0488}\inst{\ref{inst:0013}}
\and         C.~                     Fabricius\orcit{0000-0003-2639-1372}\inst{\ref{inst:0034}}
\and         Y.~                    Fr\'{e}mat\orcit{0000-0002-4645-6017}\inst{\ref{inst:0059}}
\and         A.~                      Guerrier\inst{\ref{inst:0039}}
\and         U.~                        Heiter\orcit{0000-0001-6825-1066}\inst{\ref{inst:0061}}
\and         E.~                        Masana\orcit{0000-0002-4819-329X}\inst{\ref{inst:0034}}
\and         R.~                      Messineo\inst{\ref{inst:0063}}
\and         N.~                       Mowlavi\orcit{0000-0003-1578-6993}\inst{\ref{inst:0031}}
\and         C.~                       Nicolas\inst{\ref{inst:0039}}
\and         K.~                  Nienartowicz\orcit{0000-0001-5415-0547}\inst{\ref{inst:0066},\ref{inst:0009}}
\and         F.~                       Pailler\orcit{0000-0002-4834-481X}\inst{\ref{inst:0039}}
\and         P.~                       Panuzzo\orcit{0000-0002-0016-8271}\inst{\ref{inst:0025}}
\and         F.~                        Riclet\inst{\ref{inst:0039}}
\and         W.~                          Roux\orcit{0000-0002-7816-1950}\inst{\ref{inst:0039}}
\and       G.M.~                      Seabroke\orcit{0000-0003-4072-9536}\inst{\ref{inst:0072}}
\and         R.~                         Sordo\orcit{0000-0003-4979-0659}\inst{\ref{inst:0022}}
\and         F.~                  Th\'{e}venin\inst{\ref{inst:0015}}
\and         G.~                  Gracia-Abril\inst{\ref{inst:0075},\ref{inst:0028}}
\and         J.~                       Portell\orcit{0000-0002-8886-8925}\inst{\ref{inst:0034}}
\and         M.~                       Altmann\orcit{0000-0002-0530-0913}\inst{\ref{inst:0028},\ref{inst:0079}}
\and         R.~                        Andrae\orcit{0000-0001-8006-6365}\inst{\ref{inst:0001}}
\and         M.~                        Audard\orcit{0000-0003-4721-034X}\inst{\ref{inst:0031},\ref{inst:0009}}
\and         K.~                        Benson\inst{\ref{inst:0072}}
\and         J.~                      Berthier\orcit{0000-0003-1846-6485}\inst{\ref{inst:0084}}
\and         R.~                        Blomme\orcit{0000-0002-2526-346X}\inst{\ref{inst:0059}}
\and       P.W.~                       Burgess\inst{\ref{inst:0013}}
\and         D.~                      Busonero\orcit{0000-0002-3903-7076}\inst{\ref{inst:0012}}
\and         G.~                         Busso\orcit{0000-0003-0937-9849}\inst{\ref{inst:0013}}
\and         H.~                   C\'{a}novas\orcit{0000-0001-7668-8022}\inst{\ref{inst:0002}}
\and         B.~                         Carry\orcit{0000-0001-5242-3089}\inst{\ref{inst:0015}}
\and         A.~                       Cellino\orcit{0000-0002-6645-334X}\inst{\ref{inst:0012}}
\and         N.~                         Cheek\inst{\ref{inst:0092}}
\and         G.~                    Clementini\orcit{0000-0001-9206-9723}\inst{\ref{inst:0055}}
\and         Y.~                      Damerdji\orcit{0000-0002-3107-4024}\inst{\ref{inst:0003},\ref{inst:0095}}
\and         M.~                      Davidson\inst{\ref{inst:0096}}
\and         P.~                    de Teodoro\inst{\ref{inst:0032}}
\and         M.~              Nu\~{n}ez Campos\inst{\ref{inst:0033}}
\and         A.~                      Dell'Oro\orcit{0000-0003-1561-9685}\inst{\ref{inst:0042}}
\and         P.~                        Esquej\orcit{0000-0001-8195-628X}\inst{\ref{inst:0016}}
\and         J.~   Fern\'{a}ndez-Hern\'{a}ndez\inst{\ref{inst:0101}}
\and         E.~                        Fraile\inst{\ref{inst:0016}}
\and         P.~              Garc\'{i}a-Lario\orcit{0000-0003-4039-8212}\inst{\ref{inst:0032}}
\and         E.~                        Gosset\inst{\ref{inst:0003},\ref{inst:0041}}
\and         R.~                       Haigron\inst{\ref{inst:0025}}
\and      J.-L.~                     Halbwachs\orcit{0000-0003-2968-6395}\inst{\ref{inst:0107}}
\and       N.C.~                        Hambly\orcit{0000-0002-9901-9064}\inst{\ref{inst:0096}}
\and       D.L.~                      Harrison\orcit{0000-0001-8687-6588}\inst{\ref{inst:0013},\ref{inst:0110}}
\and         J.~                 Hern\'{a}ndez\orcit{0000-0002-0361-4994}\inst{\ref{inst:0032}}
\and         D.~                    Hestroffer\orcit{0000-0003-0472-9459}\inst{\ref{inst:0084}}
\and       S.T.~                       Hodgkin\orcit{0000-0002-5470-3962}\inst{\ref{inst:0013}}
\and         B.~                          Holl\orcit{0000-0001-6220-3266}\inst{\ref{inst:0031},\ref{inst:0009}}
\and         K.~                    Jan{\ss}en\orcit{0000-0002-8163-2493}\inst{\ref{inst:0116}}
\and         G.~          Jevardat de Fombelle\inst{\ref{inst:0031}}
\and         S.~                        Jordan\orcit{0000-0001-6316-6831}\inst{\ref{inst:0028}}
\and       A.C.~                     Lanzafame\orcit{0000-0002-2697-3607}\inst{\ref{inst:0119},\ref{inst:0120}}
\and         W.~                  L\"{ o}ffler\inst{\ref{inst:0028}}
\and         O.~                       Marchal\orcit{ 0000-0001-7461-892}\inst{\ref{inst:0107}}
\and       P.M.~                       Marrese\orcit{0000-0002-8162-3810}\inst{\ref{inst:0123},\ref{inst:0124}}
\and         A.~                      Moitinho\orcit{0000-0003-0822-5995}\inst{\ref{inst:0017}}
\and         K.~                      Muinonen\orcit{0000-0001-8058-2642}\inst{\ref{inst:0126},\ref{inst:0127}}
\and         P.~                       Osborne\inst{\ref{inst:0013}}
\and         E.~                       Pancino\orcit{0000-0003-0788-5879}\inst{\ref{inst:0042},\ref{inst:0124}}
\and         T.~                       Pauwels\inst{\ref{inst:0059}}
\and         A.~                  Recio-Blanco\orcit{0000-0002-6550-7377}\inst{\ref{inst:0015}}
\and         C.~                     Reyl\'{e}\orcit{0000-0003-2258-2403}\inst{\ref{inst:0133}}
\and         M.~                        Riello\orcit{0000-0002-3134-0935}\inst{\ref{inst:0013}}
\and         T.~                      Roegiers\orcit{0000-0002-1231-4440}\inst{\ref{inst:0135}}
\and         J.~                       Rybizki\orcit{0000-0002-0993-6089}\inst{\ref{inst:0001}}
\and       L.M.~                         Sarro\orcit{0000-0002-5622-5191}\inst{\ref{inst:0137}}
\and         C.~                        Siopis\orcit{0000-0002-6267-2924}\inst{\ref{inst:0040}}
\and         M.~                         Smith\inst{\ref{inst:0072}}
\and         A.~                      Sozzetti\orcit{0000-0002-7504-365X}\inst{\ref{inst:0012}}
\and         E.~                       Utrilla\inst{\ref{inst:0033}}
\and         M.~                   van Leeuwen\orcit{0000-0001-9698-2392}\inst{\ref{inst:0013}}
\and         U.~                         Abbas\orcit{0000-0002-5076-766X}\inst{\ref{inst:0012}}
\and         P.~               \'{A}brah\'{a}m\orcit{0000-0001-6015-646X}\inst{\ref{inst:0144},\ref{inst:0145}}
\and         A.~                Abreu Aramburu\inst{\ref{inst:0101}}
\and         C.~                         Aerts\orcit{0000-0003-1822-7126}\inst{\ref{inst:0147},\ref{inst:0148},\ref{inst:0001}}
\and       J.J.~                        Aguado\inst{\ref{inst:0137}}
\and         M.~                          Ajaj\inst{\ref{inst:0025}}
\and         F.~                 Aldea-Montero\inst{\ref{inst:0032}}
\and         G.~                     Altavilla\orcit{0000-0002-9934-1352}\inst{\ref{inst:0123},\ref{inst:0124}}
\and       M.A.~                   \'{A}lvarez\orcit{0000-0002-6786-2620}\inst{\ref{inst:0005}}
\and         J.~                         Alves\orcit{0000-0002-4355-0921}\inst{\ref{inst:0156}}
\and       R.I.~                      Anderson\orcit{0000-0001-8089-4419}\inst{\ref{inst:0157}}
\and         E.~                Anglada Varela\orcit{0000-0001-7563-0689}\inst{\ref{inst:0101}}
\and         T.~                        Antoja\orcit{0000-0003-2595-5148}\inst{\ref{inst:0034}}
\and         D.~                        Baines\orcit{0000-0002-6923-3756}\inst{\ref{inst:0002}}
\and       S.G.~                         Baker\orcit{0000-0002-6436-1257}\inst{\ref{inst:0072}}
\and         L.~        Balaguer-N\'{u}\~{n}ez\orcit{0000-0001-9789-7069}\inst{\ref{inst:0034}}
\and         E.~                      Balbinot\orcit{0000-0002-1322-3153}\inst{\ref{inst:0163}}
\and         Z.~                         Balog\orcit{0000-0003-1748-2926}\inst{\ref{inst:0028},\ref{inst:0001}}
\and         C.~                       Barache\inst{\ref{inst:0079}}
\and         D.~                       Barbato\inst{\ref{inst:0031},\ref{inst:0012}}
\and         M.~                        Barros\orcit{0000-0002-9728-9618}\inst{\ref{inst:0017}}
\and       M.A.~                       Barstow\orcit{0000-0002-7116-3259}\inst{\ref{inst:0170}}
\and         S.~                 Bartolom\'{e}\orcit{0000-0002-6290-6030}\inst{\ref{inst:0034}}
\and      J.-L.~                     Bassilana\inst{\ref{inst:0172}}
\and         N.~                       Bauchet\inst{\ref{inst:0025}}
\and         U.~                      Becciani\orcit{0000-0002-4389-8688}\inst{\ref{inst:0119}}
\and         M.~                    Bellazzini\orcit{0000-0001-8200-810X}\inst{\ref{inst:0055}}
\and         A.~                     Berihuete\orcit{0000-0002-8589-4423}\inst{\ref{inst:0176}}
\and         M.~                        Bernet\orcit{0000-0001-7503-1010}\inst{\ref{inst:0034}}
\and         S.~                       Bertone\orcit{0000-0001-9885-8440}\inst{\ref{inst:0178},\ref{inst:0179},\ref{inst:0012}}
\and         L.~                       Bianchi\orcit{0000-0002-7999-4372}\inst{\ref{inst:0181}}
\and         A.~                    Binnenfeld\orcit{0000-0002-9319-3838}\inst{\ref{inst:0182}}
\and         S.~               Blanco-Cuaresma\orcit{0000-0002-1584-0171}\inst{\ref{inst:0183}}
\and         T.~                          Boch\orcit{0000-0001-5818-2781}\inst{\ref{inst:0107}}
\and         A.~                       Bombrun\inst{\ref{inst:0185}}
\and         D.~                       Bossini\orcit{0000-0002-9480-8400}\inst{\ref{inst:0186}}
\and         S.~                    Bouquillon\inst{\ref{inst:0079},\ref{inst:0188}}
\and         A.~                     Bragaglia\orcit{0000-0002-0338-7883}\inst{\ref{inst:0055}}
\and         L.~                      Bramante\inst{\ref{inst:0063}}
\and         E.~                        Breedt\orcit{0000-0001-6180-3438}\inst{\ref{inst:0013}}
\and         A.~                       Bressan\orcit{0000-0002-7922-8440}\inst{\ref{inst:0192}}
\and         N.~                     Brouillet\orcit{0000-0002-3274-7024}\inst{\ref{inst:0004}}
\and         E.~                    Brugaletta\orcit{0000-0003-2598-6737}\inst{\ref{inst:0119}}
\and         B.~                   Bucciarelli\orcit{0000-0002-5303-0268}\inst{\ref{inst:0012},\ref{inst:0052}}
\and         A.~                       Burlacu\inst{\ref{inst:0197}}
\and       A.G.~                     Butkevich\orcit{0000-0002-4098-3588}\inst{\ref{inst:0012}}
\and         R.~                         Buzzi\orcit{0000-0001-9389-5701}\inst{\ref{inst:0012}}
\and         E.~                        Caffau\orcit{0000-0001-6011-6134}\inst{\ref{inst:0025}}
\and         R.~                   Cancelliere\orcit{0000-0002-9120-3799}\inst{\ref{inst:0201}}
\and         T.~                 Cantat-Gaudin\orcit{0000-0001-8726-2588}\inst{\ref{inst:0034},\ref{inst:0001}}
\and         T.~                      Carlucci\inst{\ref{inst:0079}}
\and       J.M.~                      Carrasco\orcit{0000-0002-3029-5853}\inst{\ref{inst:0034}}
\and         L.~                   Casamiquela\orcit{0000-0001-5238-8674}\inst{\ref{inst:0004},\ref{inst:0025}}
\and         M.~                    Castellani\orcit{0000-0002-7650-7428}\inst{\ref{inst:0123}}
\and         A.~                 Castro-Ginard\orcit{0000-0002-9419-3725}\inst{\ref{inst:0021}}
\and         L.~                        Chaoul\inst{\ref{inst:0039}}
\and         P.~                       Charlot\orcit{0000-0002-9142-716X}\inst{\ref{inst:0004}}
\and         L.~                        Chemin\orcit{0000-0002-3834-7937}\inst{\ref{inst:0212}}
\and         V.~                    Chiaramida\inst{\ref{inst:0063}}
\and         A.~                     Chiavassa\orcit{0000-0003-3891-7554}\inst{\ref{inst:0015}}
\and         N.~                       Chornay\orcit{0000-0002-8767-3907}\inst{\ref{inst:0013}}
\and         G.~                     Comoretto\inst{\ref{inst:0002},\ref{inst:0217}}
\and         G.~                      Contursi\orcit{0000-0001-5370-1511}\inst{\ref{inst:0015}}
\and       W.J.~                        Cooper\orcit{0000-0003-3501-8967}\inst{\ref{inst:0219},\ref{inst:0012}}
\and         T.~                        Cornez\inst{\ref{inst:0172}}
\and         S.~                        Cowell\inst{\ref{inst:0013}}
\and         F.~                         Crifo\inst{\ref{inst:0025}}
\and         M.~                       Cropper\orcit{0000-0003-4571-9468}\inst{\ref{inst:0072}}
\and         M.~                        Crosta\orcit{0000-0003-4369-3786}\inst{\ref{inst:0012},\ref{inst:0226}}
\and         C.~                       Crowley\inst{\ref{inst:0185}}
\and         C.~                       Dafonte\orcit{0000-0003-4693-7555}\inst{\ref{inst:0005}}
\and         A.~                    Dapergolas\inst{\ref{inst:0007}}
\and         P.~                         David\inst{\ref{inst:0084}}
\and         P.~                    de Laverny\orcit{0000-0002-2817-4104}\inst{\ref{inst:0015}}
\and         F.~                      De Luise\orcit{0000-0002-6570-8208}\inst{\ref{inst:0232}}
\and         R.~                      De March\orcit{0000-0003-0567-842X}\inst{\ref{inst:0063}}
\and         J.~                     De Ridder\orcit{0000-0001-6726-2863}\inst{\ref{inst:0147}}
\and         R.~                      de Souza\inst{\ref{inst:0020}}
\and         A.~                     de Torres\inst{\ref{inst:0185}}
\and       E.F.~                    del Peloso\inst{\ref{inst:0028}}
\and         E.~                      del Pozo\inst{\ref{inst:0033}}
\and         M.~                         Delbo\orcit{0000-0002-8963-2404}\inst{\ref{inst:0015}}
\and         A.~                       Delgado\inst{\ref{inst:0016}}
\and      J.-B.~                       Delisle\orcit{0000-0001-5844-9888}\inst{\ref{inst:0031}}
\and         C.~                      Demouchy\inst{\ref{inst:0242}}
\and       T.E.~                 Dharmawardena\orcit{0000-0002-9583-5216}\inst{\ref{inst:0001}}
\and         S.~                       Diakite\inst{\ref{inst:0244}}
\and         E.~                     Distefano\orcit{0000-0002-2448-2513}\inst{\ref{inst:0119}}
\and         C.~                       Dolding\inst{\ref{inst:0072}}
\and         H.~                          Enke\orcit{0000-0002-2366-8316}\inst{\ref{inst:0116}}
\and         C.~                         Fabre\inst{\ref{inst:0248}}
\and         M.~                      Fabrizio\orcit{0000-0001-5829-111X}\inst{\ref{inst:0123},\ref{inst:0124}}
\and         S.~                       Faigler\orcit{0000-0002-8368-5724}\inst{\ref{inst:0251}}
\and         G.~                      Fedorets\orcit{0000-0002-8418-4809}\inst{\ref{inst:0126},\ref{inst:0253}}
\and         P.~                      Fernique\orcit{0000-0002-3304-2923}\inst{\ref{inst:0107},\ref{inst:0255}}
\and         F.~                      Figueras\orcit{0000-0002-3393-0007}\inst{\ref{inst:0034}}
\and         Y.~                      Fournier\orcit{0000-0002-6633-9088}\inst{\ref{inst:0116}}
\and         C.~                        Fouron\inst{\ref{inst:0197}}
\and         F.~                     Fragkoudi\orcit{0000-0002-0897-3013}\inst{\ref{inst:0259},\ref{inst:0260},\ref{inst:0261}}
\and         M.~                           Gai\orcit{0000-0001-9008-134X}\inst{\ref{inst:0012}}
\and         A.~              Garcia-Gutierrez\inst{\ref{inst:0034}}
\and         M.~              Garcia-Reinaldos\inst{\ref{inst:0032}}
\and         M.~             Garc\'{i}a-Torres\orcit{0000-0002-6867-7080}\inst{\ref{inst:0265}}
\and         A.~                      Garofalo\orcit{0000-0002-5907-0375}\inst{\ref{inst:0055}}
\and         A.~                         Gavel\orcit{0000-0002-2963-722X}\inst{\ref{inst:0061}}
\and         E.~                       Gerlach\orcit{0000-0002-9533-2168}\inst{\ref{inst:0008}}
\and         R.~                         Geyer\orcit{0000-0001-6967-8707}\inst{\ref{inst:0008}}
\and         P.~                      Giacobbe\orcit{0000-0001-7034-7024}\inst{\ref{inst:0012}}
\and         G.~                       Gilmore\orcit{0000-0003-4632-0213}\inst{\ref{inst:0013}}
\and         S.~                        Girona\orcit{0000-0002-1975-1918}\inst{\ref{inst:0272}}
\and         G.~                     Giuffrida\inst{\ref{inst:0123}}
\and         R.~                         Gomel\inst{\ref{inst:0251}}
\and         A.~                         Gomez\orcit{0000-0002-3796-3690}\inst{\ref{inst:0005}}
\and         J.~    Gonz\'{a}lez-N\'{u}\~{n}ez\orcit{0000-0001-5311-5555}\inst{\ref{inst:0092},\ref{inst:0277}}
\and         I.~   Gonz\'{a}lez-Santamar\'{i}a\orcit{0000-0002-8537-9384}\inst{\ref{inst:0005}}
\and       J.J.~            Gonz\'{a}lez-Vidal\inst{\ref{inst:0034}}
\and         M.~                       Granvik\orcit{0000-0002-5624-1888}\inst{\ref{inst:0126},\ref{inst:0281}}
\and         P.~                      Guillout\inst{\ref{inst:0107}}
\and         J.~                       Guiraud\inst{\ref{inst:0039}}
\and         R.~     Guti\'{e}rrez-S\'{a}nchez\inst{\ref{inst:0002}}
\and       L.P.~                           Guy\orcit{0000-0003-0800-8755}\inst{\ref{inst:0009},\ref{inst:0286}}
\and         M.~                        Hauser\inst{\ref{inst:0001},\ref{inst:0288}}
\and         M.~                       Haywood\orcit{0000-0003-0434-0400}\inst{\ref{inst:0025}}
\and         A.~                        Helmer\inst{\ref{inst:0172}}
\and         A.~                         Helmi\orcit{0000-0003-3937-7641}\inst{\ref{inst:0163}}
\and       M.H.~                     Sarmiento\orcit{0000-0003-4252-5115}\inst{\ref{inst:0033}}
\and       S.L.~                       Hidalgo\orcit{0000-0002-0002-9298}\inst{\ref{inst:0293},\ref{inst:0294}}
\and         T.~Hilger\orcit{0000-0003-1646-0063}\inst{\ref{inst:0007}}
\and         N.~                   H\l{}adczuk\orcit{0000-0001-9163-4209}\inst{\ref{inst:0032},\ref{inst:0296}}
\and         D.~                         Hobbs\orcit{0000-0002-2696-1366}\inst{\ref{inst:0036}}
\and         G.~                       Holland\inst{\ref{inst:0013}}
\and       H.E.~                        Huckle\inst{\ref{inst:0072}}
\and         K.~                       Jardine\inst{\ref{inst:0300}}
\and         G.~                    Jasniewicz\inst{\ref{inst:0301}}
\and         A.~          Jean-Antoine Piccolo\orcit{0000-0001-8622-212X}\inst{\ref{inst:0039}}
\and     \'{O}.~            Jim\'{e}nez-Arranz\orcit{0000-0001-7434-5165}\inst{\ref{inst:0034}}
\and         J.~             Juaristi Campillo\inst{\ref{inst:0028}}
\and         F.~                         Julbe\inst{\ref{inst:0034}}
\and         L.~                     Karbevska\inst{\ref{inst:0009},\ref{inst:0307}}
\and         P.~                      Kervella\orcit{0000-0003-0626-1749}\inst{\ref{inst:0308}}
\and         S.~                        Khanna\orcit{0000-0002-2604-4277}\inst{\ref{inst:0163},\ref{inst:0012}}
\and         M.~                      Kontizas\orcit{0000-0001-7177-0158}\inst{\ref{inst:0006}}
\and         G.~                    Kordopatis\orcit{0000-0002-9035-3920}\inst{\ref{inst:0015}}
\and       A.J.~                          Korn\orcit{0000-0002-3881-6756}\inst{\ref{inst:0061}}
\and      \'{A}~                K\'{o}sp\'{a}l\orcit{'{u}t 15-17, 1121 B}\inst{\ref{inst:0144},\ref{inst:0001},\ref{inst:0145}}
\and         Z.~           Kostrzewa-Rutkowska\inst{\ref{inst:0021},\ref{inst:0318}}
\and         K.~                Kruszy\'{n}ska\orcit{0000-0002-2729-5369}\inst{\ref{inst:0319}}
\and         M.~                           Kun\orcit{0000-0002-7538-5166}\inst{\ref{inst:0144}}
\and         P.~                       Laizeau\inst{\ref{inst:0321}}
\and         S.~                       Lambert\orcit{0000-0001-6759-5502}\inst{\ref{inst:0079}}
\and       A.F.~                         Lanza\orcit{0000-0001-5928-7251}\inst{\ref{inst:0119}}
\and         Y.~                         Lasne\inst{\ref{inst:0172}}
\and      J.-F.~                    Le Campion\inst{\ref{inst:0004}}
\and         Y.~                      Lebreton\orcit{0000-0002-4834-2144}\inst{\ref{inst:0308},\ref{inst:0327}}
\and         T.~                     Lebzelter\orcit{0000-0002-0702-7551}\inst{\ref{inst:0156}}
\and         S.~                        Leccia\orcit{0000-0001-5685-6930}\inst{\ref{inst:0329}}
\and         N.~                       Leclerc\inst{\ref{inst:0025}}
\and         I.~                 Lecoeur-Taibi\orcit{0000-0003-0029-8575}\inst{\ref{inst:0009}}
\and         S.~                          Liao\orcit{0000-0002-9346-0211}\inst{\ref{inst:0332},\ref{inst:0012},\ref{inst:0334}}
\and       E.L.~                        Licata\orcit{0000-0002-5203-0135}\inst{\ref{inst:0012}}
\and     H.E.P.~                  Lindstr{\o}m\inst{\ref{inst:0012},\ref{inst:0337},\ref{inst:0338}}
\and       T.A.~                        Lister\orcit{0000-0002-3818-7769}\inst{\ref{inst:0339}}
\and         E.~                       Livanou\orcit{0000-0003-0628-2347}\inst{\ref{inst:0006}}
\and         A.~                         Lobel\orcit{0000-0001-5030-019X}\inst{\ref{inst:0059}}
\and         A.~                         Lorca\inst{\ref{inst:0033}}
\and         C.~                          Loup\inst{\ref{inst:0107}}
\and         P.~                 Madrero Pardo\inst{\ref{inst:0034}}
\and         A.~               Magdaleno Romeo\inst{\ref{inst:0197}}
\and         S.~                       Managau\inst{\ref{inst:0172}}
\and       R.G.~                          Mann\orcit{0000-0002-0194-325X}\inst{\ref{inst:0096}}
\and         M.~                      Manteiga\orcit{0000-0002-7711-5581}\inst{\ref{inst:0348}}
\and       J.M.~                      Marchant\orcit{0000-0002-3678-3145}\inst{\ref{inst:0349}}
\and         M.~                       Marconi\orcit{0000-0002-1330-2927}\inst{\ref{inst:0329}}
\and         J.~                        Marcos\inst{\ref{inst:0002}}
\and     M.M.S.~                 Marcos Santos\inst{\ref{inst:0092}}
\and         D.~                Mar\'{i}n Pina\orcit{0000-0001-6482-1842}\inst{\ref{inst:0034}}
\and         S.~                      Marinoni\orcit{0000-0001-7990-6849}\inst{\ref{inst:0123},\ref{inst:0124}}
\and         F.~                       Marocco\orcit{0000-0001-7519-1700}\inst{\ref{inst:0356}}
\and       D.J.~                      Marshall\orcit{0000-0003-3956-3524}\inst{\ref{inst:0357}}
\and         L.~                   Martin Polo\inst{\ref{inst:0092}}
\and       J.M.~            Mart\'{i}n-Fleitas\orcit{0000-0002-8594-569X}\inst{\ref{inst:0033}}
\and         G.~                        Marton\orcit{0000-0002-1326-1686}\inst{\ref{inst:0144}}
\and         N.~                          Mary\inst{\ref{inst:0172}}
\and         A.~                         Masip\orcit{0000-0003-1419-0020}\inst{\ref{inst:0034}}
\and         D.~                       Massari\orcit{0000-0001-8892-4301}\inst{\ref{inst:0055}}
\and         A.~          Mastrobuono-Battisti\orcit{0000-0002-2386-9142}\inst{\ref{inst:0025}}
\and         T.~                         Mazeh\orcit{0000-0002-3569-3391}\inst{\ref{inst:0251}}
\and       P.J.~                      McMillan\orcit{0000-0002-8861-2620}\inst{\ref{inst:0036}}
\and         S.~                       Messina\orcit{0000-0002-2851-2468}\inst{\ref{inst:0119}}
\and         D.~                      Michalik\orcit{0000-0002-7618-6556}\inst{\ref{inst:0023}}
\and       N.R.~                        Millar\inst{\ref{inst:0013}}
\and         A.~                         Mints\orcit{0000-0002-8440-1455}\inst{\ref{inst:0116}}
\and         D.~                        Molina\orcit{0000-0003-4814-0275}\inst{\ref{inst:0034}}
\and         R.~                      Molinaro\orcit{0000-0003-3055-6002}\inst{\ref{inst:0329}}
\and         L.~                    Moln\'{a}r\orcit{0000-0002-8159-1599}\inst{\ref{inst:0144},\ref{inst:0374},\ref{inst:0145}}
\and         G.~                        Monari\orcit{0000-0002-6863-0661}\inst{\ref{inst:0107}}
\and         M.~                   Mongui\'{o}\orcit{0000-0002-4519-6700}\inst{\ref{inst:0034}}
\and         P.~                   Montegriffo\orcit{0000-0001-5013-5948}\inst{\ref{inst:0055}}
\and         A.~                       Montero\inst{\ref{inst:0033}}
\and         R.~                           Mor\orcit{0000-0002-8179-6527}\inst{\ref{inst:0034}}
\and         A.~                          Mora\inst{\ref{inst:0033}}
\and         R.~                    Morbidelli\orcit{0000-0001-7627-4946}\inst{\ref{inst:0012}}
\and         T.~                         Morel\orcit{0000-0002-8176-4816}\inst{\ref{inst:0003}}
\and         D.~                        Morris\inst{\ref{inst:0096}}
\and         T.~                      Muraveva\orcit{0000-0002-0969-1915}\inst{\ref{inst:0055}}
\and       C.P.~                        Murphy\inst{\ref{inst:0032}}
\and         I.~                       Musella\orcit{0000-0001-5909-6615}\inst{\ref{inst:0329}}
\and         Z.~                          Nagy\orcit{0000-0002-3632-1194}\inst{\ref{inst:0144}}
\and         L.~                         Noval\inst{\ref{inst:0172}}
\and         F.~                     Oca\~{n}a\inst{\ref{inst:0002},\ref{inst:0390}}
\and         A.~                         Ogden\inst{\ref{inst:0013}}
\and         C.~                     Ordenovic\inst{\ref{inst:0015}}
\and       J.O.~                        Osinde\inst{\ref{inst:0016}}
\and         C.~                        Pagani\orcit{0000-0001-5477-4720}\inst{\ref{inst:0170}}
\and         I.~                        Pagano\orcit{0000-0001-9573-4928}\inst{\ref{inst:0119}}
\and         L.~                     Palaversa\orcit{0000-0003-3710-0331}\inst{\ref{inst:0396},\ref{inst:0013}}
\and       P.A.~                       Palicio\orcit{0000-0002-7432-8709}\inst{\ref{inst:0015}}
\and         L.~               Pallas-Quintela\orcit{0000-0001-9296-3100}\inst{\ref{inst:0005}}
\and         A.~                        Panahi\orcit{0000-0001-5850-4373}\inst{\ref{inst:0251}}
\and         S.~               Payne-Wardenaar\inst{\ref{inst:0028}}
\and         X.~         Pe\~{n}alosa Esteller\inst{\ref{inst:0034}}
\and         A.~                 Penttil\"{ a}\orcit{0000-0001-7403-1721}\inst{\ref{inst:0126}}
\and         B.~                        Pichon\orcit{0000 0000 0062 1449}\inst{\ref{inst:0015}}
\and       A.M.~                    Piersimoni\orcit{0000-0002-8019-3708}\inst{\ref{inst:0232}}
\and      F.-X.~                        Pineau\orcit{0000-0002-2335-4499}\inst{\ref{inst:0107}}
\and         E.~                        Plachy\orcit{0000-0002-5481-3352}\inst{\ref{inst:0144},\ref{inst:0374},\ref{inst:0145}}
\and         G.~                          Plum\inst{\ref{inst:0025}}
\and         E.~                        Poggio\orcit{0000-0003-3793-8505}\inst{\ref{inst:0015},\ref{inst:0012}}
\and         A.~                      Pr\v{s}a\orcit{0000-0002-1913-0281}\inst{\ref{inst:0412}}
\and         L.~                        Pulone\orcit{0000-0002-5285-998X}\inst{\ref{inst:0123}}
\and         E.~                        Racero\orcit{0000-0002-6101-9050}\inst{\ref{inst:0092},\ref{inst:0390}}
\and         S.~                       Ragaini\inst{\ref{inst:0055}}
\and         M.~                        Rainer\orcit{0000-0002-8786-2572}\inst{\ref{inst:0042},\ref{inst:0418}}
\and         P.~                         Ramos\orcit{0000-0002-5080-7027}\inst{\ref{inst:0034},\ref{inst:0107}}
\and         M.~                  Ramos-Lerate\inst{\ref{inst:0002}}
\and         P.~                  Re Fiorentin\orcit{0000-0002-4995-0475}\inst{\ref{inst:0012}}
\and         S.~                        Regibo\inst{\ref{inst:0147}}
\and       P.J.~                      Richards\inst{\ref{inst:0424}}
\and         C.~                     Rios Diaz\inst{\ref{inst:0016}}
\and         V.~                        Ripepi\orcit{0000-0003-1801-426X}\inst{\ref{inst:0329}}
\and         A.~                          Riva\orcit{0000-0002-6928-8589}\inst{\ref{inst:0012}}
\and      H.-W.~                           Rix\orcit{0000-0003-4996-9069}\inst{\ref{inst:0001}}
\and         G.~                         Rixon\orcit{0000-0003-4399-6568}\inst{\ref{inst:0013}}
\and         N.~                      Robichon\orcit{0000-0003-4545-7517}\inst{\ref{inst:0025}}
\and       A.C.~                         Robin\orcit{0000-0001-8654-9499}\inst{\ref{inst:0133}}
\and         C.~                         Robin\inst{\ref{inst:0172}}
\and         M.~                       Roelens\orcit{0000-0003-0876-4673}\inst{\ref{inst:0031}}
\and     H.R.O.~                        Rogues\inst{\ref{inst:0242}}
\and         L.~                    Rohrbasser\inst{\ref{inst:0009}}
\and         M.~              Romero-G\'{o}mez\orcit{0000-0003-3936-1025}\inst{\ref{inst:0034}}
\and         N.~                        Rowell\orcit{0000-0003-3809-1895}\inst{\ref{inst:0096}}
\and         F.~                         Royer\orcit{0000-0002-9374-8645}\inst{\ref{inst:0025}}
\and         D.~                    Ruz Mieres\orcit{0000-0002-9455-157X}\inst{\ref{inst:0013}}
\and       K.A.~                       Rybicki\orcit{0000-0002-9326-9329}\inst{\ref{inst:0319}}
\and         G.~                      Sadowski\orcit{0000-0002-3411-1003}\inst{\ref{inst:0040}}
\and         A.~        S\'{a}ez N\'{u}\~{n}ez\inst{\ref{inst:0034}}
\and         A.~       Sagrist\`{a} Sell\'{e}s\orcit{0000-0001-6191-2028}\inst{\ref{inst:0028}}
\and         J.~                      Sahlmann\orcit{0000-0001-9525-3673}\inst{\ref{inst:0016}}
\and         E.~                      Salguero\inst{\ref{inst:0101}}
\and         N.~                       Samaras\orcit{0000-0001-8375-6652}\inst{\ref{inst:0059},\ref{inst:0447}}
\and         V.~               Sanchez Gimenez\orcit{0000-0003-1797-3557}\inst{\ref{inst:0034}}
\and         N.~                         Sanna\orcit{0000-0001-9275-9492}\inst{\ref{inst:0042}}
\and         R.~                 Santove\~{n}a\orcit{0000-0002-9257-2131}\inst{\ref{inst:0005}}
\and         M.~                       Sarasso\orcit{0000-0001-5121-0727}\inst{\ref{inst:0012}}
\and       M.~                    Schultheis\orcit{0000-0002-6590-1657}\inst{\ref{inst:0015}}
\and         E.~                       Sciacca\orcit{0000-0002-5574-2787}\inst{\ref{inst:0119}}
\and         M.~                         Segol\inst{\ref{inst:0242}}
\and       J.C.~                       Segovia\inst{\ref{inst:0092}}
\and         D.~                 S\'{e}gransan\orcit{0000-0003-2355-8034}\inst{\ref{inst:0031}}
\and         D.~                        Semeux\inst{\ref{inst:0248}}
\and         S.~                        Shahaf\orcit{0000-0001-9298-8068}\inst{\ref{inst:0458}}
\and       H.I.~                      Siddiqui\orcit{0000-0003-1853-6033}\inst{\ref{inst:0459}}
\and         A.~                       Siebert\orcit{0000-0001-8059-2840}\inst{\ref{inst:0107},\ref{inst:0255}}
\and         L.~                       Siltala\orcit{0000-0002-6938-794X}\inst{\ref{inst:0126}}
\and         A.~                       Silvelo\orcit{0000-0002-5126-6365}\inst{\ref{inst:0005}}
\and         E.~                        Slezak\inst{\ref{inst:0015}}
\and         I.~                        Slezak\inst{\ref{inst:0015}}
\and       R.L.~                         Smart\orcit{0000-0002-4424-4766}\inst{\ref{inst:0012}}
\and       O.N.~                        Snaith\inst{\ref{inst:0025}}
\and         E.~                        Solano\inst{\ref{inst:0468}}
\and         F.~                       Solitro\inst{\ref{inst:0063}}
\and         D.~                        Souami\orcit{0000-0003-4058-0815}\inst{\ref{inst:0308},\ref{inst:0471}}
\and         J.~                       Souchay\inst{\ref{inst:0079}}
\and         A.~                        Spagna\orcit{0000-0003-1732-2412}\inst{\ref{inst:0012}}
\and         L.~                         Spina\orcit{0000-0002-9760-6249}\inst{\ref{inst:0022}}
\and         F.~                         Spoto\orcit{0000-0001-7319-5847}\inst{\ref{inst:0183}}
\and       I.A.~                        Steele\orcit{0000-0001-8397-5759}\inst{\ref{inst:0349}}
\and         H.~            Steidelm\"{ u}ller\inst{\ref{inst:0008}}
\and       C.A.~                    Stephenson\inst{\ref{inst:0002},\ref{inst:0479}}
\and         M.~                  S\"{ u}veges\orcit{0000-0003-3017-5322}\inst{\ref{inst:0480}}
\and         J.~                        Surdej\orcit{0000-0002-7005-1976}\inst{\ref{inst:0003},\ref{inst:0482}}
\and         L.~                      Szabados\orcit{0000-0002-2046-4131}\inst{\ref{inst:0144}}
\and         E.~                  Szegedi-Elek\orcit{0000-0001-7807-6644}\inst{\ref{inst:0144}}
\and         F.~                         Taris\inst{\ref{inst:0079}}
\and       M.B.~                        Taylor\orcit{0000-0002-4209-1479}\inst{\ref{inst:0486}}
\and         L.~                       Tolomei\orcit{0000-0002-3541-3230}\inst{\ref{inst:0063}}
\and         N.~                       Tonello\orcit{0000-0003-0550-1667}\inst{\ref{inst:0272}}
\and         F.~                         Torra\orcit{0000-0002-8429-299X}\inst{\ref{inst:0056}}
\and         J.~               Torra$^\dagger$\inst{\ref{inst:0034}}
\and         G.~                Torralba Elipe\orcit{0000-0001-8738-194X}\inst{\ref{inst:0005}}
\and         M.~                     Trabucchi\orcit{0000-0002-1429-2388}\inst{\ref{inst:0492},\ref{inst:0031}}
\and       A.T.~                       Tsounis\inst{\ref{inst:0494}}
\and         C.~                         Turon\orcit{0000-0003-1236-5157}\inst{\ref{inst:0025}}
\and         A.~                          Ulla\orcit{0000-0001-6424-5005}\inst{\ref{inst:0496}}
\and         N.~                         Unger\orcit{0000-0003-3993-7127}\inst{\ref{inst:0031}}
\and       M.V.~                      Vaillant\inst{\ref{inst:0172}}
\and         E.~                    van Dillen\inst{\ref{inst:0242}}
\and         W.~                    van Reeven\inst{\ref{inst:0500}}
\and         O.~                         Vanel\orcit{0000-0002-7898-0454}\inst{\ref{inst:0025}}
\and         A.~                     Vecchiato\orcit{0000-0003-1399-5556}\inst{\ref{inst:0012}}
\and         Y.~                         Viala\inst{\ref{inst:0025}}
\and         D.~                       Vicente\orcit{0000-0002-1584-1182}\inst{\ref{inst:0272}}
\and         S.~                     Voutsinas\inst{\ref{inst:0096}}
\and         M.~                        Weiler\inst{\ref{inst:0034}}
\and         T.~                        Wevers\orcit{0000-0002-4043-9400}\inst{\ref{inst:0013},\ref{inst:0508}}
\and      \L{}.~                   Wyrzykowski\orcit{0000-0002-9658-6151}\inst{\ref{inst:0319}}
\and         A.~                        Yoldas\inst{\ref{inst:0013}}
\and         P.~                         Yvard\inst{\ref{inst:0242}}
\and         H.~                          Zhao\orcit{0000-0003-2645-6869}\inst{\ref{inst:0015}}
\and         J.~                         Zorec\inst{\ref{inst:0513}}
\and         S.~                        Zucker\orcit{0000-0003-3173-3138}\inst{\ref{inst:0182}}
\and         T.~                       Zwitter\orcit{0000-0002-2325-8763}\inst{\ref{inst:0515}}
}
\institute{
     Max Planck Institute for Astronomy, K\"{ o}nigstuhl 17, 69117 Heidelberg, Germany\relax                                                                                                                                                                                                                                                                       \label{inst:0001}
\and Telespazio UK S.L. for European Space Agency (ESA), Camino bajo del Castillo, s/n, Urbanizacion Villafranca del Castillo, Villanueva de la Ca\~{n}ada, 28692 Madrid, Spain\relax                                                                                                                                                                              \label{inst:0002}\vfill
\and Institut d'Astrophysique et de G\'{e}ophysique, Universit\'{e} de Li\`{e}ge, 19c, All\'{e}e du 6 Ao\^{u}t, B-4000 Li\`{e}ge, Belgium\relax                                                                                                                                                                                                                    \label{inst:0003}\vfill
\and Laboratoire d'astrophysique de Bordeaux, Univ. Bordeaux, CNRS, B18N, all{\'e}e Geoffroy Saint-Hilaire, 33615 Pessac, France\relax                                                                                                                                                                                                                             \label{inst:0004}\vfill
\and CIGUS CITIC - Department of Computer Science and Information Technologies, University of A Coru\~{n}a, Campus de Elvi\~{n}a s/n, A Coru\~{n}a, 15071, Spain\relax                                                                                                                                                                                             \label{inst:0005}\vfill
\and Department of Astrophysics, Astronomy and Mechanics, National and Kapodistrian University of Athens, Panepistimiopolis, Zografos, 15783 Athens, Greece\relax                                                                                                                                                                                                  \label{inst:0006}\vfill
\and National Observatory of Athens, I. Metaxa and Vas. Pavlou, Palaia Penteli, 15236 Athens, Greece\relax                                                                                                                                                                                                                                                         \label{inst:0007}\vfill
\and Lohrmann Observatory, Technische Universit\"{ a}t Dresden, Mommsenstra{\ss}e 13, 01062 Dresden, Germany\relax                                                                                                                                                                                                                                                 \label{inst:0008}\vfill
\and Department of Astronomy, University of Geneva, Chemin d'Ecogia 16, 1290 Versoix, Switzerland\relax                                                                                                                                                                                                                                                            \label{inst:0009}\vfill
\and Dpto. de Matem\'{a}tica Aplicada y Ciencias de la Computaci\'{o}n, Univ. de Cantabria, ETS Ingenieros de Caminos, Canales y Puertos, Avda. de los Castros s/n, 39005 Santander, Spain\relax                                                                                                                                                                   \label{inst:0011}\vfill
\and INAF - Osservatorio Astrofisico di Torino, via Osservatorio 20, 10025 Pino Torinese (TO), Italy\relax                                                                                                                                                                                                                                                         \label{inst:0012}\vfill
\and Institute of Astronomy, University of Cambridge, Madingley Road, Cambridge CB3 0HA, United Kingdom\relax                                                                                                                                                                                                                                                      \label{inst:0013}\vfill
\and Universit\'{e} C\^{o}te d'Azur, Observatoire de la C\^{o}te d'Azur, CNRS, Laboratoire Lagrange, Bd de l'Observatoire, CS 34229, 06304 Nice Cedex 4, France\relax                                                                                                                                                                                              \label{inst:0015}\vfill
\and RHEA for European Space Agency (ESA), Camino bajo del Castillo, s/n, Urbanizacion Villafranca del Castillo, Villanueva de la Ca\~{n}ada, 28692 Madrid, Spain\relax                                                                                                                                                                                            \label{inst:0016}\vfill
\and CENTRA, Faculdade de Ci\^{e}ncias, Universidade de Lisboa, Edif. C8, Campo Grande, 1749-016 Lisboa, Portugal\relax                                                                                                                                                                                                                                            \label{inst:0017}\vfill
\and Department of Informatics, Donald Bren School of Information and Computer Sciences, University of California, Irvine, 5226 Donald Bren Hall, 92697-3440 CA Irvine, United States\relax                                                                                                                                                                        \label{inst:0018}\vfill
\and Instituto de Astronomia, Geof\`{i}sica e Ci\^{e}ncias Atmosf\'{e}ricas, Universidade de S\~{a}o Paulo, Rua do Mat\~{a}o, 1226, Cidade Universitaria, 05508-900 S\~{a}o Paulo, SP, Brazil\relax                                                                                                                                                                \label{inst:0020}\vfill
\and Leiden Observatory, Leiden University, Niels Bohrweg 2, 2333 CA Leiden, The Netherlands\relax                                                                                                                                                                                                                                                                 \label{inst:0021}\vfill
\and INAF - Osservatorio astronomico di Padova, Vicolo Osservatorio 5, 35122 Padova, Italy\relax                                                                                                                                                                                                                                                                   \label{inst:0022}\vfill
\and European Space Agency (ESA), European Space Research and Technology Centre (ESTEC), Keplerlaan 1, 2201AZ, Noordwijk, The Netherlands\relax                                                                                                                                                                                                                    \label{inst:0023}\vfill
\and GEPI, Observatoire de Paris, Universit\'{e} PSL, CNRS, 5 Place Jules Janssen, 92190 Meudon, France\relax                                                                                                                                                                                                                                                      \label{inst:0025}\vfill
\and Univ. Grenoble Alpes, CNRS, IPAG, 38000 Grenoble, France\relax                                                                                                                                                                                                                                                                                                \label{inst:0026}\vfill
\and Astronomisches Rechen-Institut, Zentrum f\"{ u}r Astronomie der Universit\"{ a}t Heidelberg, M\"{ o}nchhofstr. 12-14, 69120 Heidelberg, Germany\relax                                                                                                                                                                                                         \label{inst:0028}\vfill
\and Department of Astronomy, University of Geneva, Chemin Pegasi 51, 1290 Versoix, Switzerland\relax                                                                                                                                                                                                                                                              \label{inst:0031}\vfill
\and European Space Agency (ESA), European Space Astronomy Centre (ESAC), Camino bajo del Castillo, s/n, Urbanizacion Villafranca del Castillo, Villanueva de la Ca\~{n}ada, 28692 Madrid, Spain\relax                                                                                                                                                             \label{inst:0032}\vfill
\and Aurora Technology for European Space Agency (ESA), Camino bajo del Castillo, s/n, Urbanizacion Villafranca del Castillo, Villanueva de la Ca\~{n}ada, 28692 Madrid, Spain\relax                                                                                                                                                                               \label{inst:0033}\vfill
\and Institut de Ci\`{e}ncies del Cosmos (ICCUB), Universitat  de  Barcelona  (IEEC-UB), Mart\'{i} i  Franqu\`{e}s  1, 08028 Barcelona, Spain\relax                                                                                                                                                                                                                \label{inst:0034}\vfill
\and Lund Observatory, Department of Astronomy and Theoretical Physics, Lund University, Box 43, 22100 Lund, Sweden\relax                                                                                                                                                                                                                                          \label{inst:0036}\vfill
\and CNES Centre Spatial de Toulouse, 18 avenue Edouard Belin, 31401 Toulouse Cedex 9, France\relax                                                                                                                                                                                                                                                                \label{inst:0039}\vfill
\and Institut d'Astronomie et d'Astrophysique, Universit\'{e} Libre de Bruxelles CP 226, Boulevard du Triomphe, 1050 Brussels, Belgium\relax                                                                                                                                                                                                                       \label{inst:0040}\vfill
\and F.R.S.-FNRS, Rue d'Egmont 5, 1000 Brussels, Belgium\relax                                                                                                                                                                                                                                                                                                     \label{inst:0041}\vfill
\and INAF - Osservatorio Astrofisico di Arcetri, Largo Enrico Fermi 5, 50125 Firenze, Italy\relax                                                                                                                                                                                                                                                                  \label{inst:0042}\vfill
\and European Space Agency (ESA, retired)\relax                                                                                                                                                                                                                                                                                                                    \label{inst:0049}\vfill
\and University of Turin, Department of Physics, Via Pietro Giuria 1, 10125 Torino, Italy\relax                                                                                                                                                                                                                                                                    \label{inst:0052}\vfill
\and INAF - Osservatorio di Astrofisica e Scienza dello Spazio di Bologna, via Piero Gobetti 93/3, 40129 Bologna, Italy\relax                                                                                                                                                                                                                                      \label{inst:0055}\vfill
\and DAPCOM for Institut de Ci\`{e}ncies del Cosmos (ICCUB), Universitat  de  Barcelona  (IEEC-UB), Mart\'{i} i  Franqu\`{e}s  1, 08028 Barcelona, Spain\relax                                                                                                                                                                                                     \label{inst:0056}\vfill
\and Royal Observatory of Belgium, Ringlaan 3, 1180 Brussels, Belgium\relax                                                                                                                                                                                                                                                                                        \label{inst:0059}\vfill
\and Observational Astrophysics, Division of Astronomy and Space Physics, Department of Physics and Astronomy, Uppsala University, Box 516, 751 20 Uppsala, Sweden\relax                                                                                                                                                                                           \label{inst:0061}\vfill
\and ALTEC S.p.a, Corso Marche, 79,10146 Torino, Italy\relax                                                                                                                                                                                                                                                                                                       \label{inst:0063}\vfill
\and S\`{a}rl, Geneva, Switzerland\relax                                                                                                                                                                                                                                                                                                                           \label{inst:0066}\vfill
\and Mullard Space Science Laboratory, University College London, Holmbury St Mary, Dorking, Surrey RH5 6NT, United Kingdom\relax                                                                                                                                                                                                                                  \label{inst:0072}\vfill
\and Gaia DPAC Project Office, ESAC, Camino bajo del Castillo, s/n, Urbanizacion Villafranca del Castillo, Villanueva de la Ca\~{n}ada, 28692 Madrid, Spain\relax                                                                                                                                                                                                  \label{inst:0075}\vfill
\and SYRTE, Observatoire de Paris, Universit\'{e} PSL, CNRS,  Sorbonne Universit\'{e}, LNE, 61 avenue de l'Observatoire 75014 Paris, France\relax                                                                                                                                                                                                                  \label{inst:0079}\vfill
\and IMCCE, Observatoire de Paris, Universit\'{e} PSL, CNRS, Sorbonne Universit{\'e}, Univ. Lille, 77 av. Denfert-Rochereau, 75014 Paris, France\relax                                                                                                                                                                                                             \label{inst:0084}\vfill
\and Serco Gesti\'{o}n de Negocios for European Space Agency (ESA), Camino bajo del Castillo, s/n, Urbanizacion Villafranca del Castillo, Villanueva de la Ca\~{n}ada, 28692 Madrid, Spain\relax                                                                                                                                                                   \label{inst:0092}\vfill
\and CRAAG - Centre de Recherche en Astronomie, Astrophysique et G\'{e}ophysique, Route de l'Observatoire Bp 63 Bouzareah 16340 Algiers, Algeria\relax                                                                                                                                                                                                             \label{inst:0095}\vfill
\and Institute for Astronomy, University of Edinburgh, Royal Observatory, Blackford Hill, Edinburgh EH9 3HJ, United Kingdom\relax                                                                                                                                                                                                                                  \label{inst:0096}\vfill
\and ATG Europe for European Space Agency (ESA), Camino bajo del Castillo, s/n, Urbanizacion Villafranca del Castillo, Villanueva de la Ca\~{n}ada, 28692 Madrid, Spain\relax                                                                                                                                                                                      \label{inst:0101}\vfill
\and Universit\'{e} de Strasbourg, CNRS, Observatoire astronomique de Strasbourg, UMR 7550,  11 rue de l'Universit\'{e}, 67000 Strasbourg, France\relax                                                                                                                                                                                                            \label{inst:0107}\vfill
\and Kavli Institute for Cosmology Cambridge, Institute of Astronomy, Madingley Road, Cambridge, CB3 0HA\relax                                                                                                                                                                                                                                                     \label{inst:0110}\vfill
\and Leibniz Institute for Astrophysics Potsdam (AIP), An der Sternwarte 16, 14482 Potsdam, Germany\relax                                                                                                                                                                                                                                                          \label{inst:0116}\vfill
\and INAF - Osservatorio Astrofisico di Catania, via S. Sofia 78, 95123 Catania, Italy\relax                                                                                                                                                                                                                                                                       \label{inst:0119}\vfill
\and Dipartimento di Fisica e Astronomia "Ettore Majorana", Universit\`{a} di Catania, Via S. Sofia 64, 95123 Catania, Italy\relax                                                                                                                                                                                                                               \label{inst:0120}\vfill
\and INAF - Osservatorio Astronomico di Roma, Via Frascati 33, 00078 Monte Porzio Catone (Roma), Italy\relax                                                                                                                                                                                                                                                       \label{inst:0123}\vfill
\and Space Science Data Center - ASI, Via del Politecnico SNC, 00133 Roma, Italy\relax                                                                                                                                                                                                                                                                             \label{inst:0124}\vfill
\and Department of Physics, University of Helsinki, P.O. Box 64, 00014 Helsinki, Finland\relax                                                                                                                                                                                                                                                                     \label{inst:0126}\vfill
\and Finnish Geospatial Research Institute FGI, Geodeetinrinne 2, 02430 Masala, Finland\relax                                                                                                                                                                                                                                                                      \label{inst:0127}\vfill
\and Institut UTINAM CNRS UMR6213, Universit\'{e} Bourgogne Franche-Comt\'{e}, OSU THETA Franche-Comt\'{e} Bourgogne, Observatoire de Besan\c{c}on, BP1615, 25010 Besan\c{c}on Cedex, France\relax                                                                                                                                                                 \label{inst:0133}\vfill
\and HE Space Operations BV for European Space Agency (ESA), Keplerlaan 1, 2201AZ, Noordwijk, The Netherlands\relax                                                                                                                                                                                                                                                \label{inst:0135}\vfill
\and Dpto. de Inteligencia Artificial, UNED, c/ Juan del Rosal 16, 28040 Madrid, Spain\relax                                                                                                                                                                                                                                                                       \label{inst:0137}\vfill
\and Konkoly Observatory, Research Centre for Astronomy and Earth Sciences, E\"{ o}tv\"{ o}s Lor{\'a}nd Research Network (ELKH), MTA Centre of Excellence, Konkoly Thege Mikl\'{o}s \'{u}t 15-17, 1121 Budapest, Hungary\relax                                                                                                                                     \label{inst:0144}\vfill
\and ELTE E\"{ o}tv\"{ o}s Lor\'{a}nd University, Institute of Physics, 1117, P\'{a}zm\'{a}ny P\'{e}ter s\'{e}t\'{a}ny 1A, Budapest, Hungary\relax                                                                                                                                                                                                                 \label{inst:0145}\vfill
\and Instituut voor Sterrenkunde, KU Leuven, Celestijnenlaan 200D, 3001 Leuven, Belgium\relax                                                                                                                                                                                                                                                                      \label{inst:0147}\vfill
\and Department of Astrophysics/IMAPP, Radboud University, P.O.Box 9010, 6500 GL Nijmegen, The Netherlands\relax                                                                                                                                                                                                                                                   \label{inst:0148}\vfill
\and University of Vienna, Department of Astrophysics, T\"{ u}rkenschanzstra{\ss}e 17, A1180 Vienna, Austria\relax                                                                                                                                                                                                                                                 \label{inst:0156}\vfill
\and Institute of Physics, Laboratory of Astrophysics, Ecole Polytechnique F\'ed\'erale de Lausanne (EPFL), Observatoire de Sauverny, 1290 Versoix, Switzerland\relax                                                                                                                                                                                              \label{inst:0157}\vfill
\and Kapteyn Astronomical Institute, University of Groningen, Landleven 12, 9747 AD Groningen, The Netherlands\relax                                                                                                                                                                                                                                               \label{inst:0163}\vfill
\and School of Physics and Astronomy / Space Park Leicester, University of Leicester, University Road, Leicester LE1 7RH, United Kingdom\relax                                                                                                                                                                                                                     \label{inst:0170}\vfill
\and Thales Services for CNES Centre Spatial de Toulouse, 18 avenue Edouard Belin, 31401 Toulouse Cedex 9, France\relax                                                                                                                                                                                                                                            \label{inst:0172}\vfill
\and Depto. Estad\'istica e Investigaci\'on Operativa. Universidad de C\'adiz, Avda. Rep\'ublica Saharaui s/n, 11510 Puerto Real, C\'adiz, Spain\relax                                                                                                                                                                                                             \label{inst:0176}\vfill
\and Center for Research and Exploration in Space Science and Technology, University of Maryland Baltimore County, 1000 Hilltop Circle, Baltimore MD, USA\relax                                                                                                                                                                                                    \label{inst:0178}\vfill
\and GSFC - Goddard Space Flight Center, Code 698, 8800 Greenbelt Rd, 20771 MD Greenbelt, United States\relax                                                                                                                                                                                                                                                      \label{inst:0179}\vfill
\and EURIX S.r.l., Corso Vittorio Emanuele II 61, 10128, Torino, Italy\relax                                                                                                                                                                                                                                                                                       \label{inst:0181}\vfill
\and Porter School of the Environment and Earth Sciences, Tel Aviv University, Tel Aviv 6997801, Israel\relax                                                                                                                                                                                                                                                      \label{inst:0182}\vfill
\and Harvard-Smithsonian Center for Astrophysics, 60 Garden St., MS 15, Cambridge, MA 02138, USA\relax                                                                                                                                                                                                                                                             \label{inst:0183}\vfill
\and HE Space Operations BV for European Space Agency (ESA), Camino bajo del Castillo, s/n, Urbanizacion Villafranca del Castillo, Villanueva de la Ca\~{n}ada, 28692 Madrid, Spain\relax                                                                                                                                                                          \label{inst:0185}\vfill
\and Instituto de Astrof\'{i}sica e Ci\^{e}ncias do Espa\c{c}o, Universidade do Porto, CAUP, Rua das Estrelas, PT4150-762 Porto, Portugal\relax                                                                                                                                                                                                                    \label{inst:0186}\vfill
\and LFCA/DAS,Universidad de Chile,CNRS,Casilla 36-D, Santiago, Chile\relax                                                                                                                                                                                                                                                                                        \label{inst:0188}\vfill
\and SISSA - Scuola Internazionale Superiore di Studi Avanzati, via Bonomea 265, 34136 Trieste, Italy\relax                                                                                                                                                                                                                                                        \label{inst:0192}\vfill
\and Telespazio for CNES Centre Spatial de Toulouse, 18 avenue Edouard Belin, 31401 Toulouse Cedex 9, France\relax                                                                                                                                                                                                                                                 \label{inst:0197}\vfill
\and University of Turin, Department of Computer Sciences, Corso Svizzera 185, 10149 Torino, Italy\relax                                                                                                                                                                                                                                                           \label{inst:0201}\vfill
\and Centro de Astronom\'{i}a - CITEVA, Universidad de Antofagasta, Avenida Angamos 601, Antofagasta 1270300, Chile\relax                                                                                                                                                                                                                                          \label{inst:0212}\vfill
\and DLR Gesellschaft f\"{ u}r Raumfahrtanwendungen (GfR) mbH M\"{ u}nchener Stra{\ss}e 20 , 82234 We{\ss}ling\relax                                                                                                                                                                                                                                               \label{inst:0217}\vfill
\and Centre for Astrophysics Research, University of Hertfordshire, College Lane, AL10 9AB, Hatfield, United Kingdom\relax                                                                                                                                                                                                                                         \label{inst:0219}\vfill
\and University of Turin, Mathematical Department "G.Peano", Via Carlo Alberto 10, 10123 Torino, Italy\relax                                                                                                                                                                                                                                                     \label{inst:0226}\vfill
\and INAF - Osservatorio Astronomico d'Abruzzo, Via Mentore Maggini, 64100 Teramo, Italy\relax                                                                                                                                                                                                                                                                     \label{inst:0232}\vfill
\and APAVE SUDEUROPE SAS for CNES Centre Spatial de Toulouse, 18 avenue Edouard Belin, 31401 Toulouse Cedex 9, France\relax                                                                                                                                                                                                                                        \label{inst:0242}\vfill
\and M\'{e}socentre de calcul de Franche-Comt\'{e}, Universit\'{e} de Franche-Comt\'{e}, 16 route de Gray, 25030 Besan\c{c}on Cedex, France\relax                                                                                                                                                                                                                  \label{inst:0244}\vfill
\and ATOS for CNES Centre Spatial de Toulouse, 18 avenue Edouard Belin, 31401 Toulouse Cedex 9, France\relax                                                                                                                                                                                                                                                       \label{inst:0248}\vfill
\and School of Physics and Astronomy, Tel Aviv University, Tel Aviv 6997801, Israel\relax                                                                                                                                                                                                                                                                          \label{inst:0251}\vfill
\and Astrophysics Research Centre, School of Mathematics and Physics, Queen's University Belfast, Belfast BT7 1NN, UK\relax                                                                                                                                                                                                                                        \label{inst:0253}\vfill
\and Centre de Donn\'{e}es Astronomique de Strasbourg, Strasbourg, France\relax                                                                                                                                                                                                                                                                                    \label{inst:0255}\vfill
\and Institute for Computational Cosmology, Department of Physics, Durham University, Durham DH1 3LE, UK\relax                                                                                                                                                                                                                                                     \label{inst:0259}\vfill
\and European Southern Observatory, Karl-Schwarzschild-Str. 2, 85748 Garching, Germany\relax                                                                                                                                                                                                                                                                       \label{inst:0260}\vfill
\and Max-Planck-Institut f\"{ u}r Astrophysik, Karl-Schwarzschild-Stra{\ss}e 1, 85748 Garching, Germany\relax                                                                                                                                                                                                                                                      \label{inst:0261}\vfill
\and Data Science and Big Data Lab, Pablo de Olavide University, 41013, Seville, Spain\relax                                                                                                                                                                                                                                                                       \label{inst:0265}\vfill
\and Barcelona Supercomputing Center (BSC), Pla\c{c}a Eusebi G\"{ u}ell 1-3, 08034-Barcelona, Spain\relax                                                                                                                                                                                                                                                          \label{inst:0272}\vfill
\and ETSE Telecomunicaci\'{o}n, Universidade de Vigo, Campus Lagoas-Marcosende, 36310 Vigo, Galicia, Spain\relax                                                                                                                                                                                                                                                   \label{inst:0277}\vfill
\and Asteroid Engineering Laboratory, Space Systems, Lule\aa{} University of Technology, Box 848, S-981 28 Kiruna, Sweden\relax                                                                                                                                                                                                                                    \label{inst:0281}\vfill
\and Vera C Rubin Observatory,  950 N. Cherry Avenue, Tucson, AZ 85719, USA\relax                                                                                                                                                                                                                                                                                  \label{inst:0286}\vfill
\and TRUMPF Photonic Components GmbH, Lise-Meitner-Stra{\ss}e 13,  89081 Ulm, Germany\relax                                                                                                                                                                                                                                                                        \label{inst:0288}\vfill
\and IAC - Instituto de Astrofisica de Canarias, Via L\'{a}ctea s/n, 38200 La Laguna S.C., Tenerife, Spain\relax                                                                                                                                                                                                                                                   \label{inst:0293}\vfill
\and Department of Astrophysics, University of La Laguna, Via L\'{a}ctea s/n, 38200 La Laguna S.C., Tenerife, Spain\relax                                                                                                                                                                                                                                          \label{inst:0294}\vfill
\and Faculty of Aerospace Engineering, Delft University of Technology, Kluyverweg 1, 2629 HS Delft, The Netherlands\relax                                                                                                                                                                                                                                          \label{inst:0296}\vfill
\and Radagast Solutions\relax                                                                                                                                                                                                                                                                                                                                      \label{inst:0300}\vfill
\and Laboratoire Univers et Particules de Montpellier, CNRS Universit\'{e} Montpellier, Place Eug\`{e}ne Bataillon, CC72, 34095 Montpellier Cedex 05, France\relax                                                                                                                                                                                                 \label{inst:0301}\vfill
\and Universit\'{e} de Caen Normandie, C\^{o}te de Nacre Boulevard Mar\'{e}chal Juin, 14032 Caen, France\relax                                                                                                                                                                                                                                                     \label{inst:0307}\vfill
\and LESIA, Observatoire de Paris, Universit\'{e} PSL, CNRS, Sorbonne Universit\'{e}, Universit\'{e} de Paris, 5 Place Jules Janssen, 92190 Meudon, France\relax                                                                                                                                                                                                   \label{inst:0308}\vfill
\and SRON Netherlands Institute for Space Research, Niels Bohrweg 4, 2333 CA Leiden, The Netherlands\relax                                                                                                                                                                                                                                                         \label{inst:0318}\vfill
\and Astronomical Observatory, University of Warsaw,  Al. Ujazdowskie 4, 00-478 Warszawa, Poland\relax                                                                                                                                                                                                                                                             \label{inst:0319}\vfill
\and Scalian for CNES Centre Spatial de Toulouse, 18 avenue Edouard Belin, 31401 Toulouse Cedex 9, France\relax                                                                                                                                                                                                                                                    \label{inst:0321}\vfill
\and Universit\'{e} Rennes, CNRS, IPR (Institut de Physique de Rennes) - UMR 6251, 35000 Rennes, France\relax                                                                                                                                                                                                                                                      \label{inst:0327}\vfill
\and INAF - Osservatorio Astronomico di Capodimonte, Via Moiariello 16, 80131, Napoli, Italy\relax                                                                                                                                                                                                                                                                 \label{inst:0329}\vfill
\and Shanghai Astronomical Observatory, Chinese Academy of Sciences, 80 Nandan Road, Shanghai 200030, People's Republic of China\relax                                                                                                                                                                                                                             \label{inst:0332}\vfill
\and University of Chinese Academy of Sciences, No.19(A) Yuquan Road, Shijingshan District, Beijing 100049, People's Republic of China\relax                                                                                                                                                                                                                       \label{inst:0334}\vfill
\and Niels Bohr Institute, University of Copenhagen, Juliane Maries Vej 30, 2100 Copenhagen {\O}, Denmark\relax                                                                                                                                                                                                                                                    \label{inst:0337}\vfill
\and DXC Technology, Retortvej 8, 2500 Valby, Denmark\relax                                                                                                                                                                                                                                                                                                        \label{inst:0338}\vfill
\and Las Cumbres Observatory, 6740 Cortona Drive Suite 102, Goleta, CA 93117, USA\relax                                                                                                                                                                                                                                                                            \label{inst:0339}\vfill
\and CIGUS CITIC, Department of Nautical Sciences and Marine Engineering, University of A Coru\~{n}a, Paseo de Ronda 51, 15071, A Coru\~{n}a, Spain\relax                                                                                                                                                                                                          \label{inst:0348}\vfill
\and Astrophysics Research Institute, Liverpool John Moores University, 146 Brownlow Hill, Liverpool L3 5RF, United Kingdom\relax                                                                                                                                                                                                                                  \label{inst:0349}\vfill
\and IPAC, Mail Code 100-22, California Institute of Technology, 1200 E. California Blvd., Pasadena, CA 91125, USA\relax                                                                                                                                                                                                                                           \label{inst:0356}\vfill
\and IRAP, Universit\'{e} de Toulouse, CNRS, UPS, CNES, 9 Av. colonel Roche, BP 44346, 31028 Toulouse Cedex 4, France\relax                                                                                                                                                                                                                                        \label{inst:0357}\vfill
\and MTA CSFK Lend\"{ u}let Near-Field Cosmology Research Group, Konkoly Observatory, MTA Research Centre for Astronomy and Earth Sciences, Konkoly Thege Mikl\'{o}s \'{u}t 15-17, 1121 Budapest, Hungary\relax                                                                                                                                                    \label{inst:0374}\vfill
\and Departmento de F\'{i}sica de la Tierra y Astrof\'{i}sica, Universidad Complutense de Madrid, 28040 Madrid, Spain\relax                                                                                                                                                                                                                                        \label{inst:0390}\vfill
\and Ru{\dj}er Bo\v{s}kovi\'{c} Institute, Bijeni\v{c}ka cesta 54, 10000 Zagreb, Croatia\relax                                                                                                                                                                                                                                                                     \label{inst:0396}\vfill
\and Villanova University, Department of Astrophysics and Planetary Science, 800 E Lancaster Avenue, Villanova PA 19085, USA\relax                                                                                                                                                                                                                                 \label{inst:0412}\vfill
\and INAF - Osservatorio Astronomico di Brera, via E. Bianchi, 46, 23807 Merate (LC), Italy\relax                                                                                                                                                                                                                                                                  \label{inst:0418}\vfill
\and STFC, Rutherford Appleton Laboratory, Harwell, Didcot, OX11 0QX, United Kingdom\relax                                                                                                                                                                                                                                                                         \label{inst:0424}\vfill
\and Charles University, Faculty of Mathematics and Physics, Astronomical Institute of Charles University, V Holesovickach 2, 18000 Prague, Czech Republic\relax                                                                                                                                                                                                   \label{inst:0447}\vfill
\and Department of Particle Physics and Astrophysics, Weizmann Institute of Science, Rehovot 7610001, Israel\relax                                                                                                                                                                                                                                                 \label{inst:0458}\vfill
\and Department of Astrophysical Sciences, 4 Ivy Lane, Princeton University, Princeton NJ 08544, USA\relax                                                                                                                                                                                                                                                         \label{inst:0459}\vfill
\and Departamento de Astrof\'{i}sica, Centro de Astrobiolog\'{i}a (CSIC-INTA), ESA-ESAC. Camino Bajo del Castillo s/n. 28692 Villanueva de la Ca\~{n}ada, Madrid, Spain\relax                                                                                                                                                                                      \label{inst:0468}\vfill
\and naXys, University of Namur, Rempart de la Vierge, 5000 Namur, Belgium\relax                                                                                                                                                                                                                                                                                   \label{inst:0471}\vfill
\and CGI Deutschland B.V. \& Co. KG, Mornewegstr. 30, 64293 Darmstadt, Germany\relax                                                                                                                                                                                                                                                                               \label{inst:0479}\vfill
\and Institute of Global Health, University of Geneva\relax                                                                                                                                                                                                                                                                                                        \label{inst:0480}\vfill
\and Astronomical Observatory Institute, Faculty of Physics, Adam Mickiewicz University, Pozna\'{n}, Poland\relax                                                                                                                                                                                                                                                  \label{inst:0482}\vfill
\and H H Wills Physics Laboratory, University of Bristol, Tyndall Avenue, Bristol BS8 1TL, United Kingdom\relax                                                                                                                                                                                                                                                    \label{inst:0486}\vfill
\and Department of Physics and Astronomy G. Galilei, University of Padova, Vicolo dell'Osservatorio 3, 35122, Padova, Italy\relax                                                                                                                                                                                                                                  \label{inst:0492}\vfill
\and CERN, Geneva, Switzerland\relax                                                                                                                                                                                                                                                                                                                               \label{inst:0494}\vfill
\and Applied Physics Department, Universidade de Vigo, 36310 Vigo, Spain\relax                                                                                                                                                                                                                                                                                     \label{inst:0496}\vfill
\and Association of Universities for Research in Astronomy, 1331 Pennsylvania Ave. NW, Washington, DC 20004, USA\relax                                                                                                                                                                                                                                             \label{inst:0500}\vfill
\and European Southern Observatory, Alonso de C\'ordova 3107, Casilla 19, Santiago, Chile\relax                                                                                                                                                                                                                                                                    \label{inst:0508}\vfill
\and Sorbonne Universit\'{e}, CNRS, UMR7095, Institut d'Astrophysique de Paris, 98bis bd. Arago, 75014 Paris, France\relax                                                                                                                                                                                                                                         \label{inst:0513}\vfill
\and Faculty of Mathematics and Physics, University of Ljubljana, Jadranska ulica 19, 1000 Ljubljana, Slovenia\relax                                                                                                                                                                                                                                               \label{inst:0515}\vfill
}


\date{First submitted 31 January 2022. Resubmitted 27 April
  2022. Accepted 27 April 2022} \abstract{ The \gaia Galactic survey
  mission is designed and optimized to obtain astrometry, photometry,
  and spectroscopy of nearly two billion stars in our Galaxy. Yet as
  an all-sky multi-epoch survey, \gaia also observes several million
  extragalactic objects down to a magnitude of $\gmag\sim21$\,mag. Due
  to the nature of the \gaia onboard-selection algorithms, these are
  mostly point-source-like objects.  Using data provided by the
  satellite, we have identified quasar and galaxy candidates via
  supervised machine learning methods, and estimate their redshifts
  using the low resolution BP/RP spectra. We further characterise the
  surface brightness profiles of host galaxies of quasars and of
  galaxies from pre-defined input lists.  Here we give an overview of
  the processing of extragalactic objects, describe the data products
  in \gdr{3}, and analyse their properties. Two integrated tables
  contain the main results for a high completeness, but low purity
  \new{(50--70\%)}, set of 6.6 million candidate quasars and 4.8
  million candidate galaxies. We provide queries that select purer
  sub-samples of these containing 1.9 million probable quasars and 2.9
  million probable galaxies \new{(both $\sim$95\% purity)}. We also
  use high quality BP/RP spectra of 43 thousand high probability
  quasars over the redshift range 0.05--4.36 to construct a composite
  quasar spectrum spanning restframe wavelengths from 72--1000\,nm.  }
\keywords{}
\maketitle

\section{Introduction}\label{sec:introduction}

The primary objective of the \gaia\ mission is to study the structure and origin of our Galaxy by measuring the distribution, kinematics, and physical properties of its constituent stars  \citep{2016Prusti}. The satellite and its observing strategy were therefore designed to optimize the measurement of astrometry, photometry, and spectroscopy of point sources.
Nonetheless, by observing the entire sky multiple times down to a limiting magnitude of $G\simeq21$\,mag, \gaia\ has observed millions of extragalactic objects since it started observing in mid 2014. Various data on many of these objects are provided as part of the third \gaia\ data release (DR3), covering both previously-identified objects and new candidate objects identified using the \gaia\ data. The purpose of this paper is to summarize how extragalactic objects were identified, what their properties are, and what data on them are provided in \gdr{3}.
 
Extragalactic objects are classified or analysed by several modules in the \gaia data processing system. These modules were provided by different coordination units (CUs) within the Data Processing and Analysis Consortium (DPAC) and operate largely independently. They are as follows: CU3 Astrometry, which assembled a list of extragalactic point sources from external catalogues to use in defining the astrometric reference frame \citep{EDR3-DPACP-133};
CU4 Extended Objects (EO), which analyses the surface brightness profiles of \new{an input list of} objects to look for physical extension; CU7 Variability, which uses photometric light curves to characterise variability; CU8 Astrophysical Parameters, which uses astrometry, photometry, and the BP/RP spectra to classify objects and to estimate redshifts. 
Whereas the modules from CU3 and CU4 work on a predefined lists of extragalactic objects identified in other surveys, 
the Vari module in CU7 and the Discrete Source Classifier (DSC) module in CU8 use supervised machine learning to discover new objects. These classifiers use only \gaia\ data. The inclusion of additional data, such as infrared photometry, should improve the classification performance (sample completeness and purity). However, a key principle of the DPAC is to provide homogeneous classifications based only on the \gaia data, unaffected by issues with other catalogues, such as incompleteness.

It is important to realise that there is no common definition of quasar or galaxy across the various \gaia\ modules. A common definition is also not possible, because each module uses different data to classify or select objects, including different training sets. But broadly speaking, the term 'extragalactic' in the context of this paper refers to unresolved or barely resolved individual objects more than 50\,Mpc from the Sun.

If \gaia\ were to obtain noise-free, unbiased parallaxes, then identifying extragalactic objects would be simple: They would be all the objects with parallaxes below some threshold. Yet we do not have this luxury: Despite the high precision of \gdr{3} parallaxes -- around 0.5\,mas at $G=20$\,mag and 0.25\,mas at $G=19$\,mag \citep{2021A&A...649A...2L}
-- this is not nearly enough to reliably identify extragalactic objects through a simple cut on parallaxes (or proper motions). Indeed, 657 million objects in \gdr{3} have raw 
parallaxes below 0.25\,mas, the vast majority of which are of course stars in our Galaxy. This is not to say that parallaxes, and moreover proper motions, are not useful, however, and we do indeed make use of them in our classifications and analyses.

Most of the extragalactic candidates we have identified are bundled into two integrated tables in \gdr{3}, called {\tt qso\_candidates} and {\tt galaxy\_candidates}. As their names make clear, the construction of these tables has been driven primarily by the desire to be complete, rather than pure. 
Together these tables contain around 11.3 million unique objects and have \new{global purities of 50--70\%}, although they are significantly higher when we exclude the Galactic plane, high density regions around clusters and galaxies, and the faintest sources.
These tables are nonetheless a significant improvement over the  {\tt gaia\_source} table, which has 1.8 billion objects and an extragalactic purity of around 0.2\%. Our rationale for producing completeness-driven integrated tables is that it is easier for users to then select a sub-sample of purer objects (according to their own criteria) from our integrated tables, than it would be to find objects (in {\tt gaia\_source}) that had been removed from purity-driven tables. In Sect.~\ref{sec:howto_purer_samples} we recommend how to extract a purer sub-sample \new{($\sim$96\%)} from the two integrated tables.

This paper is not the first to deal with classifying extragalactic objects using \gaia\ data. 
Initial studies cross-matched \gaia\ positions to other catalogues to analyse the properties of quasars and galaxies \cite[e.g.][]{2018ApJS..236...37P,2019A&A...624A.145S}.
Several studies have made cuts on the astrometry \cite[e.g.][]{2018A&A...615L...8H,2018A&A...616A..14G}, sometimes combined with classification using non-\gaia data \cite[e.g.][]{2021ApJS..254....6F}, and others 
have applied machine learning methods to a number of Gaia metrics \cite[e.g.][]{2019MNRAS.490.5615B} to identify extragalactic objects. Purer samples should be attainable when combining \gaia\ data with more discriminatory data, albeit at the loss of completeness if \gaia\ is the larger survey, and some studies report good results here  \cite[e.g.][]{2021arXiv211102131W}.
Other studies have used the Gaia data to characterise specific types of extragalactic object, such as gravitational lenses \cite[e.g.][]{2018A&A...616L..11K,2019A&A...622A.165D}.

This paper is structured as follows. Section~\ref{sec:modules} summarizes the extragalactic processing modules that deliver results in \gdr{3} and Sect.~\ref{sec:tables} describes the various tables that provide these results. Section~\ref{sec:basic_properties} presents the properties of the extragalactic objects, such as sky distributions, spectra, surface brightnesses, and light curves. In Sect.~\ref{sec:internal_comparison} we provide some basic comparisons between the results of the different modules, and in Sect.~\ref{sec:external_comparison} we compare the results to  external surveys. 
In Sect.~\ref{sec:composite} we compute composite quasar spectra from individual quasar spectra at a range of redshifts. Section~\ref{sec:howto_purer_samples} describes a purer, and necessarily less complete, sub-sample of the integrated extragalactic tables. We conclude in Sect.~\ref{sec:conclusions} with some suggested use cases.

Many more details on the topics discussed here can be found in the extensive online documentation that accompanies this data release.\footnote{\url{\linktodoc}} We point in particular to the table and field descriptions there. Several other release papers provide details that are not in the documentation. These are \cite{DR3-DPACP-158} for the CU8 classification and redshift estimation modules,
\cite{DR3-DPACP-165} for the CU7 variability classifier and
\cite{DR3-DPACP-167} for the resulting selection of Active Galactic Nuclei (AGN),
\cite{DR3-DPACP-153} for the CU4 surface brightness profile analysis, and
\cite{EDR3-DPACP-133} for the \gaia-CRF3 (Celestial Reference Frame 3). Readers may also want to consult \cite{DR3-DPACP-118} for a description of the BP/RP spectrophotometry and \cite{2021A&A...649A...2L} for the astrometric (parallax and proper motion) processing (the latter unchanged from \gedr{3}).

\section{Extragalactic processing modules}\label{sec:modules}

The modules in the \gaia\ data processing system that deal explicitly with extragalactic objects are as follows.
DSC and Vari classify \gaia objects using supervised machine learning methods. Vari additionally provides characterisations of the light curves.
UGC (Unresolved Galaxy Classifier), QSOC (Quasar Classifier), and OA (Outlier Analyser) analyse the results from DSC, the first two computing redshifts. 
EO analyses the surface brightness profiles of an input source list. 
We summarize these modules here, leaving more detailed descriptions to the individual processing papers cited below.
We also include in our analysis the list of quasars identified for \gaia-CRF3.

Some sources, in particular galaxies, are partially resolved by \gaia. Their two-dimensional structure -- combined with the fact that \gaia\ observes sources over a range of position angles -- can induce a spurious (non-intrinsic) photometric variability or an apparent astrometric variability, the latter potentially being interpreted by the astrometric processing \citep{2021A&A...649A...2L} as spuriously large parallaxes and proper motions. The DSC and Vari modules take advantage of these spurious measurements to help them classify extragalactic sources.

\subsection{Discrete Source Classifier (CU8-DSC)}\label{sec:modules:dsc}

\begin{figure}
\begin{center}
\includegraphics[width=0.40\textwidth, angle=0]{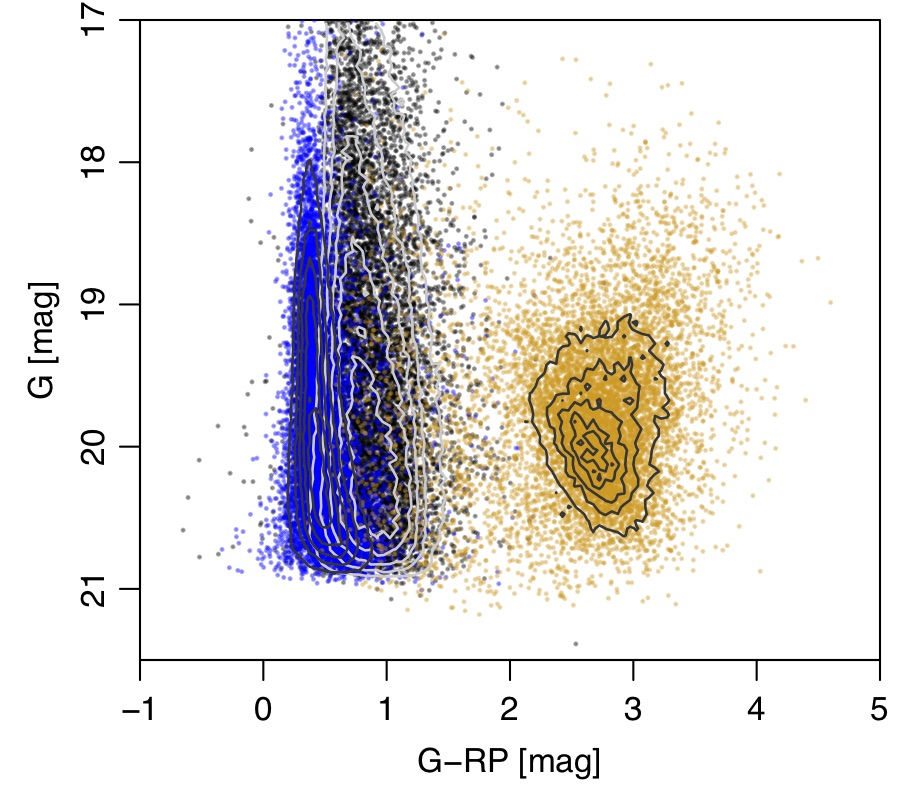}
\includegraphics[width=0.40\textwidth, angle=0]{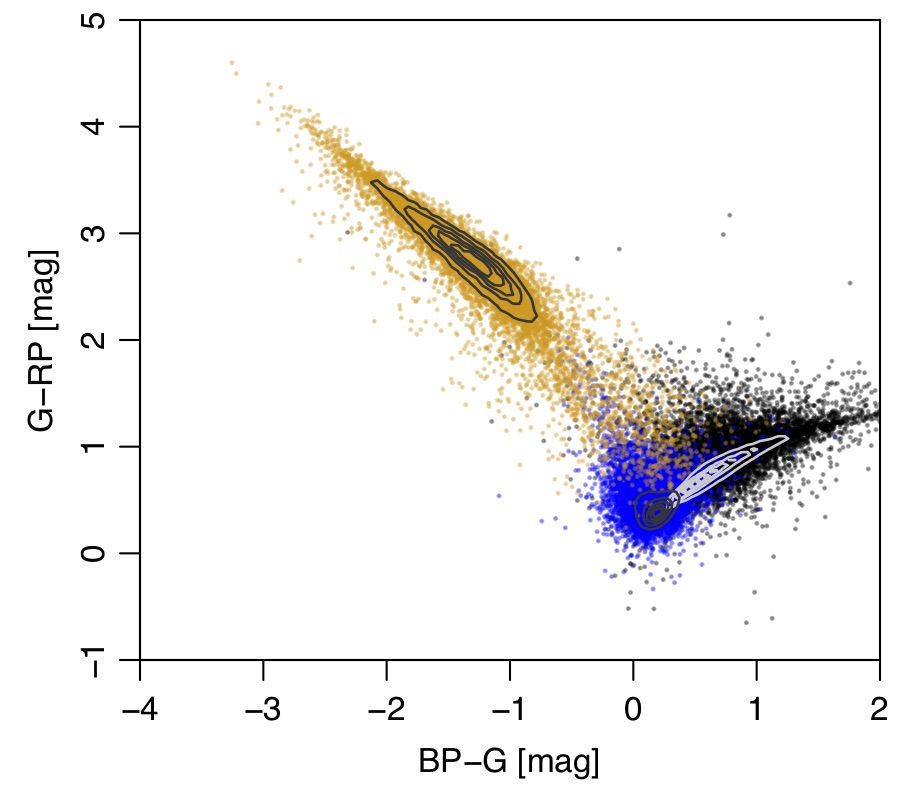}
\caption{Colour--magnitude diagram (top) and colour--colour diagram (bottom) of the DSC training data for the quasars (blue) and galaxies (orange) as well as stars  (black).  The contours in each panel show the variation in source density on a linear scale. The points are equal-sized random subsets of sources from each class. There is significant overlap, in particular between stars and quasars: in reality the former dominate by a factor of about a thousand, and so overlap much more than is shown here. Plots for each class separately are provided in the \linktosec{Data_analysis/chap_cu8par/sec_cu8par_apsis/ssec_cu8par_apsis_dsc.html}.
\label{fig:dsc_3classes_cmd}
}
\end{center}
\end{figure}

The Discrete Source Classifier uses the BP/RP spectrum together with the mean G-band magnitude, the variability in this band, the parallax, and the proper motion
to classify each \gaia source probabilistically into five classes: quasar; galaxy; anonymous (essentially single star); white dwarf; binary star. DSC is trained empirically on \gaia data with labels for the quasar and galaxy classes coming from Sloan Digital Sky Survey (SDSS) spectroscopic classifications.
The distributions of the training data in colour and magnitude are shown in Fig.~\ref{fig:dsc_3classes_cmd}.
The training data define the classes
(see Sect.~\ref{sec:external_comparison:quasars}), so these are not the same class definition adopted by other modules that contribute extragalactic source identifications to \gdr{3}.
DSC comprises three classifiers.
Specmod uses the BP/RP spectrum only and gives results for all five classes in DSC.
Allosmod uses various photometric and astrometric features and only gives results for quasars, galaxies, and single stars.
Specmod and Allosmod are nonetheless trained on a common set of data that has complete data for both classifiers. 
\new{One consequence of this is that Specmod is also applied to some types of sources it was not trained on, for example galaxies that lack measured parallaxes and proper motions.}
Combmod combines the Specmod and Allosmod classification probabilities in a Bayesian manner
to give probabilities for all five classes (using the algorithm described in the appendix of \citealt{DR3-DPACP-158}).
Probabilities from all three classifiers are provided in the {\tt astrophysical\_parameters} table.
DSC is described in more detail in~\cite{DR3-DPACP-158} and in the \linktosec{Data_analysis/chap_cu8par/sec_cu8par_apsis/ssec_cu8par_apsis_dsc.html}.

DSC incorporates a global class prior that reflects the rareness of quasars and galaxies.
This makes it hard to achieve a high purity even for a good classifier. For example, if only one in every thousand sources were extragalactic, then even if a classifier had a 99.9\% accuracy, the resulting sample would only be around 50\% pure. 
For this reason one must report results not on a balanced validation set, but on one that reflects this prior.\footnote{In practice we can use a validation set with more convenient class fractions, and then adjust the confusion matrix to reflect the prior, as explained in Sect.~3.4 of \cite{2019MNRAS.490.5615B}.}

In addition to posterior probabilities, DSC also provides two class labels. The first, {\tt classlabel\_dsc}, is assigned the name of the class that achieves the highest posterior probability in Combmod that is greater than 0.5. If none of the output probabilities are above 0.5 then this class label is {\tt unclassified}. This tends to produce a complete but impure sample of objects when we properly account for extragalactic rareness. The analyses in~\cite{DR3-DPACP-158} and~\cite{LL:CBJ-094} using SDSS spectroscopically-confirmed objects shows a completeness for quasars and galaxies objects of over 90\%, but a global purity of only about 20--25\%. For Galactic latitudes above 11.5\degree\ the purities increase to 41\%.
Additional filtering increases this further (see Sect.~\ref{sec:howto_purer_samples}).
The second class label, {\tt classlabel\_dsc\_joint} defines a purer set of quasars and galaxies, and is assigned by requiring both Specmod and Allosmod probabilities to be above 0.5 for the corresponding class. This gives 
completenesses of 38\% on quasars and 83\% on galaxies, and
purities on both classes of 63\%. For Galactic latitudes above 11.5\degree\ the purities increase to about 80\%.

\subsection{Quasar Classifier (CU8-QSOC)}\label{sec:modules:qsoc}

The Quasar Classifier module (QSOC) estimates the redshift of sources classified as quasars by DSC-Combmod using their BP/RP spectra. For this selection, QSOC uses a very loose cut on the DSC quasar probability, {\tt classprob\_dsc\_combmod\_quasar} 
$\geq 0.01$. This prioritizes completeness at the expense of purity to ensure that most of the objects that are suspected to be quasars are given a redshift estimate. The QSOC redshifts are inferred with a chi-square approach, whereby the BP and RP spectra are compared to a composite quasar spectrum taken at various trial redshifts in the range $0 \lesssim z \lesssim 6$. The composite spectrum is built upon a semi-empirical library of quasars from the SDSS DR12Q sample \citep{2017Paris}. Each SDSS spectrum is first extrapolated to the wavelength range covered by BP/RP before being converted into a BP/RP spectrum using the available instrument model. More details of the algorithm can be found in \cite{DR3-DPACP-158}. In addition to the best point estimate of the redshift, QSOC also estimates lower and upper confidence intervals, {\tt redshift\_qsoc\_lower} and  {\tt redshift\_qsoc\_upper}, which are the $15.9\%$ and $84.1\%$ quantiles of a log-normal distribution. The module also sets various processing flags in {\tt flags\_qsoc}, reflecting potential issues and/or degeneracies that may occur during the prediction phase.

\subsection{Unresolved Galaxy Classifier (CU8-UGC)}\label{sec:modules:ugc}

The Unresolved Galaxy Classifier (UGC) estimates the redshift of sources classified as galaxies by DSC-Combmod with probability {\tt classprob\_dsc\_combmod\_galaxy} $\geq 0.25$. UGC
uses the BP/RP spectrum together with a supervised machine learning algorithm, the Support Vector Machine (SVM) \citep{CortesVapnik95,CC01a}. A regression model (t-SVM) is trained on a set of 6000 sources selected from galaxies in the SDSS DR16 archive \citep{2020ApJS..249....3A, 2017AJ....154...28B} that are cross-matched to sources observed by \gaia. The BP/RP spectra and the SDSS redshifts of the sources in this set are used as training input and output, respectively. The SDSS galaxies were selected to have redshifts in the range $0 \leq z \leq 0.6$ and magnitudes $17\leq G \leq 21$. Additional conditions were applied to specific parameters that influence the quality of the observed spectra. A test set of 250\,000 galaxies, selected in a similar manner as the training set, was used to estimate the performance of the model, as reported in \cite{DR3-DPACP-158}.
This set was also used to estimate statistical uncertainties of the redshift predictions in redshift bins of width 0.02. 
The bias -- the mean difference between predicted and observed redshifts -- was found to be $-0.006$ with a root mean squared error of $0.039$ for the entire redshift range $0.0\leq{}z\leq{}0.6$. However, the uneven distribution of redshift and magnitude causes the performance to be better for lower redshifts than for higher ones.  
For each estimated {\tt redshift\_ugc} we further determine the lower and upper prediction level, {\tt redshift\_ugc\_lower} and {\tt redshift\_ugc\_upper}, corresponding to the bias and the $1\sigma$ error of the SVM model in the closest bin. See the 
\linktosec{Data_analysis/chap_cu8par/sec_cu8par_apsis/ssec_cu8par_apsis_ugc.html} for more details.

\subsection{Outlier Analysis (CU8-OA)}\label{sec:modules:oa}

The Outlier Analysis (OA) module was originally intended to analyse those sources that receive low classification probabilities for all DSC classes.
As DSC-Combmod tends to give rather extreme probabilities -- near to 0.0 or 1.0 -- we used OA to analyse all sources that 
have all DSC-Combmod probabilities less than 0.999.
This corresponds to 56 million sources.
OA uses a Self-Organizing Map  \citep{1982Kohonen}, an unsupervised neural network that groups together similar data on a two-dimensional grid of neurons, in our case $30\times 30$. The data here are the BP/RP spectra. From this we compute a prototype spectrum of each neuron as the mean of all spectra assigned to that neuron. We further compute various statistics for each neuron, such as the mean \gmag, \gbp, \grp, parallax, and Galactic latitude. We also compute a quality index that is based on the intra-neuron distance distribution; it takes seven discrete values from 0 to 6, where 0 represents the best quality neurons and 6 the poorest ones. 
The method of allocation to these is described in the \linktosec{Data_analysis/chap_cu8par/sec_cu8par_apsis/ssec_cu8par_apsis_oa.html}.
Finally, we compute a class label for each neuron by finding the best match between its prototype and a series of labelled templates, although neurons with quality index 6 are not assigned a label. This information is given in the \linktotable{sec_dm_astrophysical_parameter_tables/ssec_dm_oa_neuron_information.html}{oa\_neuron\_information} and \linktotable{sec_dm_astrophysical_parameter_tables/ssec_dm_oa_neuron_xp_spectra.html}{oa\_neuron\_xp\_spectra} tables, and an interactive visualization tool that can explore these tables is available~\citep{2021Alvarez}.

\subsection{Variability (CU7)}\label{sec:modules:cu7}


Extragalactic objects can also be identified via their photometric variability. Galaxies with active nuclei show variability in their accretion, such  as in Seyfert galaxies and quasars, and in the case of blazars variability can be intrinsic or geometrical, related to a relativistic plasma jet directed towards us.

Using a supervised classification method Vari-Classification described in \cite{DR3-DPACP-165}, we identified 1.0~million Active Galactic Nuclei (AGN) and 2.5~million galaxy candidates from the variability of the \gaia light curves. Epoch photometry in the \gmag, \gbp, and \grp bands are published for AGN candidates, and for those galaxies that are part of the \gaia Andromeda Photometric Survey \citep[GAPS;][]{DR3-DPACP-142} or that might be misclassified as real variables in \gdr{3} (and so published in one of the variability tables). Indeed, the apparent variability of galaxies in the \gaia data is mostly an artefact of their extension combined with the Gaia on-board detection algorithm and scanning law (see Sect.~\ref{subsubsec:lightcurve_galaxies}) and so does not justify the release of their time series (which are meant only for genuine variable objects). Nevertheless, the characteristics of these artificial brightness variations made it possible to identify galaxies as if they were variable objects.
Light curve statistics for all sources with light curves are published in the 
\linktotable{sec_dm_variability_tables/ssec_dm_vari_summary.html}{vari\_summary} table.

Further analysis and characterisation of the variable AGN classifications (by the module Vari-AGN) led to a higher purity selection of about 872\,000 objects \citep{DR3-DPACP-167}, whose AGN-specific metrics are published in the \texttt{vari\_agn} table, and repeated in the \texttt{qso\_candidates} table (see Sect.~\ref{sec:tables}). 
The purity of this AGN sample was estimated to be about 95\%. 
The galaxy sample in {\tt galaxy\_candidates} is perhaps even purer, estimated at 99\%, although with a lower completeness at around 40\% \citep{DR3-DPACP-165}.

\subsection{Surface brightness profile (CU4)}\label{sec:modules:cu4}

A source is recorded by \gaia only if the on-board video processing unit determines its light profile to be sufficiently steep at its centre \citep{2016Prusti}. While this is intended to accept only point sources, it does pick up some extended objects (see section~\ref{sec:modules}). The resulting selection function has been assessed theoretically by \cite{2015deBruijne} and \cite{2014deSouza}.

\new{The CU4 surface brightness profile module attempts to reconstruct the two-dimensional light profile of extragalactic sources in the following way (see \citealt{DR3-DPACP-153} for more details). Gaia scans each source at a range of transit angles during the course of its mission. These observations are mainly one-dimensional (nine one-dimensional Astro Field (AF) windows plus the two-dimensional Sky Mapper (SM) window), but after a sufficient number of transits, most of the surface of the source has been covered by these transits. The CU4 module attempts to reproduce these observed windows from a large number of simulations of images of galaxies, each with different shape parameters from which \gaia-like windows are extracted. The parameters that produce the best fit to the observations are taken as the profile of the source.}

The module is only applied to a pre-selected list of extragalactic sources (summarized below). Fits are made for the flux profiles for two types of objects: quasars and their decomposition into quasar and host galaxy; and galaxies.

\begin{figure}[t]
\centering
\includegraphics[width=0.45\textwidth,angle=0]{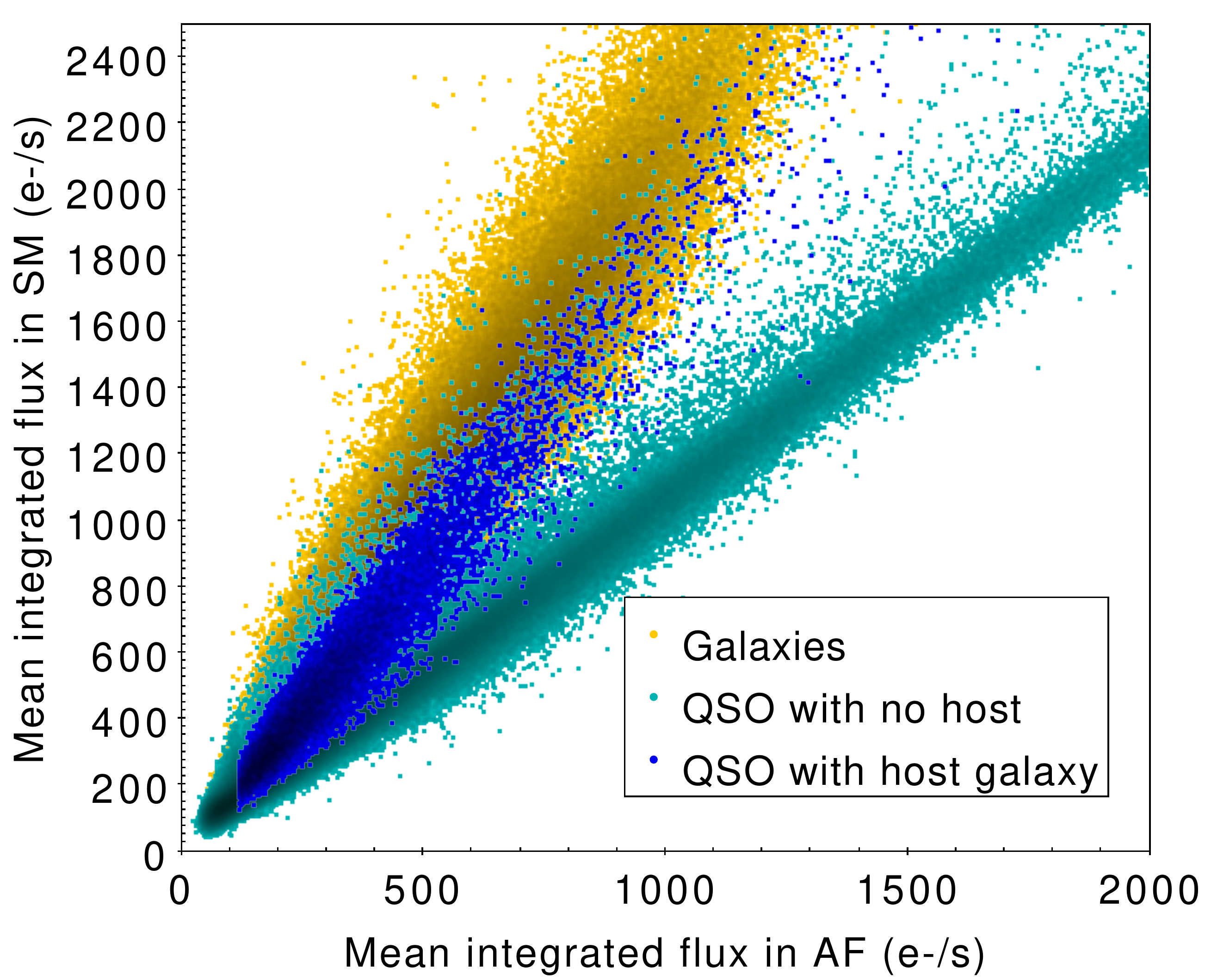}
\caption{Comparison of the flux collected in the AF and SM windows in the \gaia\ focal plane for quasars (with and without a detected host galaxy in \gaia) and for galaxies. Objects with an extension detectable by \gaia lie above the turquoise diagonal of quasars with no host galaxy.}
\label{fig:afsm}
\end{figure}

For the quasars, the module first compares the mean integrated flux of the source in the small AF window (707 mas x 2121 mas) to the mean integrated flux in the large SM window (4715 mas x 2121 mas) \citep{2016Prusti}. A larger flux in the SM window is interpreted as a detectable host galaxy, and the surface brightness profile is fit as a combination of an exponential circular profile for the central active nucleus and a S\'ersic profile (including ellipticity and position angle) for the host galaxy (see Fig.~\ref{fig:afsm}). The surface brightness profile parameters of the host galaxy are produced only when there is no other source present within 2.5\arcsec, and only for those sources with a 
half light radius
smaller than 2.5\arcsec, 
in order to avoid too large an extrapolation of the profile and so to increase the reliability of the parameters.

For the galaxies, all the objects processed exhibit flux excess in the SM window when compared to the mean flux in AF window (see Fig.~\ref{fig:afsm}), indicating that these sources are clearly extended. Two independent surface brightness profiles are fit for all objects: a S\'ersic and a de Vaucouleurs profile.

The pre-defined list of extragalactic sources for these two types of processing was determined as follows.
For quasars, several major catalogues of quasars and candidates were compiled: AllWISE \citep{2018Assef, 2015Secrest}, HMQ \citep{2015Flesch}, LQAC3 \citep{2015Souchay}, SDSS-DR12Q \citep{2017Paris}, ICRF2 \citep{2009Ma}, and a selection of unpublished classifications of \gdr{2} quasars based on photometric variability \citep{2019Rimoldini}.
Together this gave a list of 1.4 million sources.
Of these, we retained for analysis in \gdr{3} a subset of $1\,103\,691$ sources, each of which 
has at least 25 \gaia\ observations that together cover
at least 86\% of the surface area of the source.
For the galaxies, a machine learning analysis of \gdr{2} combined with the WISE survey \citep{2012Cutri}  was used to identify 1.9 million galaxy candidates \citep{2022Krone-Martins}. The same filtering of sources as for the quasars 
reduced this to 914\,837 galaxies to be analysed. 

\subsection{\gaia-CRF3 (CU3)}\label{sec:modules:cu3}

One of the outputs of the astrometric solution in \gdr{3} is the selection of a set of sources whose positions and proper motions define the celestial reference frame of \gaia~DR3, called \gaia-CRF3. These correspond to sources cross-matched between \gaia and several external quasar catalogues, and selected according to specific quality metrics. The procedure to define this source list is described in \cite{EDR3-DPACP-133} and \cite{2021klioner}. This source sample also represents an official realisation of the International Celestial Reference System (ICRS) at optical wavelengths, as acknowledged by Resolution B3 of the \cite{IAU2021}.

\section{\gdr{3} tables with extragalactic content}\label{sec:tables}

The extragalactic content of \gdr{3} is provided through a number of tables and fields. These list, among other measures, the outputs of the modules described in the previous section.

The \linktotable{sec_dm_main_source_catalogue/ssec_dm_gaia_source.html}{gaia\_source} table provides two dedicated flags (\texttt{in\_qso\_candidates} and \texttt{in\_galaxy\_candidates}) that indicate the presence of a given source in the respective tables of the same name (described below). 
It also lists the DSC-Combmod probabilities for the 
quasar and galaxy classes (Sect.~\ref{sec:modules:dsc}).
The table 
\linktotable{sec_dm_astrophysical_parameter_tables/ssec_dm_astrophysical_parameters.html}{astrophysical\_parameters} lists all the parameters produced by the modules in CU8, namely DSC, QSOC, UGC, and OA. Further results from OA (Sect.~\ref{sec:modules:oa}) are provided in the \texttt{oa\_neuron\_information} table. These tables contain all sorts of objects, not just (candidate) extragalactic ones.
The tables 
\linktotable{sec_dm_variability_tables/ssec_dm_vari_classifier_result.html}{vari\_classification\_result} and
\linktotable{sec_dm_variability_tables/ssec_dm_vari_agn.html}{vari\_agn} 
provide information on AGN identified through the photometric light-curves (Sect.~\ref{sec:modules:cu7}).
As a complement to the \gaia-CRF3 table carried over from \gaia-EDR3 (table \texttt{agn\_cross\_id}), there is a new table \linktotable{sec_dm_reference_frame/ssec_dm_gaia_crf3_xm.html}{gaia\_crf3\_xm} in \gdr{3} that provides the complete cross-match information between the \gaia-CRF3 sources and the external catalogues in which they were identified \citep{EDR3-DPACP-133}.

\subsection{Integrated tables: \texttt{qso\_candidates} and \texttt{galaxy\_candidates}}\label{sec:integrated_tables}

In addition to the above tables, two integrated tables -- \linktotable{sec_dm_extra--galactic_tables/ssec_dm_qso_candidates.html}{qso\_candidates} and \linktotable{sec_dm_extra--galactic_tables/ssec_dm_galaxy_candidates.html}{galaxy\_candidates} -- are a compilation of the results from all processing modules that have classified or analysed extragalactic objects. 
While some of their columns are copies of information available in the above-mentioned tables, the rest are provided exclusively through these integrated tables. This is the case for the DSC class labels and the redshifts stemming from QSOC and UGC,  
as well as the results from the surface brightness profile analysis.
These two integrated tables are limited to sources that are more likely to be extragalactic, and have been selected using a number of different selection rules that are defined in the \linktosec{Data_analysis/chap_cu3qso/}.
Below we provide just a summary of these rules.\\

\noindent The \texttt{qso\_candidates} table is constructed as follows.
\begin{itemize}
\item Sources for which the quasar class probability was larger than 0.5 for any of the three DSC classifiers (Specmod, Allosmod, Combmod -- see Sect.~\ref{sec:modules:dsc}) are included. 
In addition to this, QSOC sources with reliable redshifts were also added (Sect.~\ref{sec:modules:qsoc}). This reliability is determined from a combination of rules involving quality flags and \gaia photometry thresholds (for details see \citealt{DR3-DPACP-158}).
\item Sources based on the analysis of photometric light curves (Vari-Classification, Sect.~\ref{sec:modules:cu7}) were selected when their class label was set to AGN. This class label is defined in \cite{DR3-DPACP-165}. 
Almost all of these sources are also part of the Vari-AGN sample, but a handful are not and they have also been added to the integrated quasar table.
\item Quasars for which the surface brightness profile was analysed as described in Sect.~\ref{sec:modules:cu4} were included provided the presence or not of a host galaxy could be assessed with sufficient confidence. An ancillary table \texttt{qso\_catalogue\_name} provides the name of the external catalogues that were used to select the sources that entered this pipeline.
\item All sources used to define the \gaia-CRF3 (provided in table \texttt{agn\_cross\_id}, see Sect.~\ref{sec:modules:cu3} and \citealt{EDR3-DPACP-133}) are in the quasar table, and a dedicated flag, 
\linktoparam{sec_dm_extra--galactic_tables/ssec_dm_qso_candidates.html\#qso_candidates-gaia_crf_source}{gaia\_crf\_source},
identifies them.
\item OA does not contribute any additional sources to the table. We simply add class labels from OA to sources that are included by the above selections. These labels are not necessarily limited to be extragalactic source labels.
\end{itemize}

\noindent The \texttt{galaxy\_candidates} table is constructed as follows.
\begin{itemize}
\item Sources for which the galaxy class probability was larger than 0.5 for any of the three DSC classifiers (Specmod, Allosmod, and Combmod -- see Sect.~\ref{sec:modules:dsc}) are included. 
In addition to this, UGC sources with reliable redshifts were also added (Sect.~\ref{sec:modules:ugc}). The reliability is determined by a combination of two sets of rules, one concerning the quality of the \bprp spectrum of the source, the other involving the comparison of outputs from three models estimating the redshift (for details see \citealt{DR3-DPACP-158}). 
\item Sources identified by Vari-Classification  (Sect.~\ref{sec:modules:cu7}) were selected if their class label was set to GALAXY. For a description of how this class label was defined, see \cite{DR3-DPACP-165}. 
\item Galaxies for which the surface brightness profile was analysed as described in Sect.~\ref{sec:modules:cu4} were included if the light profile parameters could be derived with sufficient quality. In complement to this, an ancillary table \texttt{galaxy\_catalogue\_name} provides the name of the external catalogues that were used to select the sources that entered this pipeline. In \gdr{3} the only applicable catalogue is that described in \cite{2022Krone-Martins}.
\item As for the \texttt{qso\_candidates} table, OA does not contribute additional sources. It only provides additional columns, which are filled just for those sources that were processed by OA.
\end{itemize}

The source lists according to the above selection criteria were concatenated into two lists, one for the \texttt{qso\_candidates} table and one for the \texttt{galaxy\_candidates} table.
A complete list of the parameters (table columns) available in each table is given in the \linktosec{Gaia_archive/chap_datamodel/sec_dm_extra--galactic_tables/}.
Columns are filled for all sources regardless of how they are selected; thus a source may have a DSC probability that does not meet the above DSC selection criteria, for example (see Table~\ref{tab:contribution}). Not all parameters are available for all sources, as not all sources were treated by all modules.
There are 6\,649\,162 
sources in the \texttt{qso\_candidates} table and 4\,842\,342 in the \texttt{galaxy\_candidates} table. 
This large number of sources is mostly due to the selection rules of the DSC module, which favour completeness over purity (see section~\ref{sec:modules:dsc} and \citealt{DR3-DPACP-158}). Users should therefore be aware that there is significant stellar contamination in these tables. For DSC this can be addressed using the {\tt classlabel\_dsc\_joint} field. We address more generally how to build purer sub-samples in Sect.~\ref{sec:howto_purer_samples}. There are 174\,146
sources in common between the two tables, and their union contains 11.3 million sources. 

\begin{figure}[t!]
\centering
\includegraphics[width=0.4\textwidth,angle=0]{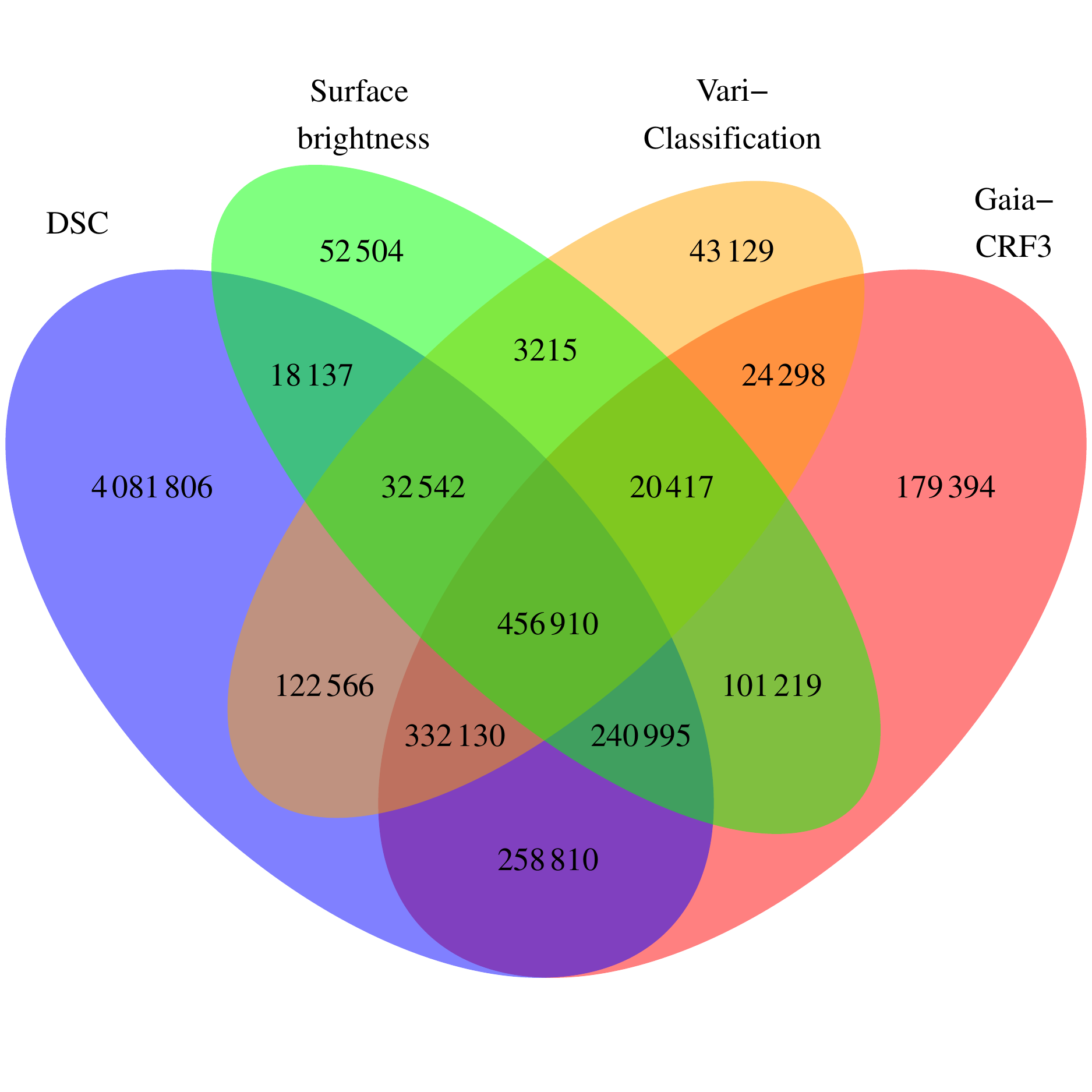}
\caption{Quadruple Venn diagram for  contributions to the \texttt{qso\_candidates} table from DSC, the Surface brightness sample, Vari-Classification, and \gaia-CRF3.
\label{fig:qsovenn1}
}
\end{figure}

\begin{figure}[t!]
\centering
\includegraphics[width=0.4\textwidth,angle=0]{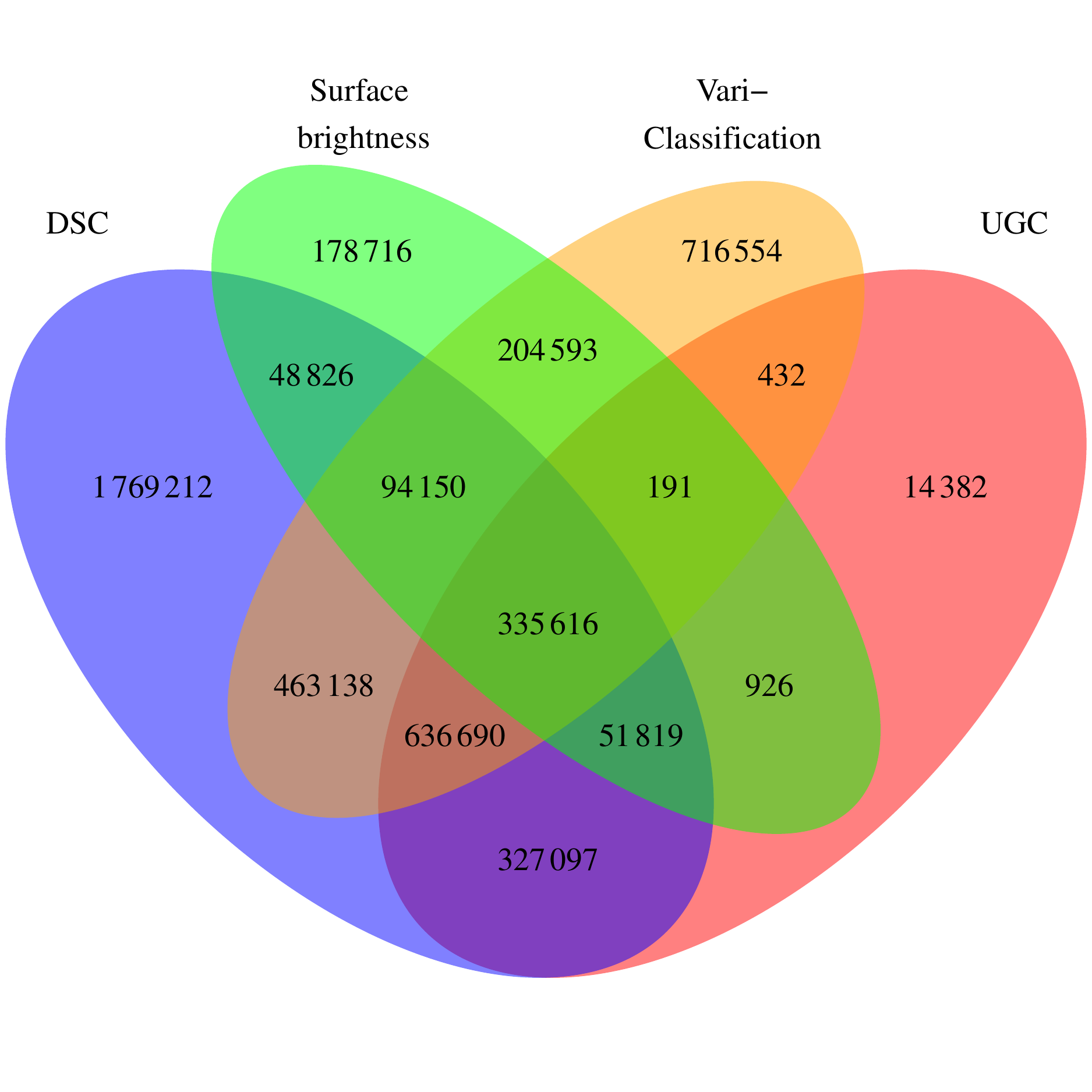}
\caption{Quadruple Venn diagram for contributions to the \texttt{galaxy\_candidates} table from DSC, the Surface brightness sample, Vari-Classification, and UGC. 
\label{fig:galvenn1}
}
\end{figure}

Table~\ref{tab:contribution} gives an overview of how many sources from each module contribute to the integrated tables. Source overlaps between the modules within each table are shown in Tables~\ref{tab:overlap:qso} and~\ref{tab:overlap:gal}, and graphically represented in the Venn diagrams in Figs.~\ref{fig:qsovenn1} and~\ref{fig:galvenn1}. Information about the distribution of the parameters featured in the tables is provided in the next section. 

\begin{table}[t]
  \caption{Number of sources from each of the extragalactic processing modules contributing to the \texttt{qso\_candidates} and \texttt{galaxy\_candidates} tables (second column), or to the set of parameters featured for the respective modules (third column). The difference between the two columns indicates the number of sources where parameters are provided despite the sources not being eligible according to the selection rules of that module. A given source can be contributed by more than one module. 
  \label{tab:contribution}}
  \centering
  \begin{tabular}{lrr}
    \hline\hline
    \noalign{\smallskip}
    Module & Selected  & Featuring \\
    & sources & parameters\\
    \noalign{\smallskip}
    \hline
    \noalign{\smallskip}
    {\tt qso\_candidates} & 6\,649\,162 & \\ \hline
    \noalign{\smallskip}
    DSC & 5 543 896  & 6 647 511 \\
    QSOC & 1 834 118  & 6 375 063 \\
    Vari-Classification & 1 035 207 & 1 122 361 \\
    Vari-AGN & 872 228 &  872 228 \\
    Surface brightness & 925 939 & 1 084 248 \\
    \gaia-CRF3 & 1 614 173 & 1 614 173 \\
    OA  & N/A & 2 803 225 \\[3pt]
    \hline 
    \noalign{\smallskip}
    {\tt galaxy\_candidates} & 4\,842\,342 & \\ \hline
    \noalign{\smallskip}
    DSC & 3 726 548  & 4 841 799 \\
    UGC & 1 367 153  & 1 367 153 \\
    Vari-Classification & 2 451 364  & 2 477 273 \\
    Surface brightness & 914 837 & 914 837 \\
    OA  & N/A & 1 901 026 \\
    \noalign{\smallskip}
    \hline
  \end{tabular}
\end{table}


\begin{table*}[t]
  \caption{Source overlaps between the modules contributing to the \texttt{qso\_candidates} table. See text for details about the module names.
  \label{tab:overlap:qso}}
  \centering
  \begin{tabular}{lrrrrrrr}
    \hline\hline
    \noalign{\smallskip}
    Module &	 Surface   		&Vari- 	& Vari-AGN  			& OA 		& QSOC & DSC \\
           	& brightness	 	&classification 	& 	&   & & \\    
    \noalign{\smallskip}
    \hline
    \noalign{\smallskip}
    \gaia-CRF3			&819 541 &	833 755 	&722 211 &	550 807 &	672 454 	&1 288 845\\
    Surface brightness		&	&	513 084 &	483 786 &	278 078 	&458 241 &	748 584\\
    Vari-classification	& &		& 872 184 & 245 318 &	477 971 &	944 148	\\
    Vari-AGN	&		&	&	&186 836 	& 442 436 	& 814 315\\
    OA  		&        	&  	& 	& 		& 896 173 & 2 085 554 \\
    QSOC				&		&	&	&		&		& 1 097 229 \\[3pt]
     \hline
  \end{tabular}
\end{table*}

\begin{table*}[t]
  \caption{As \ref{tab:overlap:qso} for modules contributing to the \texttt{galaxy\_candidates} table.
  \label{tab:overlap:gal}}
  \centering
  \begin{tabular}{lrrrrrrr}
    \hline\hline
    \noalign{\smallskip}
    Module &	 Vari-  	& UGC & OA & DSC\\ 
        &  classification & & & \\ 
    \noalign{\smallskip}
    \hline
    \noalign{\smallskip}
Surface brightness & 634\,550 &	388\,552 &	434\,880 &	530\,411 \\
Vari-Classification 	&	 	 	& 972\,929 	& 1\,070\,865 &	1\,529\,594 \\
UGC 	&		 	&  & 190\,583 &	1\,351\,222\\
OA 	&	&	&	 & 840\,409\\[3pt]
     \hline
  \end{tabular}
\end{table*}

\new{To estimate the overall purity of the integrated tables, we must be aware that modules with different purities can contribute the same source to a table. The estimation can be simplified, however, when we consider that all modules except DSC have similar high purities. Specifically, for the {\tt qso\_candidates} we assume that the modules other than DSC have an average purity of 96\%, compared to a global DSC purity of 24\%. From Fig.~\ref{fig:qsovenn1} we see that 4.1 million sources are contributed only by DSC, with the remaining 2.6 million contributed by the other modules. This gives an overall purity of the {\tt qso\_candidates} table of 52\%. In a similar way, we estimate the overall purity of the {\tt galaxy\_candidates} to be 69\%. We show how to obtain a purer sub-sample in Sect.~\ref{sec:howto_purer_samples}.}

\section{Basic properties}\label{sec:basic_properties}

\subsection{Parameter distributions}

\subsubsection{Integrated tables}


\begin{figure*}
\begin{center}
\includegraphics[width=0.49\textwidth,angle=0]{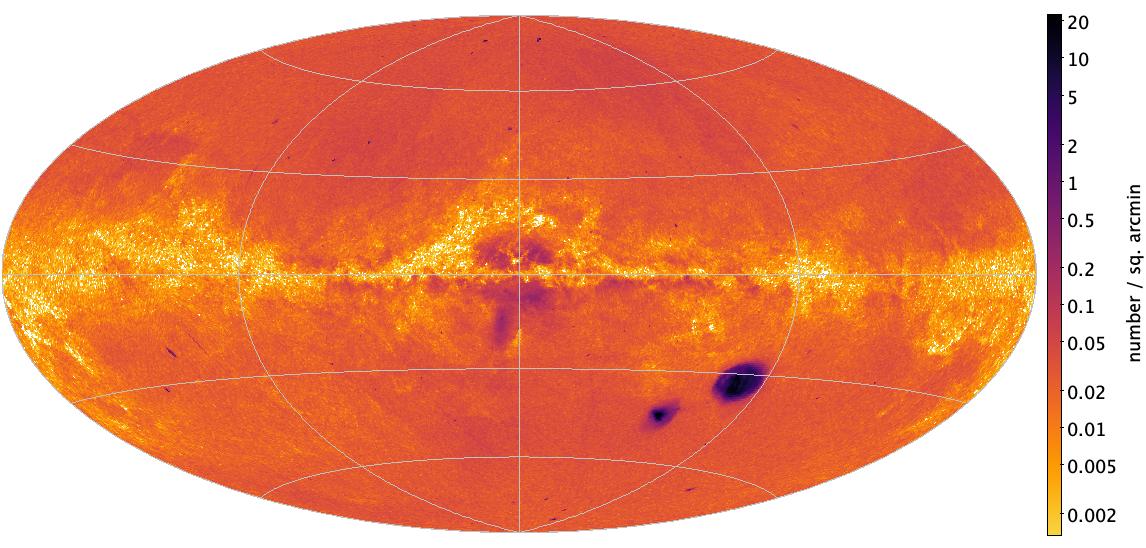}
\includegraphics[width=0.49\textwidth,angle=0]{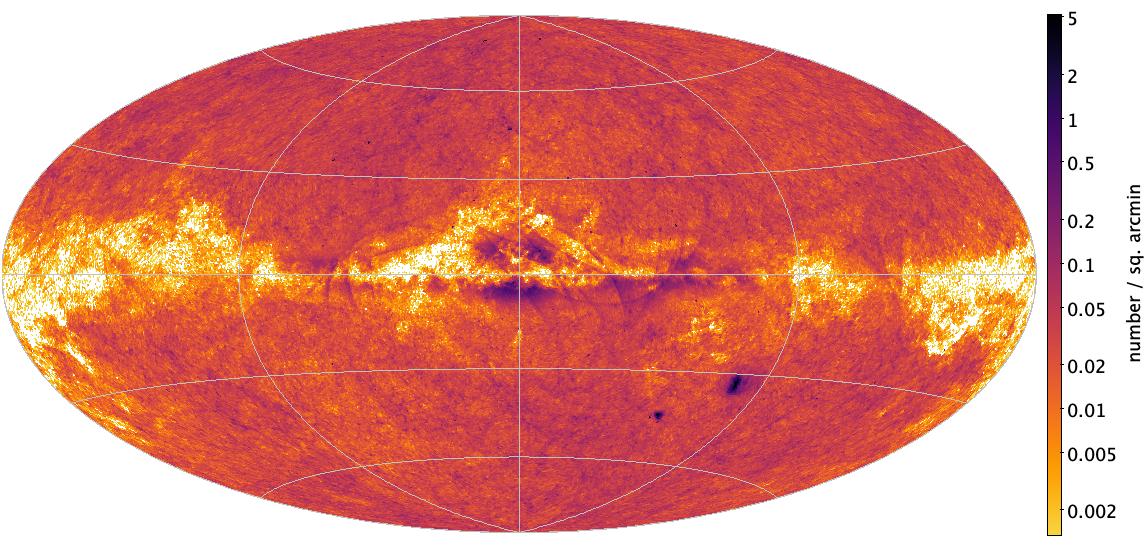}
\caption{Galactic sky distribution of all the sources in the
{\tt qso\_candidates} table (left) and {\tt galaxy\_candidates} table (right).
The plot is shown at HEALpixel level 7 (0.210 sq.\ deg.) in Hammer--Aitoff projection.
The colour scale, which is logarithmic, covers the full range for each panel, so is different for each panel. 
\label{fig:qso_galaxy_skyplots}}
\end{center}
\end{figure*}

\begin{figure}
    \centering
    \includegraphics[width=0.49\textwidth]{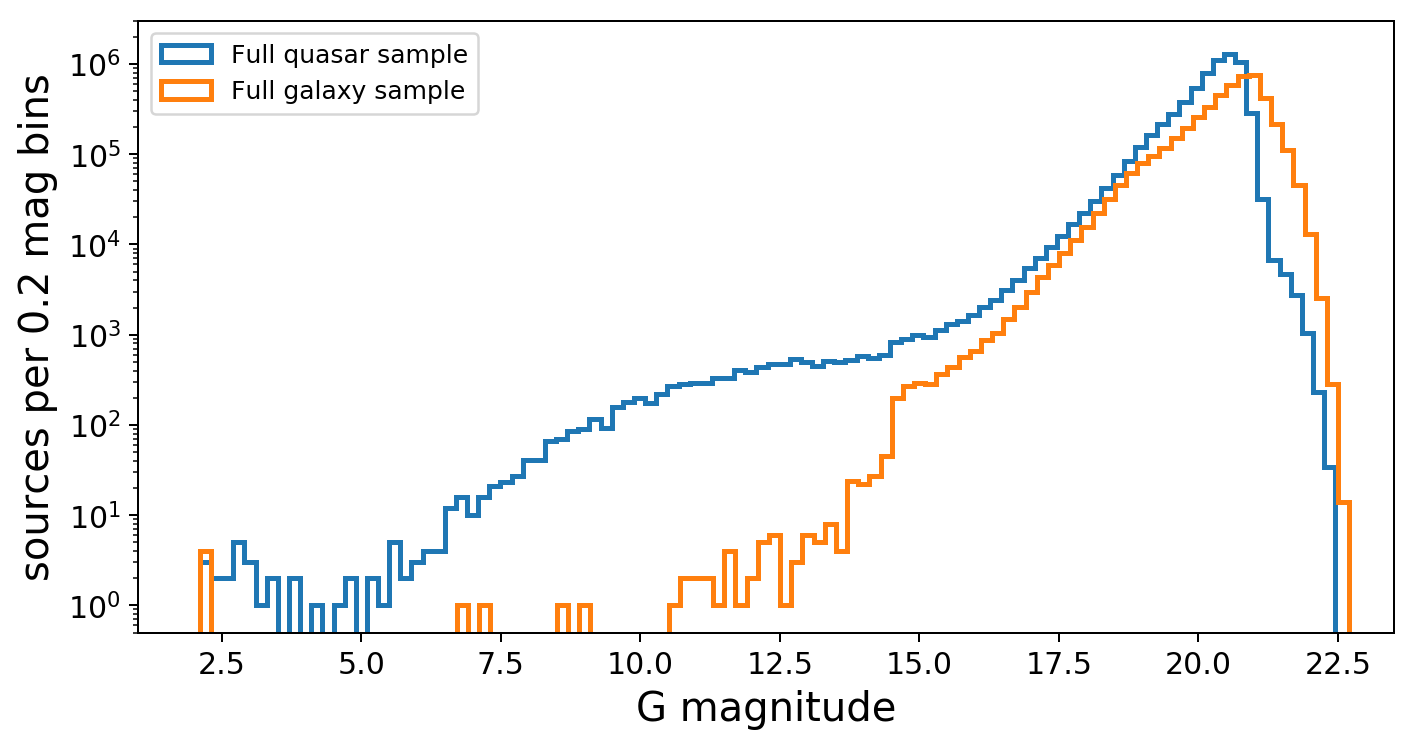}
    \caption{G-band magnitude distribution of all objects in the {\tt qso\_candidates} (blue) and {\tt galaxy\_candidates} (orange) table on a logarithmic scale. 
    The brightest known quasar (3C273 -- \texttt{source\_id} 3700386905605055360) has a G magnitude of 12.8.}
    \label{fig:qso_galaxy_gmagdist}
\end{figure}

Figure~\ref{fig:qso_galaxy_skyplots} shows the sky distribution on a logarithmic density scale of all sources in the {\tt qso\_candidates} and {\tt galaxy\_candidates} tables. As already noted, there is considerable contamination in these due to misclassifications and the completeness-driven nature of the tables (i.e.\ the absence of filtering in some modules). \new{This is apparent from the overdensities around the Large and Small Magellanic Clouds (LMC and SMC).}
If we exclude generous regions around the LMC and SMC (defined in appendix~\ref{sec:adql_queries}), then the  number of sources in the 
{\tt qso\_candidates} table drops to 3.95 million (59\% of the full table)
and the number of sources in the 
{\tt galaxy\_candidates} table drops to 4.67 million (96\% of the full table).
Some patterns are also an artefact of the use of input lists for some of the modules. Many of these sources are also faint, with poorer data in \gdr{3}, as can be seen in Fig.~\ref{fig:qso_galaxy_gmagdist}. There is also a small fraction of sources that are too bright to be genuine quasars or galaxies, which is an inevitable consequence of even a small misclassification probability and limited filtering.

\begin{figure}
\begin{center}
\includegraphics[width=0.40\textwidth,angle=0]{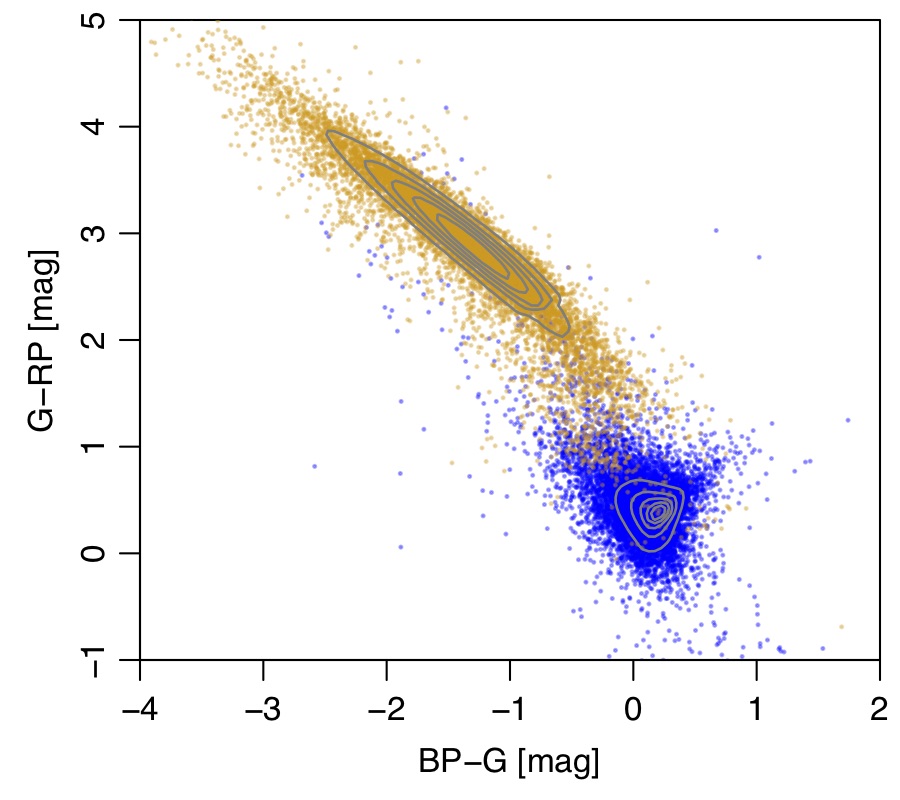}
\includegraphics[width=0.40\textwidth,angle=0]{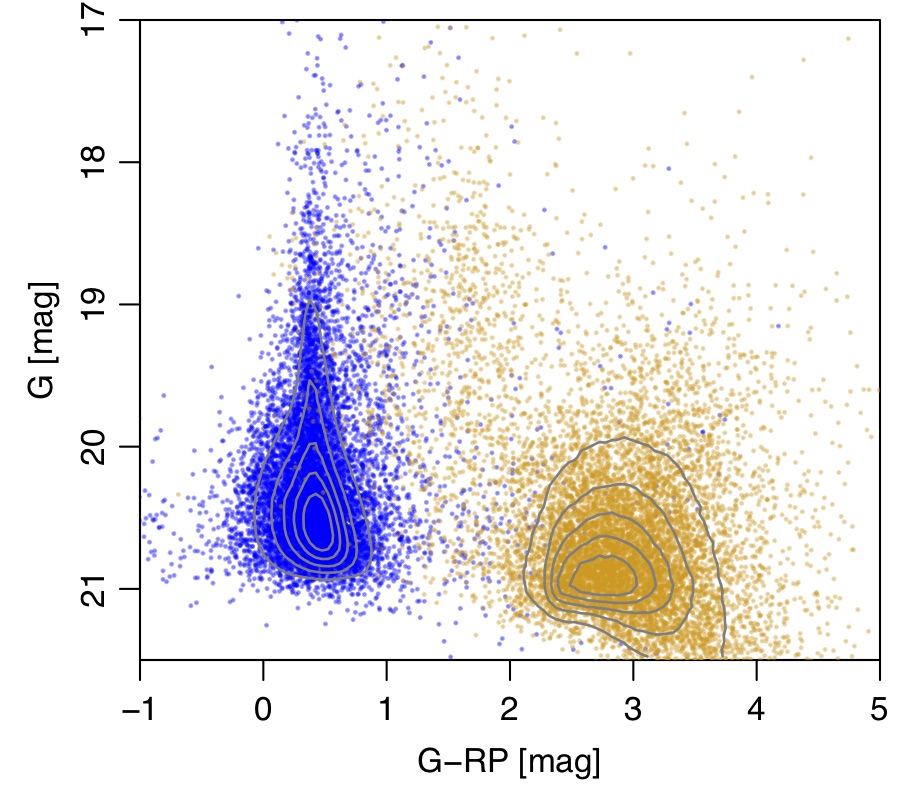}
\caption{Colour--colour diagram (top) and colour--magnitude diagram (bottom) for all sources in the {\tt qso\_candidates} table (blue) and {\tt galaxy\_candidates} table (orange). The contours show density on a linear scale. The points are a random selection of 10\,000 sources for each class. 
\label{fig:dr3int5_bothtables_ccd_cmd}
}
\end{center}
\end{figure}

The \gaia\ colour--colour diagram (CCD) and colour--magnitude diagram (CMD) are shown in Fig.~\ref{fig:dr3int5_bothtables_ccd_cmd}. 
Quasars and galaxies separate quite well, but recall that \gaia\ observes primarily those galaxies with point-source like cores.
What is not seen in these diagrams is the distribution of the stars, which outnumber true quasars and galaxies by a factor of 500--1000 in \gdr{3}, and which make it hard to identify extragalactic objects based only on their \gaia\ colours.

\subsubsection{DSC subset of the integrated tables}\label{sec:integrated_DSC_subset}

\begin{figure}[t]
\begin{center}
\includegraphics[width=0.24\textwidth,angle=0]{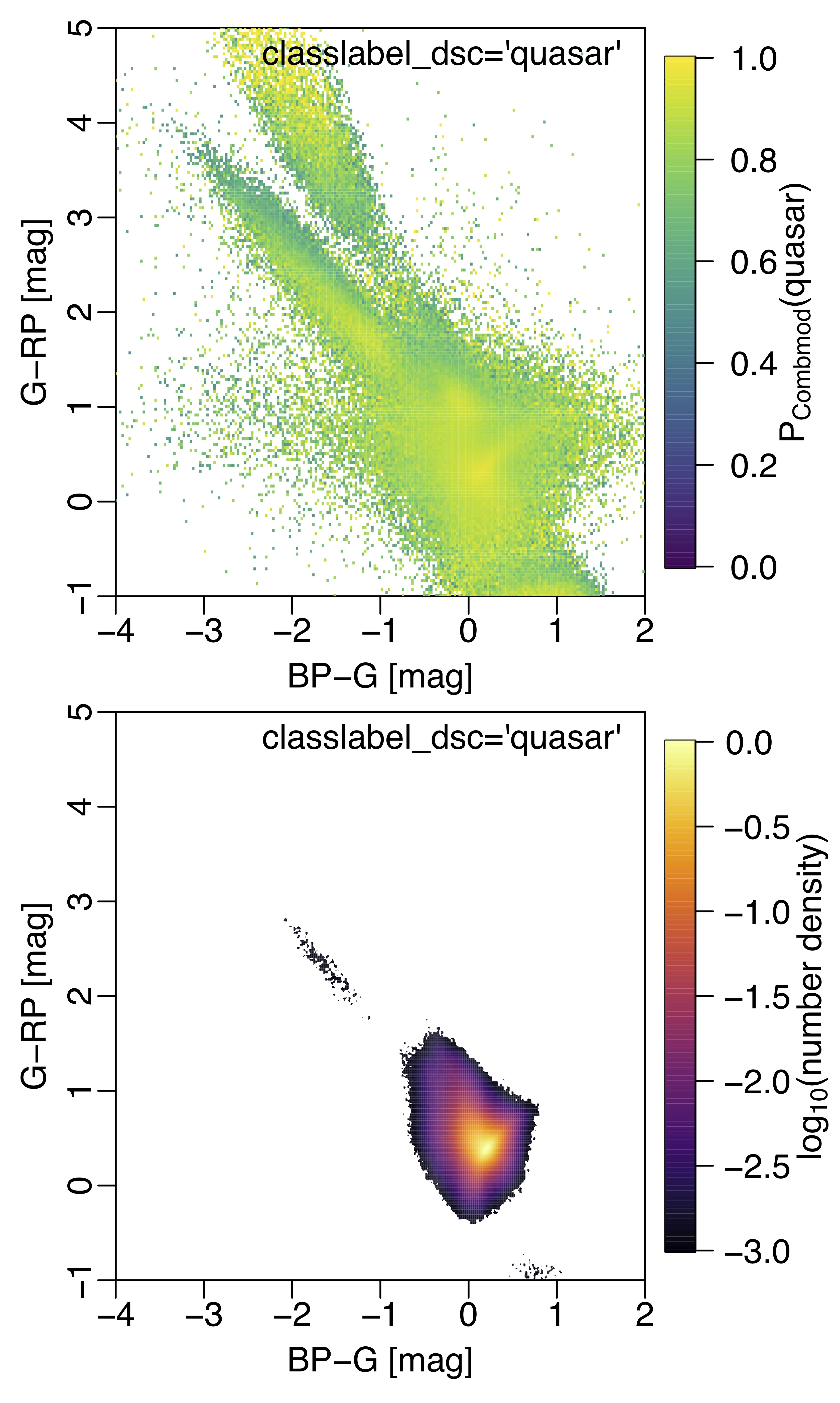}
\includegraphics[width=0.24\textwidth,angle=0]{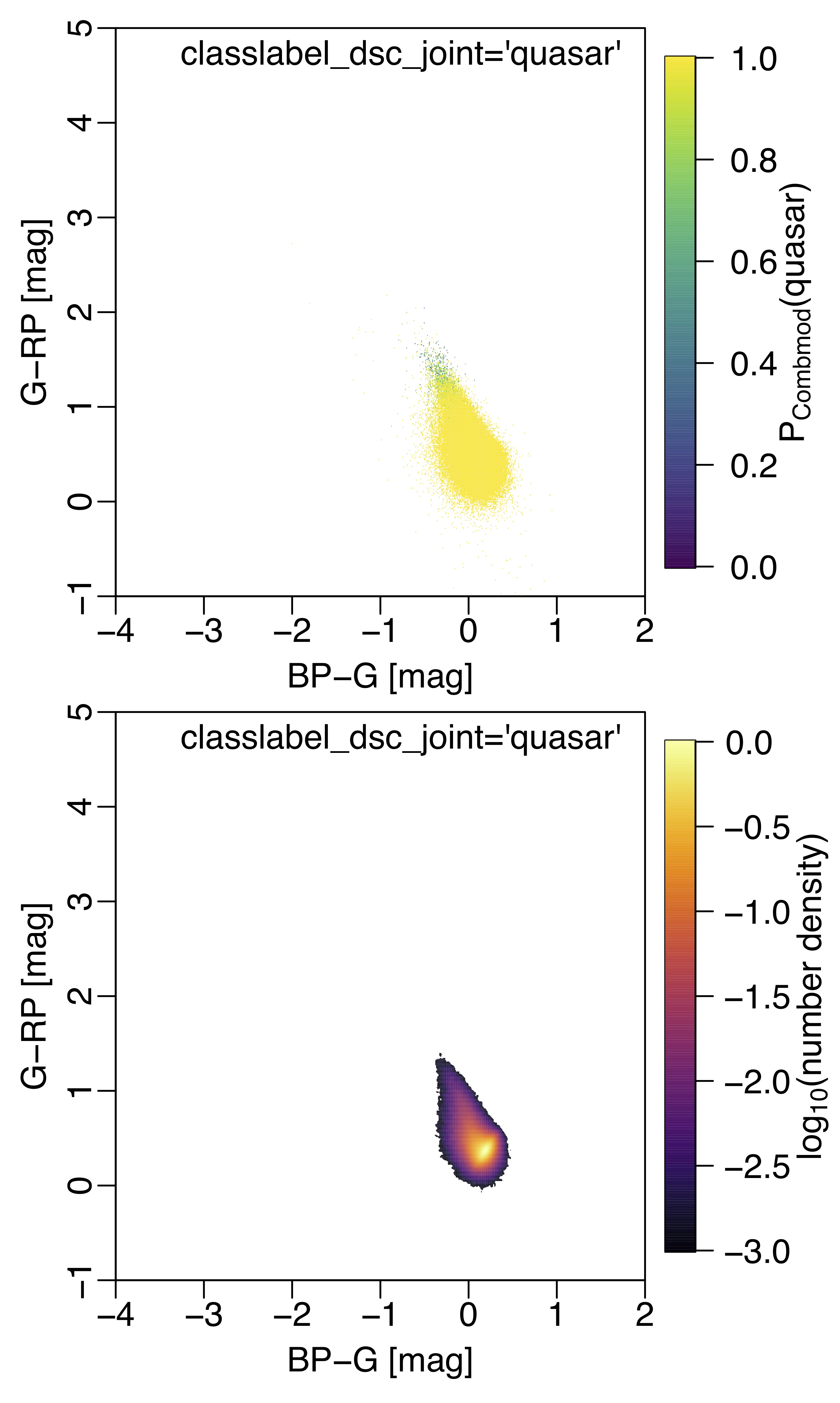}
\caption{Colour--colour diagram for sources in the {\tt qso\_candidates} table, excluding regions around the LMC and SMC. The left column shows sources with {\tt classlabel\_dsc\,=\,quasar} (2.77 million sources), the right column shows sources with {\tt classlabel\_dsc\_joint\,=\,quasar} (the purer subset, 0.52 million sources).
These numbers refer to the number of sources plotted, which are those that have all \gaia bands.
The upper panel shows the mean DSC-Combmod probability for the quasar class (the field {\tt classprob\_dsc\_combmod\_quasar}). The lower panel shows the density of sources on a log scale relative to the peak density in that panel (densities 1000 times lower than the peak are not shown). 
\label{fig:dr3int5_qsotable_ccd_grp_bpg}
}
\end{center}
\end{figure}

\begin{figure}[t]
\begin{center}
\includegraphics[width=0.24\textwidth,angle=0]{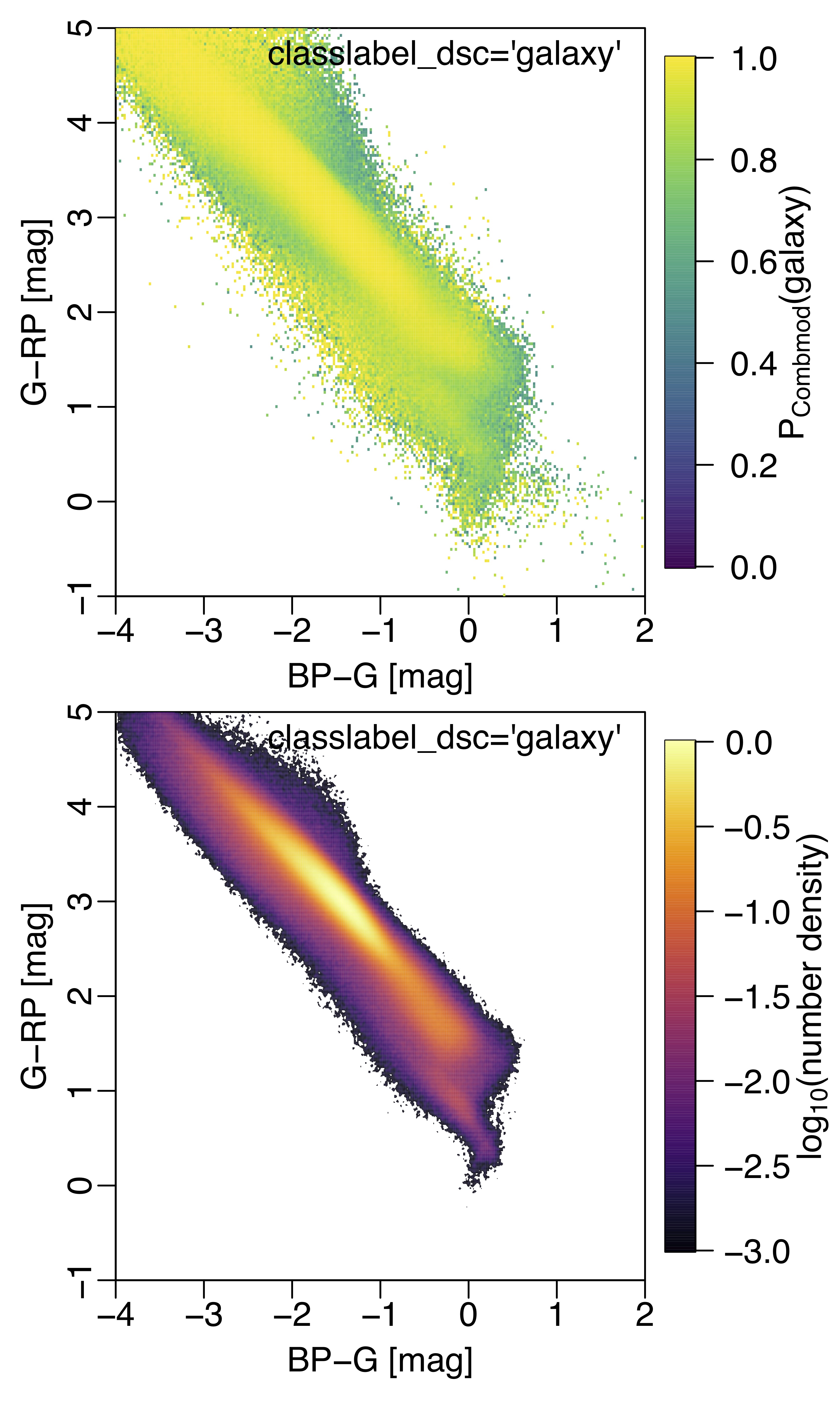}
\includegraphics[width=0.24\textwidth,angle=0]{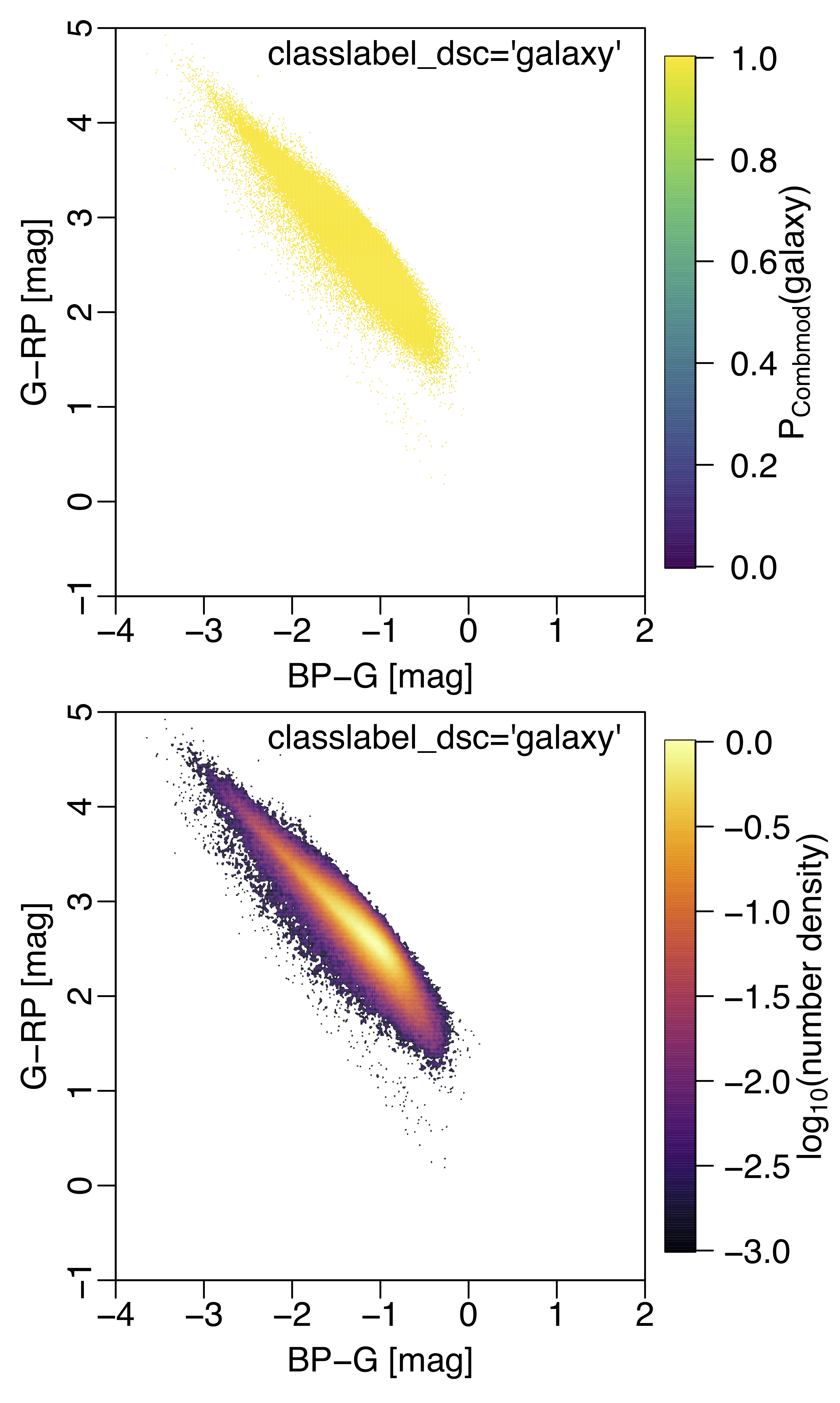}
\caption{As Fig.~\ref{fig:dr3int5_qsotable_ccd_grp_bpg}, but for sources in the {\tt galaxy\_candidates} table. There are 3.24 million sources with {\tt classlabel\_dsc\,=\,galaxy} and 0.25 million sources with {\tt classlabel\_dsc\_joint\,=\,galaxy} (in both cases excluding the regions around the LMC and SMC, and requiring all three \gaia bands).
\label{fig:dr3int5_galaxytable_ccd_grp_bpg}
}
\end{center}
\end{figure}

DSC is the dominant contributor to the {\tt qso\_candidates} and {\tt galaxy\_candidates} tables, so we look here at two subsets for each table defined by the DSC class labels (Sect.~\ref{sec:modules:dsc}).
The first is selected by 
{\tt classlabel\_dsc}, which gives 5\,243\,012 
quasars in the {\tt qso\_candidates} table (class {\tt quasar}) and 3\,566\,085 
galaxies in the {\tt galaxy\_candidates} table (class {\tt galaxy}).
Through comparison to SDSS spectroscopic classifications, and accommodating for the significant contamination by stars, we estimate these samples to have rather low purities of 24\% and 22\% respectively (see \citealt{LL:CBJ-094}, summarized in \citealt{DR3-DPACP-158}, and Sect.~\ref{sec:external_comparison:quasars} below).
The second subset is the purer one identified using {\tt classlabel\_dsc\_joint}, which gives 547\,201
quasars in the {\tt qso\_candidates} table and 251\,063 
galaxies in the {\tt galaxy\_candidates} table. These two sets are estimated to have higher purities of 62\% and 64\% respectively, and of 79\% and 82\% respectively if we look only at higher latitudes ($|b|>11.54\degree$).

Figure~\ref{fig:dr3int5_qsotable_ccd_grp_bpg} shows the \gaia\ colour--colour diagrams for quasars in the {\tt qso\_candidates} table according to these two subsets. 
The upper panels show the DSC-Combmod probabilities. In the upper left panel we see that there are sources far away from the main clump of quasars, but the lower panel reveals that there are very few of them. These are all removed in the {\tt classlabel\_dsc\_joint\,=\,quasar} set (right column), which shows only high Combmod probabilities.
Figure~\ref{fig:dr3int5_galaxytable_ccd_grp_bpg} shows the corresponding colour--colour diagrams for the {\tt galaxy\_candidates} table. Again we see how the set defined by {\tt classlabel\_dsc\_joint\,=\,galaxy} has a tighter distribution and higher Combmod probabilities than the less pure set defined by  {\tt classlabel\_dsc\,=\,galaxy}.
Similar figures showing the quasar and galaxy populations together are shown in  
\cite{DR3-DPACP-158}. These also show that 
use of the joint label preferentially removes fainter, lower signal-to-noise sources, as these are less likely to get a high probability classification in both Specmod and Allosmod. 

One thing to bear in mind is that Specmod and Allosmod do not deal with identical sets of sources, because these classifiers require different input data. In particular, Allosmod requires parallaxes and proper motions, that is 5p or 6p astrometric solutions (see \citealt{2021A&A...649A...4L} for the definition of these solutions). Galaxies often 
only get 2p solutions (no parallax or proper motion) on account of their physical extent. Of the 3\,566\,085 million sources in the {\tt galaxy\_candidates} table with {\tt classlabel\_dsc\,=\,galaxy}, 
3\,367\,211 have all three photometric bands, but of these, only 1\,015\,462 have parallaxes and proper motions and so can be classified by Allosmod (these numbers are for the whole sky, so including the LMC and SMC). As {\tt classlabel\_dsc\_joint} can only be set to galaxy when Allosmod results are present, the change in the distribution
we see in Fig.~\ref{fig:dr3int5_galaxytable_ccd_grp_bpg}
for the two class labels is partially due to this. Plots in \cite{DR3-DPACP-158} show the change when only considering the subset with 5p or 6p solutions.
Most quasars, in contrast, do have 5p or 6p solutions: Of the 5\,243\,012 
sources
in the {\tt qso\_candidates} with {\tt classlabel\_dsc\,=\,quasar}, 5\,086\,531
have all three photometric bands, of which 4\,815\,212 have parallaxes and proper motions.

Because DSC is not the only contributor to the integrated tables, some of the sources in these tables have DSC class labels that are not the class of the table. 
In the {\tt qso\_candidates} table, 156\,970 sources have {\tt classlabel\_dsc} set to {\tt galaxy}, and 12\,302 have {\tt classlabel\_dsc\_joint} set to {\tt galaxy}. 
In the {\tt galaxy\_candidates} table, the numbers with these two classlabels set to {\tt quasar} are 12\,933 and 234 respectively.

\subsection{BP/RP spectra}\label{sec:bp_rp_spectra}


\gaia\ observes all of its targets with the low resolution ($30 \leq \lambda / \Delta \lambda \leq 100$) \bprp\ slitless spectrograph \citep{2021A&A...652A..86C}.
\new{1.6 billion of these were used by DSC-Specmod for classification (section~\ref{sec:modules:dsc}; \citealt{DR3-DPACP-158}), but only a fraction of these are published in \gdr{3}.}
Spectra for all sources brighter than $G=17.35$\,mag with at least 15 retained observations in each of BP and RP are published in \gdr{3}, amounting to 220 million sources. \new{This includes few extragalactic sources, so a small set of these were added. In total, \bprp\ spectra of 163\,000 quasar candidates and 26\,500 galaxy candidates in the integrated tables are published in \gdr{3}. Of these, 119\,000 and 12\,600 respectively are in the purer sub-samples
defined in Sect.~\ref{sec:howto_purer_samples}.}

As described in \cite{DR3-DPACP-118}, spectra are published as a set of coefficients of basis functions, from which spectra at arbitrary samplings can be produced using a published software tool. Internal to CU8, the spectra were sampled using the tool SMS-gen \citep{DR3-DPACP-157}, which is what we used to produce the spectra shown in this section. In all cases the spectra are the mean (epoch-averaged) spectra over a time span of up to 34 months.

\subsubsection{Quasars}
\label{sssec:bp_rp_spectra_quasars}

\begin{figure*}
    \centering
    \includegraphics[width=\textwidth]{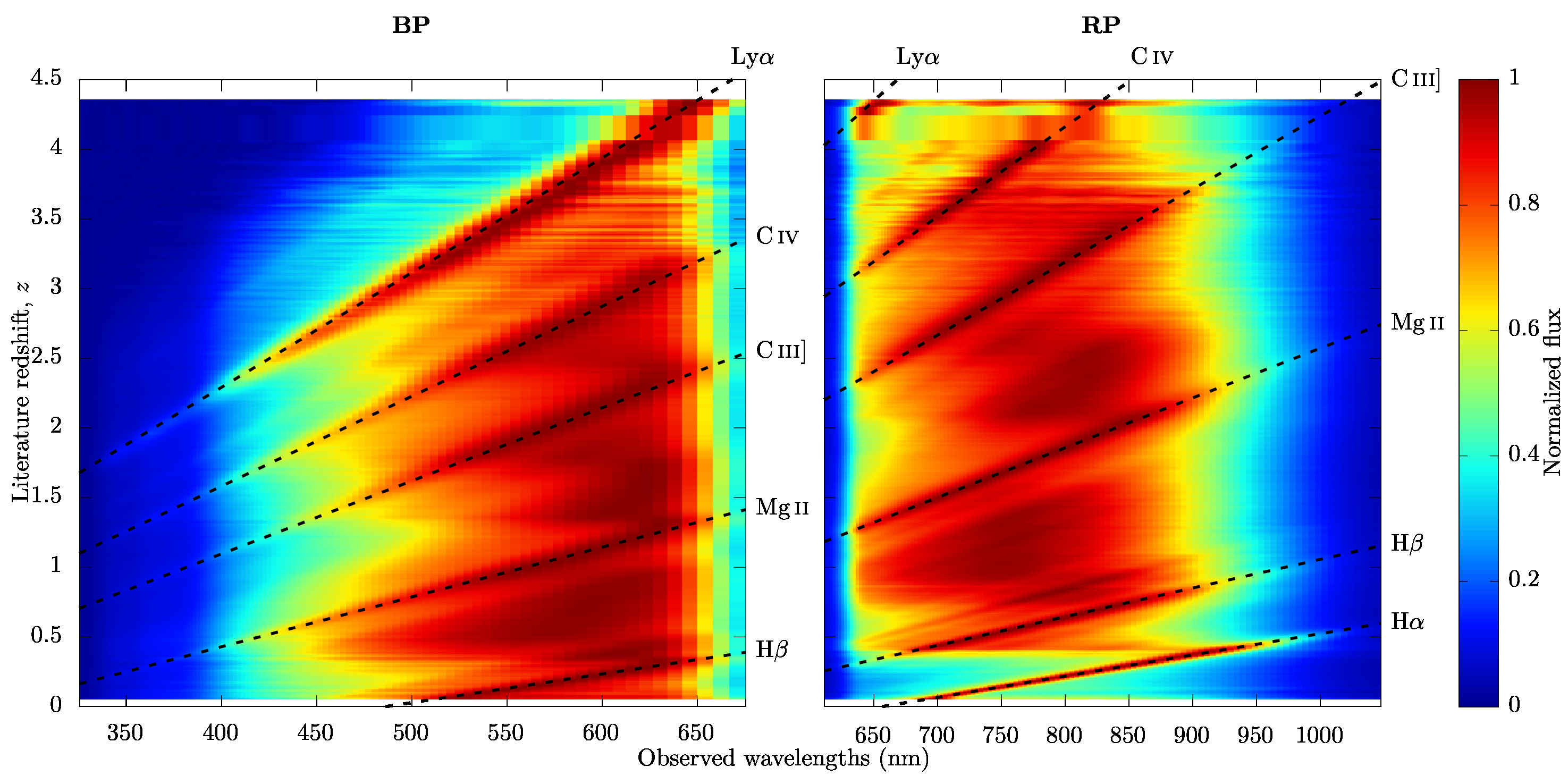}
    \caption{Distribution of the BP flux (left) and RP flux (right) as sampled by SMS-gen \citep{DR3-DPACP-157} of 42\,944 quasars published in \gdr{3} that have spectroscopically confirmed redshifts in the Milliquas 7.2 quasar catalogue of \cite{2021Flesch} ({\tt type\,=\,Q}). Dotted lines show the dominant quasar emission lines. Spectra are individually normalized in order to have a maximum flux of 1.0 and are then averaged in redshift bins of 0.01, with the inverse variance of the sampled fluxes used as the weight during the computation of the mean.}
    \label{fig:CU8_QSOs_obs}
\end{figure*}

Figure~\ref{fig:CU8_QSOs_obs} shows the \bprp spectra for 42\,944 quasars with published \bprp coefficients (field
\linktoparam{sec_dm_main_source_catalogue/ssec_dm_gaia_source.html\#gaia_source-has_xp_continuous}{has\_xp\_continuous}\,=\,{\tt true} in 
\linktotable{sec_dm_main_source_catalogue/ssec_dm_gaia_source.html}{gaia\_source}), {\tt classprob\_dsc\_combmod\_quasar}\,$> 0.01$ 
and spectroscopically confirmed redshift in the Milliquas 7.2 quasar catalogue of \cite{2021Flesch} ({\tt type\,=\,Q}). A search radius of 1$\arcsec$ was used to match the \gaia sources to their Milliquas counterparts, leading to a redshift coverage of $0.052 \leq z \leq 4.358$. The cut on the DSC Combmod quasar probability ensures that obvious stellar contaminants contained in our cross-match are discarded.
The median magnitude of the sources in Figure~\ref{fig:CU8_QSOs_obs}  is $G=18.53$\,mag. \gaia observes much fainter quasars, but the \bprp\ spectra of many of these will only be released in \gdr{4}.
While we clearly see common quasar emission lines in this averaged plot, they are not necessarily visible in the low signal-to-noise ratio (S/N) spectra of individual faint quasars. Similarly, wiggles that are an artefact of the Hermite spline representation of the \bprp\ spectra \citep{DR3-DPACP-118} tend to lower the contrast of these emission lines compared to the continuum. These wiggles smooth out faint spectral features, and can be confused with emission lines, as both have comparable strength in low S/N spectra. Typically, though, the strongest spectral features -- Ly$\alpha$, \ion{C}{iv}, H$\beta$, and H$\alpha$ -- are retained in $G < 20$\,mag spectra. 
We also see in Fig.~\ref{fig:CU8_QSOs_obs} that regions at wavelengths below 430\,nm and above 650\,nm in BP, and below 630\,nm and above 950\,nm in RP, contain little flux: spectral features in these regions generally have low S/N, complicating their detection by the DSC and QSOC algorithms.

\subsubsection{Galaxies}

\begin{figure*}
    \centering
    \includegraphics[width=\textwidth]{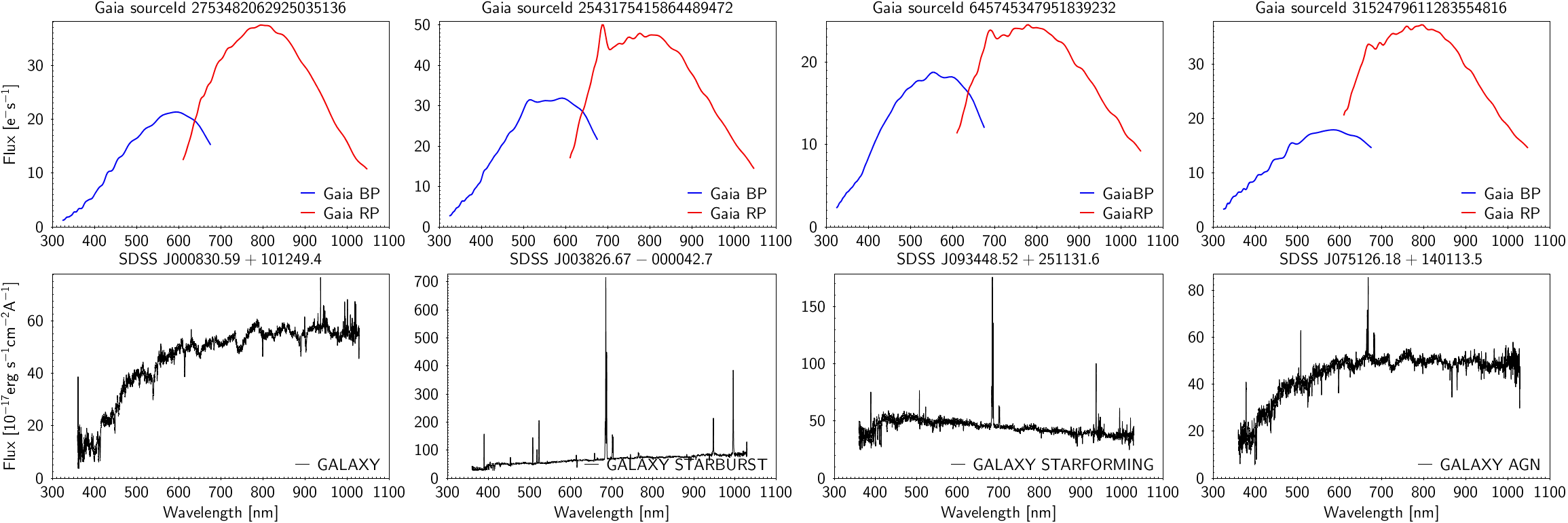}
    \caption{Galaxy spectra. Top row: Representative mean BP and RP \gaia spectra for four galaxies. Bottom row: The spectra for the same galaxies as observed with the SDSS-BOSS spectrograph (the SDSS class and subclass, if defined, are shown).  }
    \label{fig:CU8_gal_spectra}
\end{figure*}

Figure \ref{fig:CU8_gal_spectra} shows four representative spectra of galaxies as observed by \gaia (top row) and their corresponding SDSS spectra (bottom row). 
The first SDSS spectrum  on the left shows only absorption lines, suggesting an early type galaxy with little or no star formation activity (the few spikes are caused by cosmic rays). These lines are barely detectable, if at all, in the low-resolution \bprp spectrum. The two middle spectra show strong emission lines characteristic of active star formation. The strongest is the H$\alpha$ emission with [\ion{N}{II}] 
lines on either side. This set of three lines is unresolved in the RP spectrum where it merges into a single and wide emission feature. Similarly, in the BP spectrum the H$\beta$ and [\ion{O}{III}]
emission lines are merged into another wide peak. The last spectrum on the right is classified as a 'GALAXY AGN' in SDSS. The corresponding \bprp spectrum, due to the much lower resolution -- and the already mentioned wiggles -- 
shows much less prominent features.

\subsection{Surface brightness profiles}\label{subsec:cu4eo_properties}

\subsubsection{Quasars}\label{subsubsec:cu4eo_qso}

The majority of the $1\,103\,691$ quasars analysed in terms of surface brightness lie in the diagonal of Fig.~\ref{fig:afsm}. These sources are considered point-like with no host galaxy detectable by \gaia. A group of 64\,498 exhibit a clear extension, indicative of a host galaxy, as evidenced by larger fluxes in the SM window than in the AF window (Sect.~\ref{sec:modules:cu4}).
For these sources the flag 
\linktoparam{sec_dm_extra--galactic_tables/ssec_dm_qso_candidates.html\#qso_candidates-host_galaxy_detected}{host\_galaxy\_detected}\,=\,{\tt true} is set.
Among these, a robust solution from the fitting process was derived for 15\,867 sources and their surface brightness profile is given in the catalogue. \new{The flag 
\linktoparam{sec_dm_extra--galactic_tables/ssec_dm_qso_candidates.html\#qso_candidates-host_galaxy_flag}{host\_galaxy\_flag}
indicates the outcome of the fitting process for all sources considered.
Values of 1 and 2 are good fits, indicating detection of a host galaxy. 3 indicates that no host could be found, whereas 4 is a poor fit.
Sources with \texttt{host\_galaxy\_flag}\,=\,5 or 6 show no evidence of a host galaxy in our analysis,
due to non-convergence of the algorithm or the presence of a close neighbour, respectively.}

\begin{figure}[t]
\begin{center}
\includegraphics[width=0.49\textwidth,angle=0]{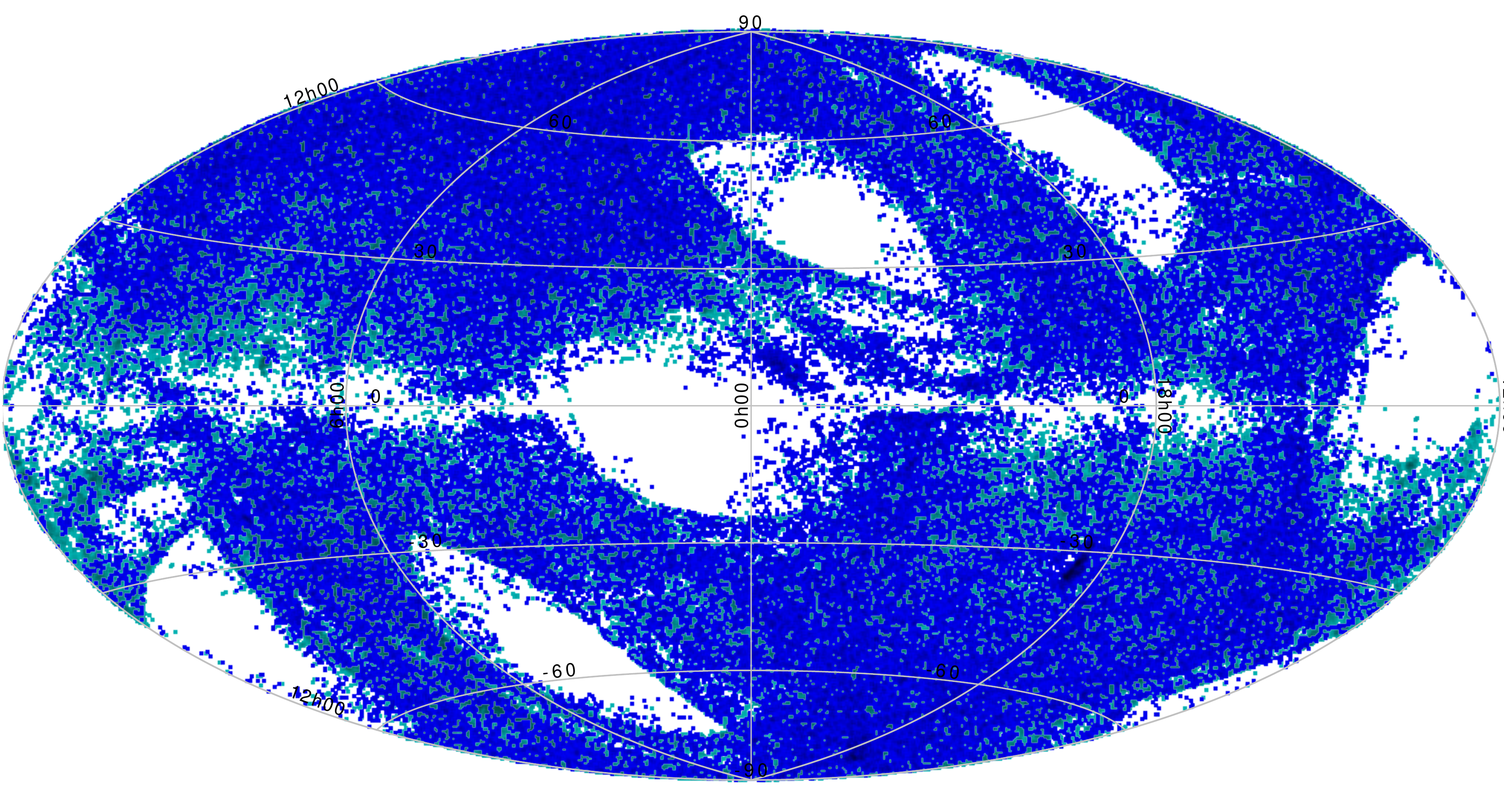}
\caption{
Distribution in Galactic coordinates (Hammer--Aitoff projection) of the quasars processed by the surface brightness profile module.
Blue points are quasars with a host galaxy detected (\texttt{host\_galaxy\_detected\,=\,true}) and turquoise points are those without a host galaxy. 
\label{cu4_qso_lb}}
\end{center}
\end{figure}

Figure~\ref{cu4_qso_lb} shows the spatial distribution on the sky of the $1\,103\,691$ quasars analysed. The coverage is inhomogeneous due to the limited sky coverage of the catalogues that constitute the quasar input list (Sect.~\ref{sec:modules:cu4}) but it also reflects the scanning law of \gaia,
as we only analyse sources that have at least 25 focal plane transits. The empty zones correspond either to the Galactic plane or to zones of lower frequency of scanning in \gdr{3}.

The distribution of the S\'ersic index for all the quasars has a mode at 0.9 and a mean of 1.9. These values are consistent with quasars  hosted by galaxies with disk-like light profiles, in agreement with a recent study of the surface brightness of host galaxies from the Hyper Suprime-Cam Subaru Strategic Program \citep{2021Li}.

The distribution of position angles of host galaxies is roughly uniform, as expected, although there is a small excess at around 90\degree.
These are sources with negligible ellipticity for which the position angle is meaningless. In such cases, our fitting algorithm favours a 90\degree position angle. The same is true for the galaxy sample discussed in the next section.

\begin{figure}[t]
\centering
\includegraphics[width=0.4\textwidth,angle=0]{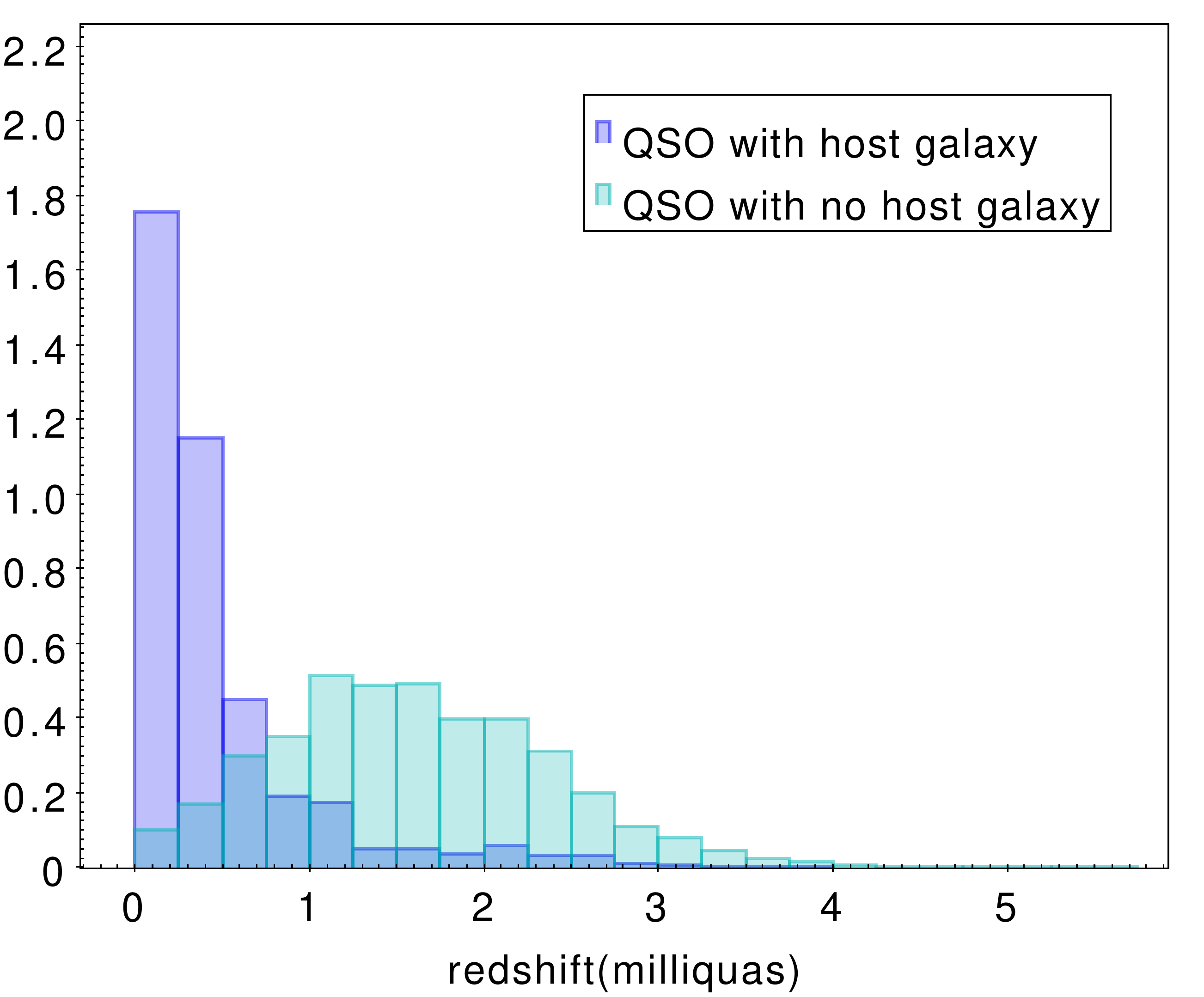}
\caption{Normalized distribution of the redshifts 
from Milliquas v7.2aa \citep{2021Flesch} 
of quasars that were analysed for surface brightness profiles. 
Blue shows the 2000 quasars for which a host galaxy was detected by \gaia, and turquoise the remaining 224\,000 quasars for which no host galaxy was detected.}
\label{fig:qso_red}
\end{figure}

226\,160 of the quasars processed
have a spectroscopic redshift listed in Milliquas 7.2 \citep{2021Flesch} (selection \texttt{TYPE\,=\,Q}).
2084 of these have a host galaxy detected by \gaia. Figure~\ref{fig:qso_red} shows the distribution of these redshifts. As expected, the quasars with a host galaxy have small redshifts (mean z=0.54) whereas those without a visible host galaxy have larger redshifts (mean z=1.71). In a few cases the host is detected for larger redshifts. These sources are usually very faint (G$>$20\,mag) and suffer either from uncertainties in the light profile fit or in the redshift measurement. 
The host galaxies resolved by \gaia\ have an effective radius (encompassing half of the total light) distribution with a peak at around 800\,mas. 

\subsubsection{Galaxies}

The surface brightness profile module processed 914\,837
galaxies. We see from Fig.~\ref{fig:afsm} that all of these have a clear spatial extension.

\begin{figure}[ht]
\centering
\includegraphics[width=0.49\textwidth,angle=0]{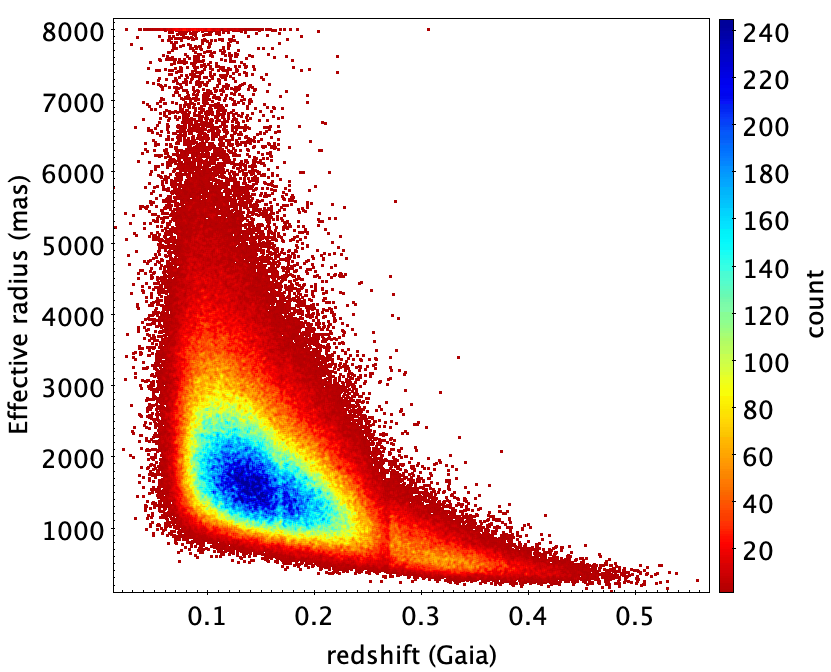}
\caption{Distribution of the effective radius (de Vaucouleurs profile) of galaxies processed by the surface brightness profile module as function of the redshifts measured by \gaia (\texttt{redshift\_ugc}).
\label{fig:gal_rad_red}
}
\end{figure}

The distribution of the effective radius of the de Vaucouleurs profile as measured by \gaia\ is shown in Fig.~\ref{fig:gal_rad_red} as a function of the \gaia\ redshifts (given by 
\linktoparam{sec_dm_extra--galactic_tables/ssec_dm_galaxy_candidates.html\#galaxy_candidates-redshift_ugc}{redshift\_ugc}
in table \linktotable{sec_dm_extra--galactic_tables/ssec_dm_galaxy_candidates.html}{galaxy\_candidates}). The redshifts are all below about 0.5, with a mean value of 0.16. 
As expected, the closer a source is to us, the larger its effective radius. There is a slight accumulation of effective radii at 8000\,mas, which corresponds to the bound of the parameter search domain, with the results that for larger galaxies the radius would remain at 8000\,mas. 

\begin{figure}
\begin{center}
\includegraphics[width=0.49\textwidth,angle=0]{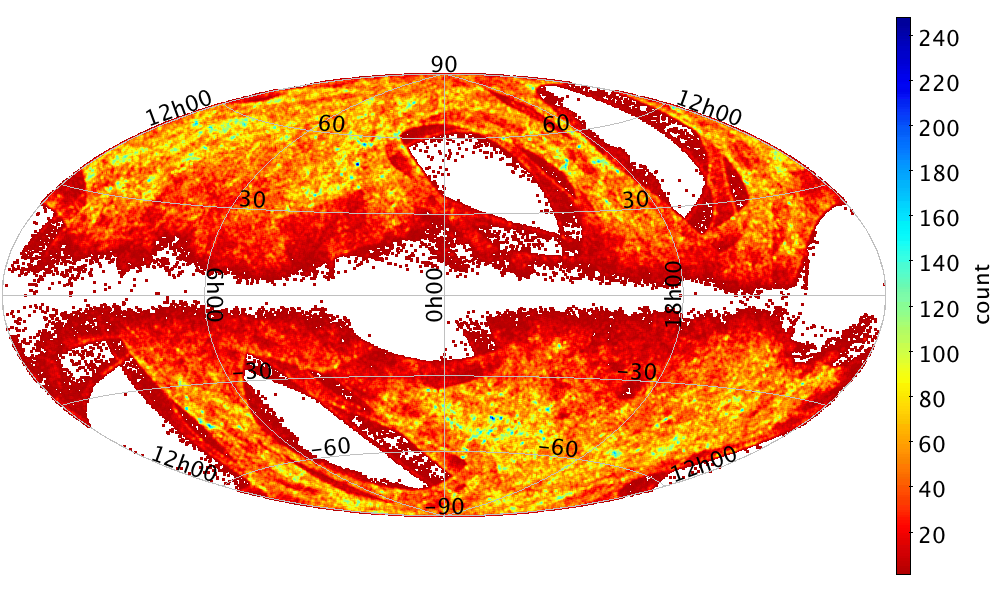}
\caption{Distribution in Galactic coordinates (Hammer-Aitoff projection) of the galaxies processed by the surface brightness profile module.  The colours show the density on a linear scale.
\label{cu4_gal_lb}}
\end{center}
\end{figure}

Figure~\ref{cu4_gal_lb} shows the distribution of these galaxies on the sky. As with the quasars in the previous section, we see an uneven distribution due primarily to the required minimum number of observations.

The distribution of the S\'ersic index peaks at around 4.5, which is consistent with the fact that the on-board detection algorithm favours elliptical types \citep{2014deSouza}. A few thousand galaxies have a S\'ersic index below 2, indicative of disk galaxies.
A visual inspection of a fraction of these reveals that most of them exhibit a compact bright bulge. 
The effective radius of the S\'ersic profile has a peak value around 1800\,mas and a de Vaucouleurs radius of around 1000\,mas, which is typical of sources with a mean redshift of 0.13.

The ellipticities derived from \gaia exhibit a peak value around 0.25. This is more or less what is expected from the projection of oblate ellipsoids (representative of elliptical) onto the plane of the sky and is also observed in other surveys, such as \cite{2008Padilla}.

\subsection{Light curves}\label{sec:light_curves}

\subsubsection{AGN}

\gdr{3} includes about a million variable AGN candidates in the \texttt{vari\_classifier\_result} table, which were selected mainly on the basis of their variability properties. For these, the epoch photometry in the \gmag, \gbp, and \grp bands is published in the {\tt light\_curve} datalink table. A complete description of the selection methods can be found in \citet{DR3-DPACP-165}, and are summarized below. More restrictive criteria were applied to achieve the higher purity sample comprising 872\,228 candidates in the \texttt{vari\_agn} table (Sect.~\ref{sec:modules:cu7}), the characteristics of which are analysed in \citet{DR3-DPACP-167}.

\begin{figure}[t]
\centering
\includegraphics[width=0.50\textwidth,angle=0]{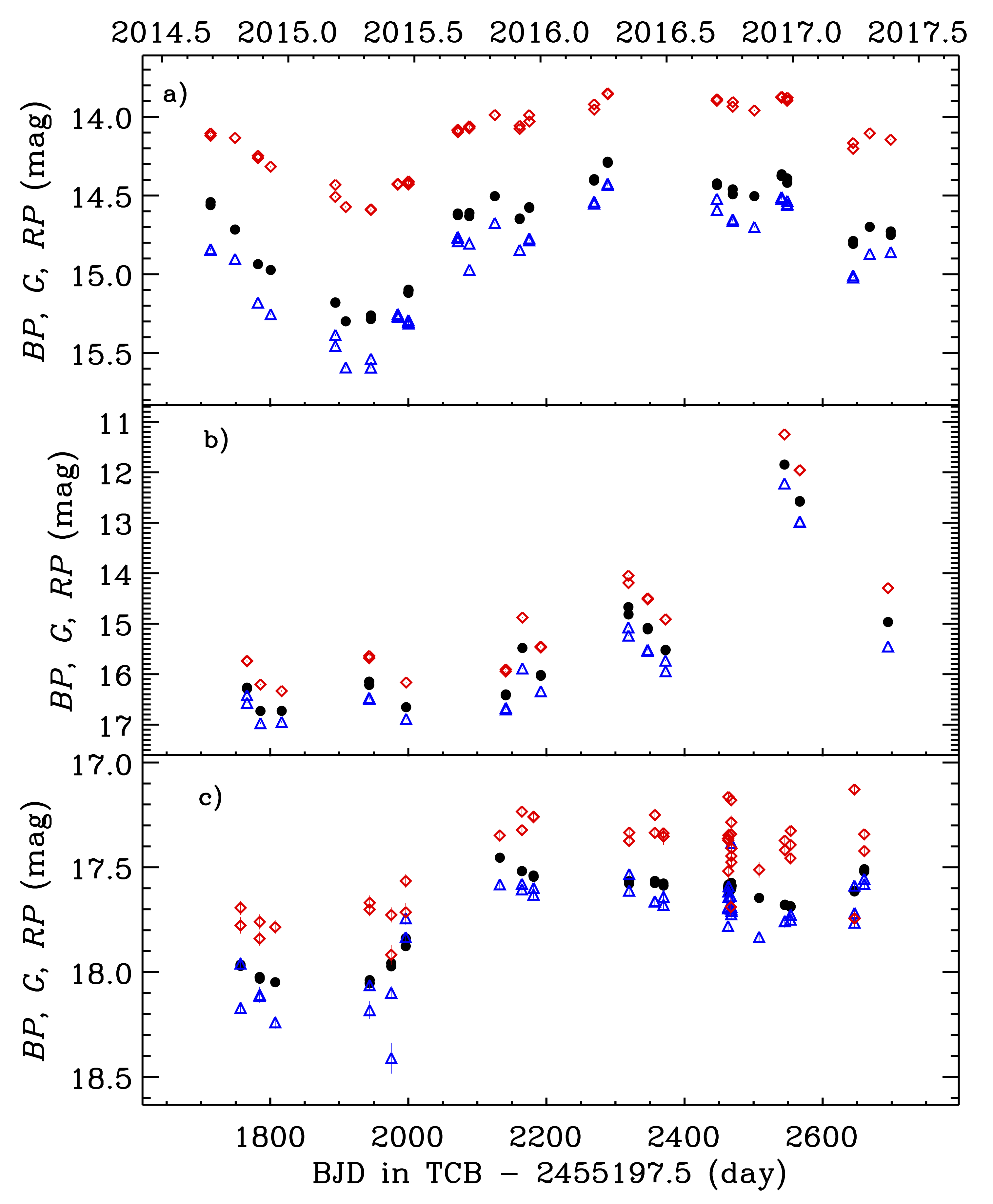}
\caption[Light curves of some variable AGN sources]
{Light curves in the \gmag (black dots), \gbp (blue triangles), and \grp (red diamonds) bands of some variable AGN sources. From top to bottom: a)~the type~1 Seyfert galaxy PG~0921+525 (source\_id 1019788071166861952); b)~the blazar CTA~102 (source\_id 2730046556694317312) caught during its historical 2016--2017 outburst; c)~the quasar B2~0945+22 (source\_id 640411921988216576).}
\label{fig:light_curves}
\end{figure}

Of the one million \gmag-band light curves of the variable AGN, 90\% contain between 20 and 244 focal plane transits 
covering 795 to 1038\,days (after applying time series filters described in Sect.~10.2 of the online documentation). On average they have 39 focal plane transits over 925\,days, which is sufficient to follow the long-term variability of most AGN. 
Figure~\ref{fig:light_curves} shows the light curves in the \gmag, \gbp, and \grp bands of three sources belonging to different AGN classes: a)~the type~1 Seyfert galaxy PG~0921+525; b)~the blazar CTA~102, which was observed during its historical 2016--2017 maximum \citep{2017raiteri}, resulting in the most variable object of the sample; c)~the quasar B2~0945+22.

Photometrically-variable AGN candidates from supervised classification were verified and further down-selected by a series of filters that use \gaia-CRF3 \citep[][]{EDR3-DPACP-133} as a reference sample. 
Variability-related constraints were set on: the index of the structure function \citep[][{\tt structure\_function\_index} in the table {\tt vari\_agn}]{1985simonetti};
quasar versus non-quasar metrics \citep[][{\tt qso\_variability} and {\tt non\_qso\_variability} in table {\tt vari\_agn}]{2011butler};
the Abbe (also called von~Neumann) parameter ({\tt abbe\_mag\_g\_fov} in table {\tt vari\_summary}) in the \gmag band versus the renormalized unit weight error of the astrometric solution ({\tt ruwe} in table {\tt gaia\_source}).
Additional cuts were made in the \bpming versus \gminrp colour space,
on parallaxes and proper motions, on the environment source number density (to avoid crowded sky regions), on the scan angle correlation with photometric variation \citep[to remove artificial effects; see][]{DR3-DPACP-164}, and finally on the variability probability (to deal with clearly variable objects).

\subsubsection{Galaxies} \label{subsubsec:lightcurve_galaxies}

\begin{figure}[t]
\centering
\includegraphics[width=0.45\textwidth,angle=0]{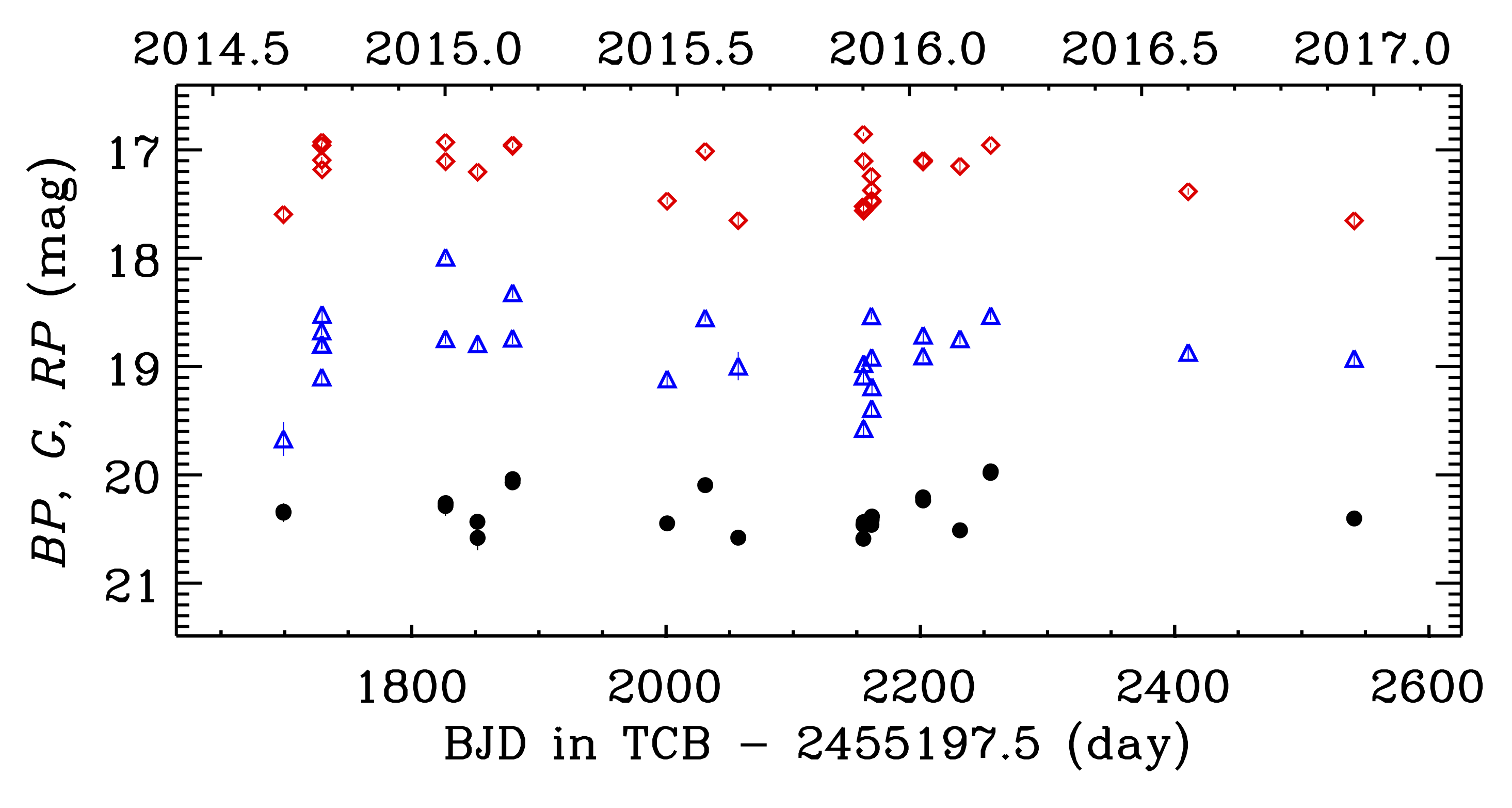}
\caption[Light curves of galaxy LEDA 2268723]
{Light curves in the \gmag\ (black dots), \gbp\ (blue triangles), and \grp\ (red diamonds) bands of the known galaxy LEDA~2268723 (source\_id 377643902971151872).}
\label{fig:light_curves_gal}
\end{figure}

\begin{figure}[t]
\centering
\includegraphics[width=0.45\textwidth, angle=0]{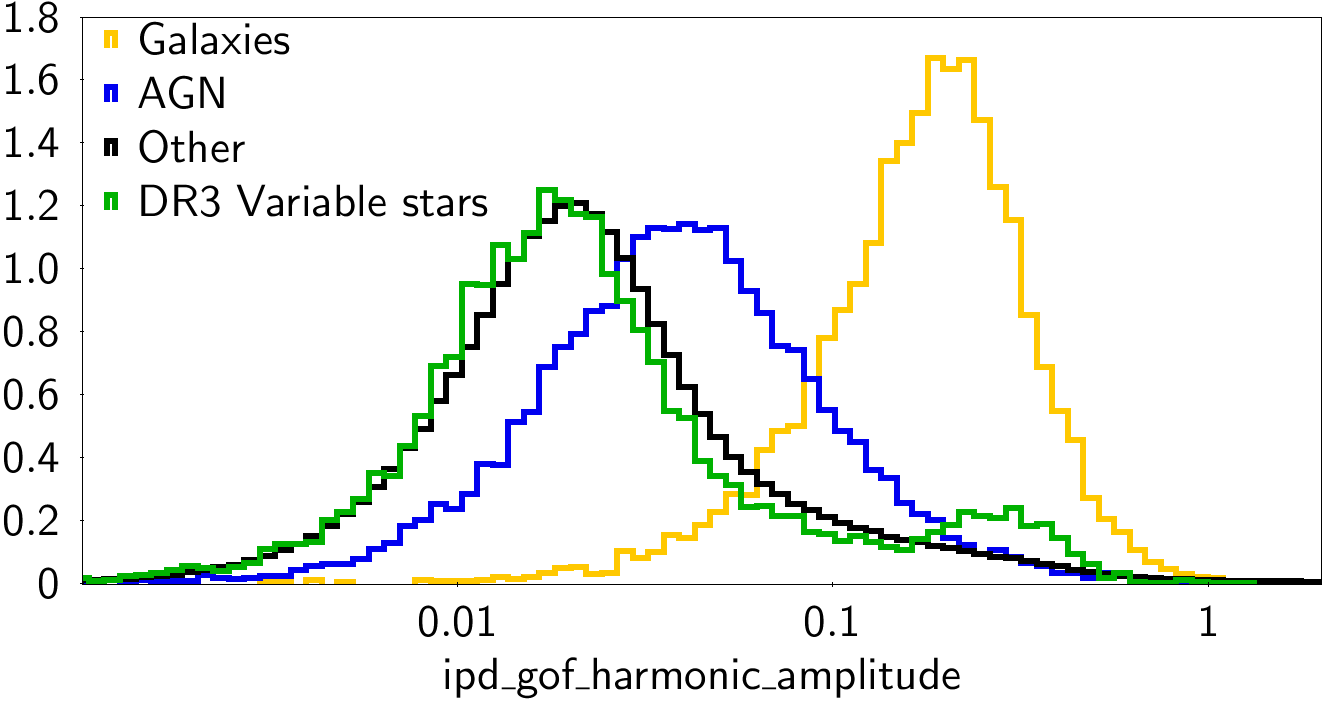}
\caption{Distributions (normalized by area) of the field {\tt  ipd\_gof\_harmonic\_amplitude} for sources of various classes in the \gaia Andromeda Photometric Survey. \new{'Other' includes constants and those variable objects that were not targeted in \gdr{3}}.}
\label{fig:gal_ipd}
\end{figure}

\begin{figure}[t]
\begin{center}
\includegraphics[width=0.45\textwidth, angle=0]{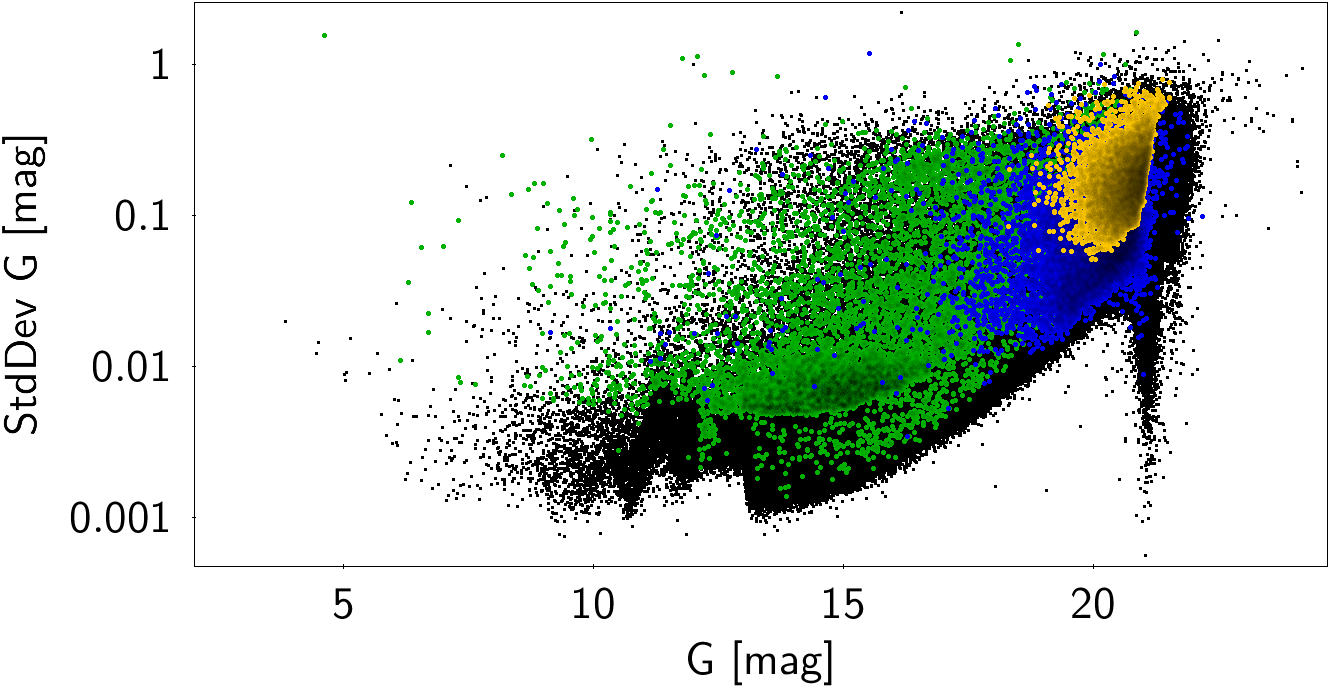}\\
\vspace{2mm}
\includegraphics[width=0.45\textwidth, angle=0]{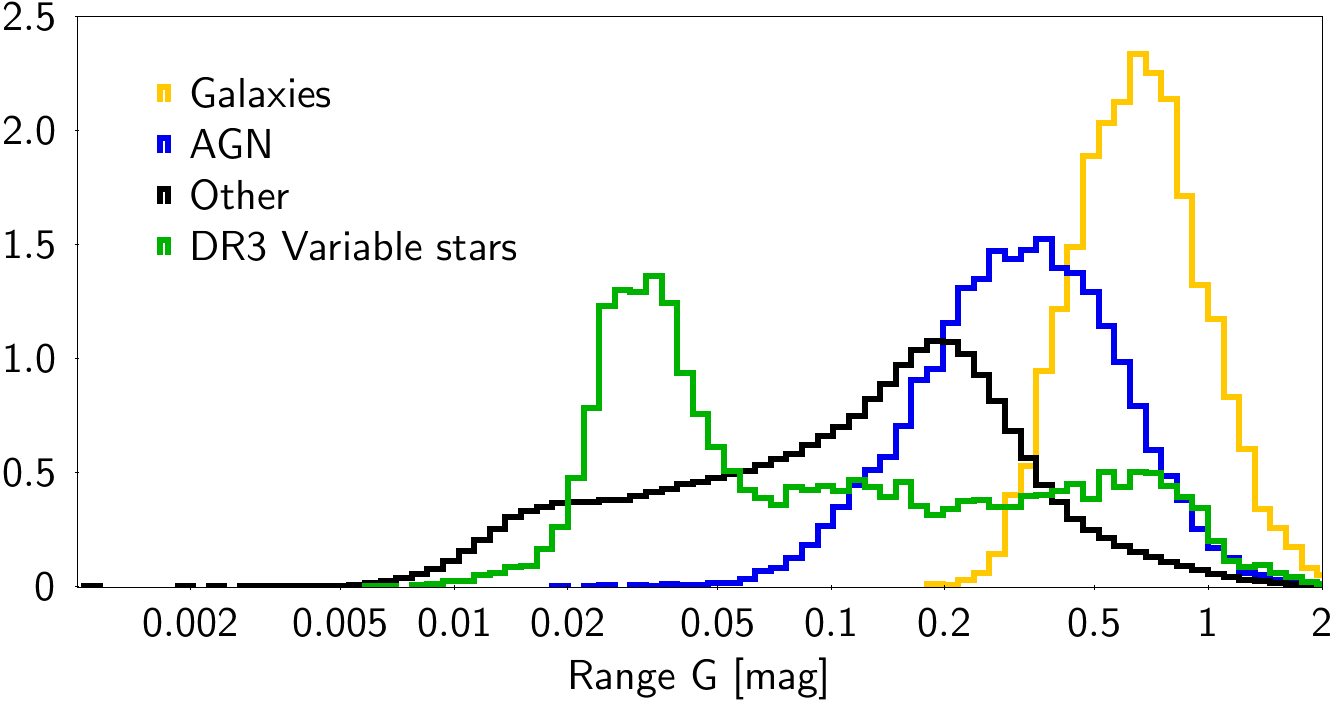}
\caption{Statistics of light curves of objects in the \gaia Andromeda Photometric Survey. Top: Standard deviation versus median \gmag magnitude. Bottom: Normalized distribution of minimum-to-maximum variability range for \gmag~band light curves. Both panels are colour-coded as in Fig.~\ref{fig:gal_ipd}. 
\new{The distributions overlap in the upper panel, with galaxies covering AGN, for example.}
\label{fig:gal_lc_stats}
}
\end{center}
\end{figure}

About 2.5~million galaxies in the \texttt{galaxy\_candidates} table were selected based on the properties of their light curves. (Only a subset of these light curves are published in \gdr{3}; see Sect.~\ref{sec:modules:cu7}.)
\gaia scans individual objects multiple times at different position angles. For extended objects this can produce an apparent -- but spurious -- photometric variability, because on each scan only part of the total flux is collected by the limited size in the allocated window \citep{DR3-DPACP-164}. Figure~\ref{fig:light_curves_gal} shows the light curve of a known galaxy, in which we see variations in excess of 0.6~mag in \gmag. Figure~\ref{fig:gal_ipd} shows the  distribution (normalized by area) of the parameter
\linktoparam{sec_dm_main_source_catalogue/ssec_dm_gaia_source.html\#gaia_source-ipd_gof_harmonic_amplitude}{ipd\_gof\_harmonic\_amplitude}
\new{for galaxies, AGN, \gdr{3} variable stars, and other objects in the \gaia Andromeda Photometric Survey}.
\new{This parameter measures the amplitude of the variation of the Image Parameters Determination goodness-of-fit statistic as function of the scan direction angle. Because galaxies are often extended objects at the \gaia resolution, they tend to have a larger value of this parameter than other types of objects.}
The galaxies that are detected by variability are based on this type of spurious signal. 
Figure~\ref{fig:gal_lc_stats} shows
how the magnitude variability distribution of galaxies within 5.5$^{\circ}$ of the Andromeda Galaxy (M31) compares to 
that of other sources \new{in the same classes as in Fig.~\ref{fig:gal_ipd}. We note that \gdr{3} variable stars  still amount to a relatively small fraction of all variables detected in \gaia}. 
The brightness variations of galaxies overlap with  high-amplitude tails of the distributions of other classes.

\section{Internal comparison}\label{sec:internal_comparison}

\subsection{Classification}\label{subsec:classif}

\begin{table*}[t]
\caption{Comparison of the classes of sources in the {\tt qso\_candidates} table according by its contributing modules. Each element gives the number of sources with different classifications between any two modules, expressed as a fraction of the number of sources in common between those two tables. 
Sources labelled unclassified in 
\linktoparam{sec_dm_extra--galactic_tables/ssec_dm_qso_candidates.html\#qso_candidates-classlabel_dsc}{classlabel\_dsc}
and
\linktoparam{sec_dm_extra--galactic_tables/ssec_dm_qso_candidates.html\#qso_candidates-classlabel_dsc_joint}{classlabel\_dsc\_joint}
are excluded. 
The columns list all the modules that provide classifications.
The rows list all modules that add sources to the table: the last four of these are not classifiers, but provide sources based on other labels.
  \label{tab:intcomparison:misclass_qso}}
  \centering  
  \begin{tabular}{ccccccc}
 \hline\hline 
 \noalign{\smallskip}
  \multirow{2}{*}{\diagbox{QSO}{Other} } & DSC & DSC Joint & Variability  & OA \\ 
                         & classification & classification & classification & classification\\
\noalign{\smallskip}
\hline
DSC classification &  & 0.0 & ~5.7 & 59.5    \\ 
DSC Joint classification & ~0.0  &  & ~0.9 & 53.9 \\ 
Variability classification & ~9.5 & 0.5 &  & 44.6  \\ 
OA classification & 30.2 & 1.1 & ~7.1 &  \\ 
Vari-AGN & ~7.3 & 0.5 & ~0.0 & 46.2  \\ 
Surface brightness & 20.0 & 2.3 & ~0.4 & 46.6  \\
\gaia-CRF3 & 20.8 & 2.0 & ~0.2 & 42.3  \\ 
QSOC & 43.2 & 0.1 & 10.8 & 61.4 \\
\hline 
\end{tabular} 
\end{table*}

\begin{table*}[t]
  \caption{As Table \ref{tab:intcomparison:misclass_qso} but for the {\tt galaxy\_candidates} table.
  \label{tab:intcomparison:misclass_gal}}
  \centering  
  \begin{tabular}{ccccccc}
 \hline\hline 
 \noalign{\smallskip}
  \multirow{2}{*}{\diagbox{Galaxy}{Other} } & DSC & DSC Joint & Variability  & OA \\ 
                         & classification & classification & classification & classification\\
\noalign{\smallskip}
\hline
DSC classification &  & 0.1 & 1.5 & 13.1 &  &  \\ 
DSC Joint classification & ~0.0  &  & 2.8 & ~2.9 \\ 
Variability classification & 37.6 & 0.0 &  & ~0.7  \\ 
OA classification & 59.7 & 6.4 & 1.3 &  \\ 
UGC & ~1.2 & 0.1 & 1.0 & ~7.3  \\ 
Surface brightness & 42.0 & 0.0 & 0.0 & ~2.3  \\ 
\hline 
\end{tabular} 
\end{table*}

The various extragalactic modules (Sect.~\ref{sec:modules})
use different methods and data. This leads to a given source being classified differently in different modules, which is apparent in the \texttt{qso\_candidates} and \texttt{galaxy\_candidates} tables that collate results from all modules.
On top of this comes the fact that different modules use different definitions of quasar and galaxy, in particular in the case of supervised learning algorithms, where the class is defined by the training data set.
Tables \ref{tab:intcomparison:misclass_qso} and \ref{tab:intcomparison:misclass_gal} show the 
percentage of different classifications of the overlapping sources between modules based on the class labels where they exist (DSC, DSC-Joint, OA, Vari-Classification) or 
the existence of parameters (from UGC, QSOC, Vari-AGN, Surface brightness).
For example, {\tt classlabel\_dsc} and Variability give different classes for 9.5\% of their common sources. 
Such disagreements also come about because some modules focus more on high completeness, whereas others focus more on high purity (partially achieved by filtering). Recall also that the classification from a module appears in the table even if that source would not have been selected for inclusion in the table by that particular module (see Table~\ref{tab:contribution}).
QSOC for quasars and UGC for galaxies are subsets of DSC selected with the properties described in Sections~\ref{sec:modules:qsoc} and~\ref{sec:modules:ugc}. Both use much lower thresholds on the DSC probabilities than do the DSC class labels.

\gaia-CRF3 does not distinguish between galaxies and quasars. Most are expected to be quasars so all are all in the {\tt qso\_candidates} table. 
OA works with a small fraction of sources that are generally faint and noisy so the comparison between OA and other modules should be carefully interpreted. 

\begin{figure} \begin{center}
\includegraphics[width=0.48\textwidth,angle=0]{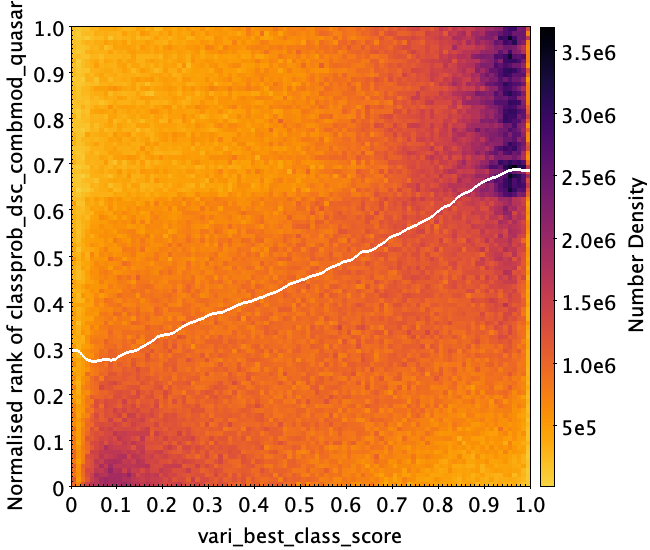}
\caption{Comparison of DSC quasar classification probabilities (transformed to normalized ranks) with scores from the variability analysis. Darker colours depict higher densities,
and the white line indicates the median rank.
We see a broad agreement between the highest and lowest ranked quasars. 
\label{fig:cu7_cu8_score_qso}} 
\end{center} \end{figure}

DSC provides posterior class probabilities. {\tt vari\_best\_class\_score} (from Vari) provides the median normalized rank, which also increases from 0 to 1 with increasing reliability, but it is not a probability.
To compare these quantities, we map the DSC probabilities into normalized ranks. Figure~\ref{fig:cu7_cu8_score_qso} compares this for {\tt classprob\_dsc\_combmod\_quasar} 
to {\tt vari\_best\_class\_score}. The deviation from a perfect correlation reflects the difference in input data types, training sets and class definitions, and classification methods in general.

\subsection{Redshift}\label{subsec:redshift}

\begin{figure} 
\begin{center}
\includegraphics[width=0.5\textwidth,angle=0]{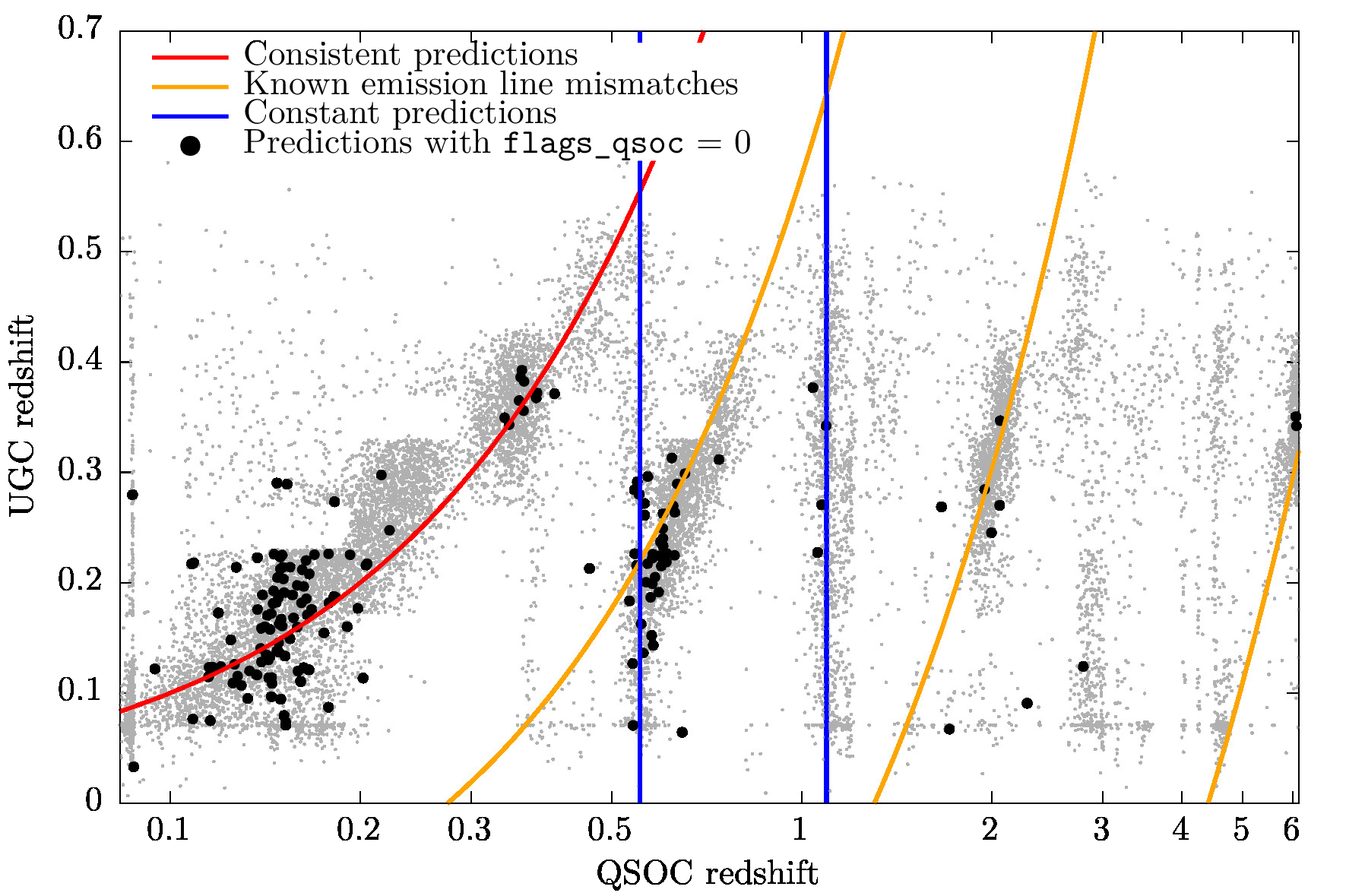}
\caption{Comparison between the UGC and QSOC redshifts. Grey dots correspond to all redshifts in common between the two tables, while black dots are restricted to those with \texttt{flags\_qsoc\,=\,0}, which corresponds to a higher reliability subset. The red curve denotes identical predictions in the two modules. Yellow curves highlight mismatches between common quasar or AGN emission lines, as explained in \cite{DR3-DPACP-158}, while the blue vertical lines show constant predictions by QSOC.
\label{fig:qsoc_vs_ugc_full}
} 
\end{center} 
\end{figure}

Redshifts are derived by two modules, the results of which are reported in the \texttt{qso\_candidates} table (from QSOC) and the \texttt{galaxy\_candidates} table (from UGC). Of the 174\,146
sources in common between the two tables, 16\,534 
have a redshift derived by both modules. These are compared in Fig.~\ref{fig:qsoc_vs_ugc_full}.
7\,469
of these sources have predictions with $| \Delta z | < 0.1$, and the correlation improves when restricting the comparison to QSOC redshifts with higher reliability (black dots in Fig.~\ref{fig:qsoc_vs_ugc_full}): In this subset, 105 of 166 
sources have $| \Delta z | < 0.1$. Specific discrepancies arise from emission line mismatches in the QSOC redshift determination. As QSOC aims to be complete, it processes galaxies, even though UGC -- by design -- generally gets better predictions on these objects (see \citealt{DR3-DPACP-158} for a more detailed explanation of these emission line mismatches). UGC, in contrast, aims to be pure and is accordingly not expected to process a significant number of quasars.
Figure~\ref{fig:qsoc_vs_ugc_full} shows loci of
constant QSOC redshifts. These are probably erroneous matches at the BP/RP spectral borders, where wiggles from the Hermite polynomials are confused with quasar emission lines in the templates.

\begin{figure}[t]
    \centering
    \includegraphics[width=0.40\textwidth,angle=0]{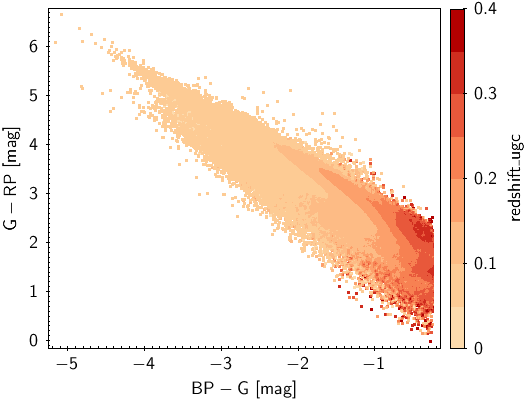}
    \caption{Colour-colour diagram for the 1\,367\,153 galaxies for which redshifts are provided by UGC, colour-coded by redshift. A small number of sources have redshifts extending up to 0.6}
    \label{fig:ugc_ccd}
\end{figure}

Figure~\ref{fig:ugc_ccd} shows the colour--colour diagram for all sources for which UGC provides a redshift value, colour-coded by redshift. We see that galaxies generally become redder in \bpming, but bluer in \gminrp as redshift increases from 0 to 0.4.

\subsection{Sources with stellar parameters}\label{subsec:ap_rv}

The extragalactic tables contain sources for which stellar astrophysical parameters are also reported in \gdr{3}. This is expected, because stellar parameters were inferred for sources independently of their classification status \citep{DR3-DPACP-157}. 
There are 255\,948 
sources in the \texttt{qso\_candidates} table and 7069 
sources in the
\texttt{galaxy\_candidates} table that have effective temperatures derived by the CU8 GSP-Phot module \citep{DR3-DPACP-160}. Checking a variety of metrics such as magnitude, sky distribution, and effective temperature itself, there is nothing apparently peculiar with these sources. Their presence is an inevitable consequence of the known stellar contamination. It is also important to remember that DSC, which is the single largest contributor to these integrated tables, did not filter out sources simply because they were bright (only DSC-Allosmod classifies sources with $G<14.5$\,mag to be stars).

There are also 4027
sources with valid radial velocities in the \texttt{qso\_candidates} table, and 160
in the \texttt{galaxy\_candidates} table. Considering that the extragalactic tables are mostly populated with faint sources, these small numbers are essentially due to the intrinsic magnitude limit of sources for which radial velocities could be derived in \gdr3 \citep{DR3-DPACP-159}. Those featuring valid radial velocities have magnitudes that are usually incompatible with extragalactic sources, so it is fair to assume that they are stars. 

\subsection{Astrometric selection}\label{sec:astrometric_analysis}

Additional insight into the classified sources can be gained by analysing their astrometric parameters. As has been demonstrated in \cite{EDR3-DPACP-133} and \cite{2021klioner}, astrometry can be used to improve the purity of a sample of quasar candidates. It is clear, however, that this can only be achieved at the cost of reducing the completeness.

The procedure here is similar to that used in the construction of \gaia-CRF3, namely a two-step astrometric filtering of a sample of candidates \citep{EDR3-DPACP-133}. In the case of \gaia-CRF3, the sample was obtained by cross-matching the \gedr{3} catalogue with several external quasar catalogues. In the present study, each of the \gaia classifiers contributing to the {\tt qso\_candidates} table is considered as an additional catalogue, and the same procedure is applied to all the external and \gaia-own selections of quasar candidates.

The first step of the astrometric filtering is to select individual sources that have high-quality astrometric solutions in \gedr3 and statistically insignificant parallaxes and proper motions \citep[see][Sect. 2.1 for the
  exact mathematical formulations]{EDR3-DPACP-133}.
This step alone is insufficient to find genuine quasars (or
extragalactic objects), as about 214 million sources in \gedr3, dubbed
'confusion sources', satisfy these astrometric criteria. These are
mostly stars of our Galaxy \citep[][appendix C]{EDR3-DPACP-133}. At least at this stage of the \gaia project,
astrometry cannot be used as an independent quasar classifier, although this may change in the future \citep[see e.g.][]{2015A&A...578A..91H,2018A&A...615L...8H}.

A second step of filtering is therefore needed. In this step, only those samples of sources are retained that show near-Gaussian distributions in the uncertainty-normalized parallaxes and proper motions. Since extragalactic sources are faint, the typical uncertainties of their astrometric parameters in \gdr3 are about two orders of magnitude larger than either the known level of systematic errors in \gdr3 \citep{2021A&A...649A...4L} or the known physical systematic effects \citep{2021klioner}. Bearing in mind that the true parallaxes and
proper motions of genuine extragalactic sources should be zero, one expects Gaussian distributions of the normalized parameters. This
requirement had proven to be very useful to distinguish genuine quasars from the confusion sources.

Both steps of the astrometric filtering obviously reject some genuine quasars that have considerable measured, but spurious, proper motions due to time-varying source structure (see Sect.~\ref{sec:modules}). A prominent example here is 3C273, which is not part of \gaia-CRF3 for this reason. The samples considered at the second
step of the astrometric filtering can come from a particular external catalogue or from one of the \gaia classifiers, but could also be selections according to various criteria (e.g.\ avoiding the crowded areas on
the sky) or intersections of such selections (e.g.\ sources
that were found to be quasars by two classifiers).

An additional characteristic of a sample of genuine extragalactic objects is that its sky distribution should  not show overdensities in known stellar structures in our Galaxy and its environments, such as clusters, although it could still be influenced by such structures, for example variable Galactic extinction. This can also be used to help decide whether a particular sample of sources should be retained.

\begin{figure}
\begin{center}
\includegraphics[width=0.48\textwidth,angle=0]{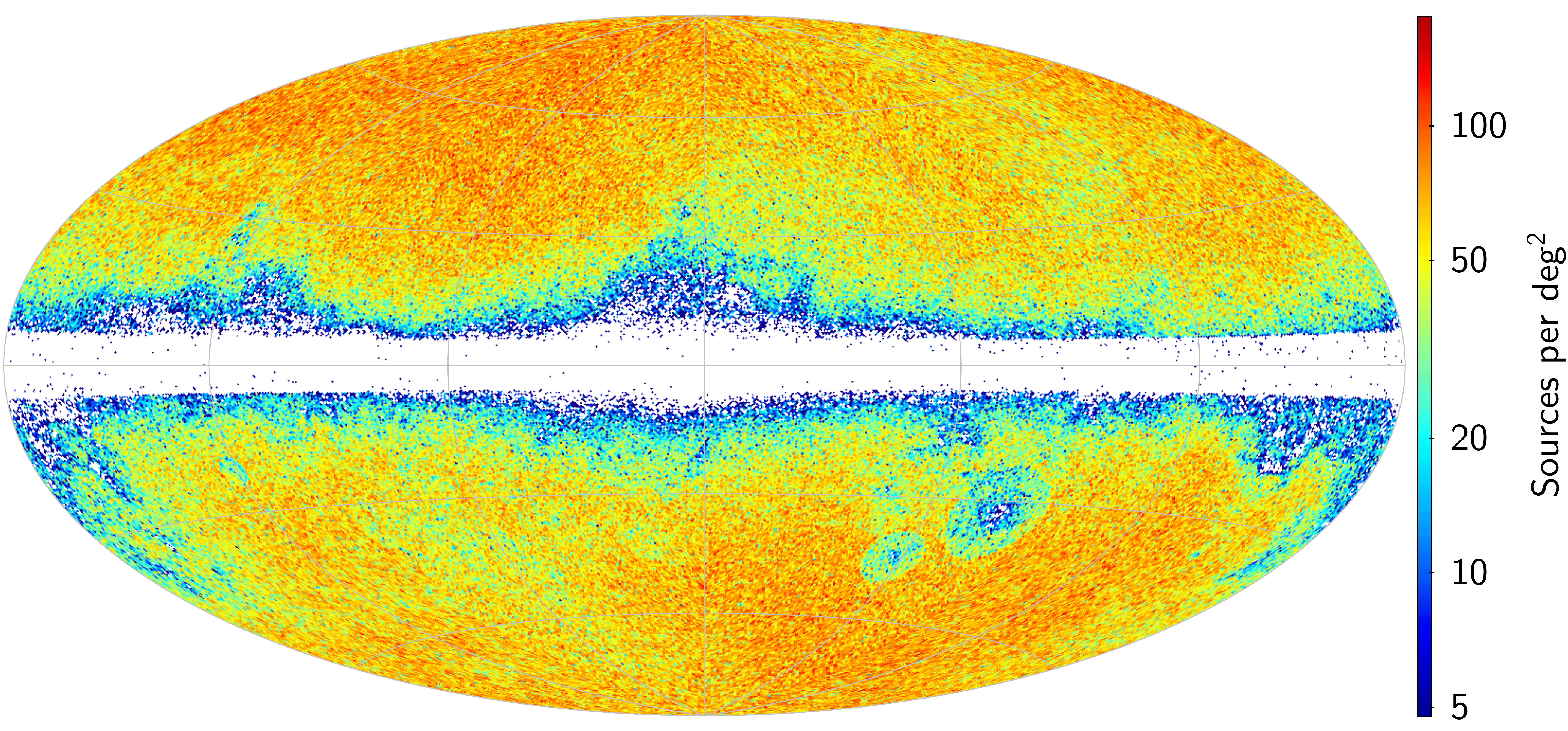}
\caption{Distribution Galactic coordinates (Hammer--Aitoff projection) of the 1\,897\,754 
sources from the astrometric selection (i.e.\ sources in the {\tt qso\_candidates} table with {\tt astrometric\_selection\_flag} set).
 The plot shows the density of sources per square degree computed from the source counts per pixel at HEALPix level 7 (pixel size $\simeq 0.21$\,deg$^2$)
\label{fig:astrometry-density}} 
\end{center}
\end{figure}

Using this two-step selection procedure we have identified a set of 1\,897\,754
quasar candidates, which we refer to as the 'astrometric selection'. They are indicated by the 
\linktoparam{sec_dm_extra--galactic_tables/ssec_dm_qso_candidates.html\#qso_candidates-astrometric_selection_flag}{astrometric\_selection\_flag} in the {\tt qso\_candidates} table.
The purity of this sample is difficult to estimate, but we believe it to be 98\% or perhaps better. 
The vast majority of these sources were identified as quasars by at least two independent external catalogues and/or \gaia classifiers. 
The density distribution
of these sources on the sky is shown on
Fig.~\ref{fig:astrometry-density}.  This set contains 1\,406\,729 sources
(74\%) with 5p astrometric solution and 491\,025 sources (26\%) with 6p solutions in \gdr3. The avoidance zone in the Galactic plane as well as the lower density of sources around the LMC and SMC result from the difficulty in reliably identifying quasars in those crowded areas. This concerns both the external catalogues and the \gaia classifiers.

\begin{figure}[t]
  \begin{center}
  \includegraphics[width=0.9\hsize]{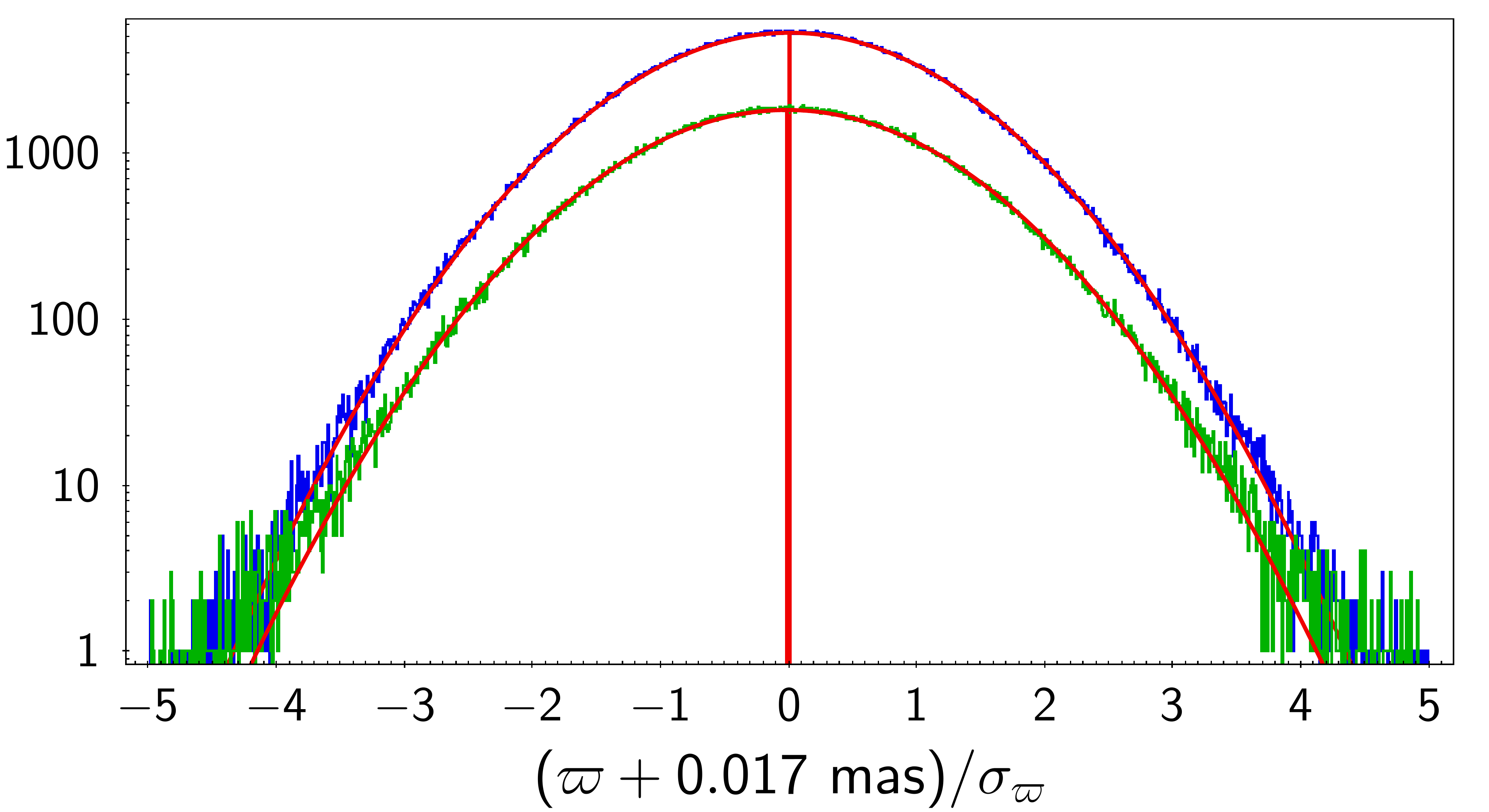} 
  
  \includegraphics[width=0.9\hsize]{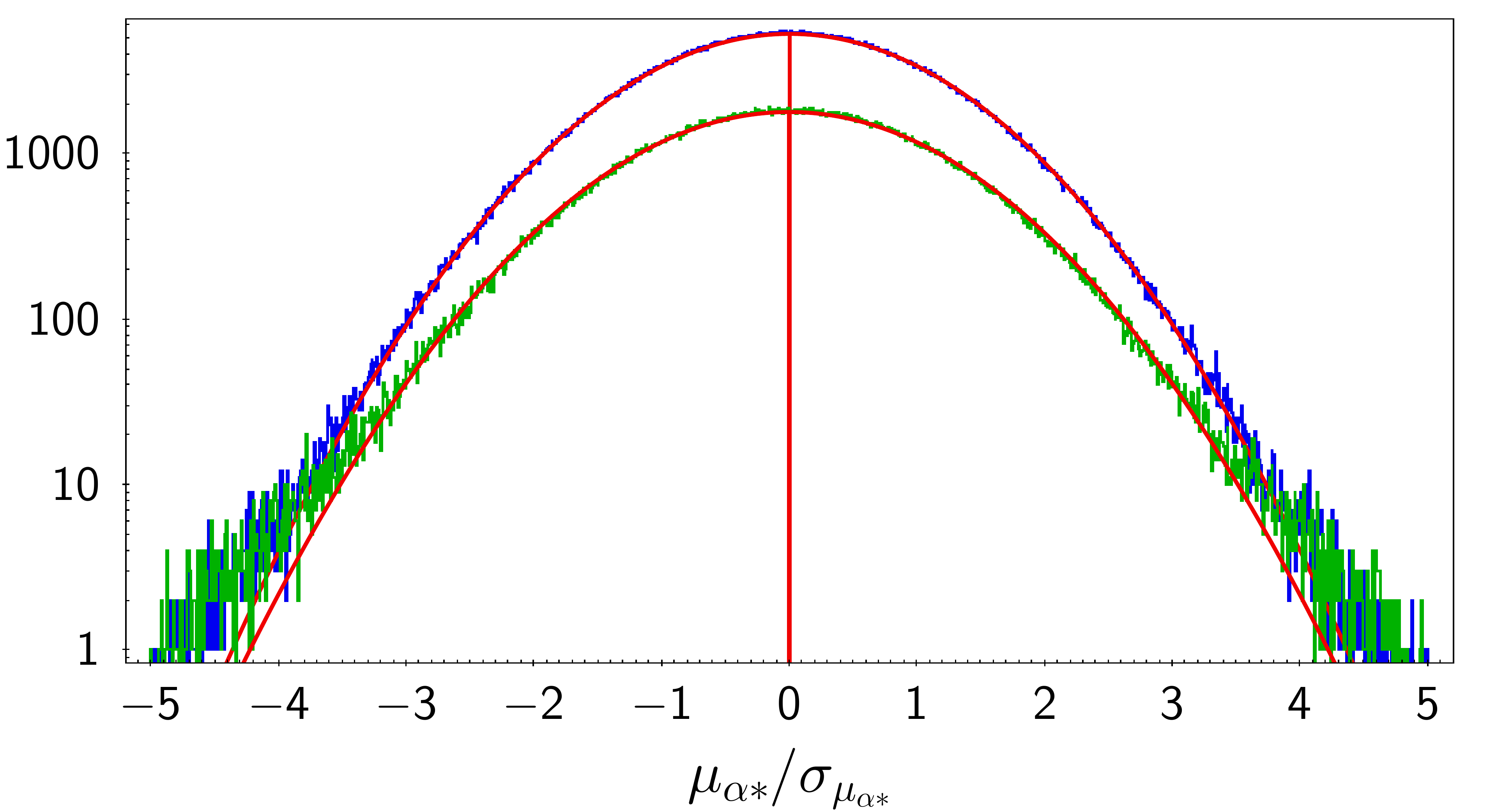}
  
  \includegraphics[width=0.9\hsize]{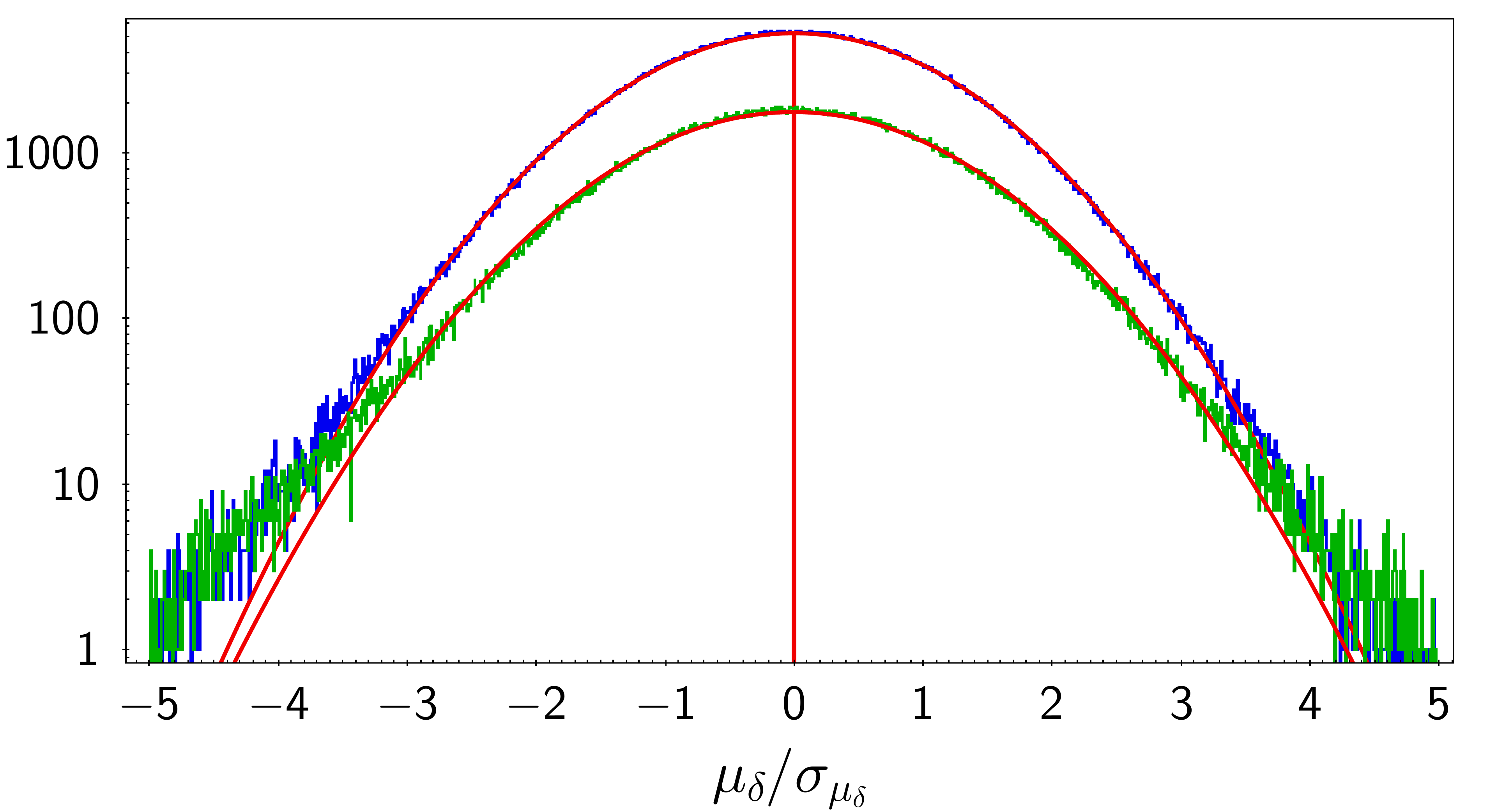}
  \caption{Distributions of the normalized parallaxes and proper motion components 
    for the sources in the astrometric selection with 5p (blue) and 6p (green) solutions. 
    The red curves show the corresponding best-fit Gaussian distributions. The global parallax zero point of
    $-0.017$~mas of \gdr3 is taken into account \citep{2021A&A...649A...2L,2021A&A...649A...4L}. The standard deviations of the best-fit Gaussian distributions for the sources with 5p (6p) solutions are 1.048 (1.068), 1.054 (1.092) and 1.063 (1.109) for the parallaxes, and proper motions in right ascension and declination, respectively. As usual in \gaia, the asterisk in $\alpha*$ in the middle panel indicates the implicit factor $\cos\delta$, that is $\mu_{\alpha*}=\dot\alpha\,\cos\delta$.
  }
\label{fig:normalizedAstrometry-histograms}
\end{center}
\end{figure}

Fig.~\ref{fig:normalizedAstrometry-histograms} shows the distributions of
the normalized parallaxes and proper motions of the astrometric selection. They are 
close to Gaussian, which suggests a reasonably low level of stellar contamination.
The standard deviations of the best-fit Gaussian distributions range from 
$1.05$ to $1.11$ and indicate by how much the formal
uncertainties of the corresponding astrometric parameters may be 
underestimated in \gdr3.



We attempted a similar astrometric selection  for the {\tt galaxy\_candidates}
table. However, since most of the galaxies have only two-parameter
astrometry in \gdr3, and as more problems in \gaia astrometry can be
expected for extended sources, the astrometric selection for the galaxy table turned out to
be less useful, so we decided not to publish it.
Nonetheless, this analysis did reveal the properties of the population of sources in the astrometric selection that were classified as both quasars and galaxies: the astrometric selection from the {\tt qso\_candidates} table contains 54\,892 sources that are also present in the {\tt galaxy\_candidates} table (cf.\ overall overlap of these tables of 174\,146 
sources). 99\% of those sources have 6p astrometric solutions. The normalized parallaxes and proper motions of this set of sources also have near-Gaussian distributions, but with standard deviations
of 1.13--1.25, which is about 10\% larger than for the  astrometric selection as a whole. This set of sources is probably dominated by AGN for which
source structure (i.e.\ the host galaxy) notably affects the astrometric
solution. Indeed, a host galaxy was detected by \gaia for 23\,805 of
these sources (43\%). Similar statistics of the normalized astrometric parameters can also be found for the set of sources in the astrometric selection for which a host galaxy was detected by \gaia (see Sect.~\ref{subsec:cu4eo_properties}), which contains 51\,586 sources.   

Thus we encounter the problem in the optical
that is well known in radio astrometry \citep[e.g.][]{2020A&A...644A.159C}, namely the influence
of the source structure on the quality of the astrometry. This topic
will need a special attention in the future \gaia data releases.

\subsection{Analysis of objects with lower probability classifications}\label{sec:weird}

\begin{figure}[t] 
\begin{center}
\includegraphics[width=0.45\textwidth,angle=0]{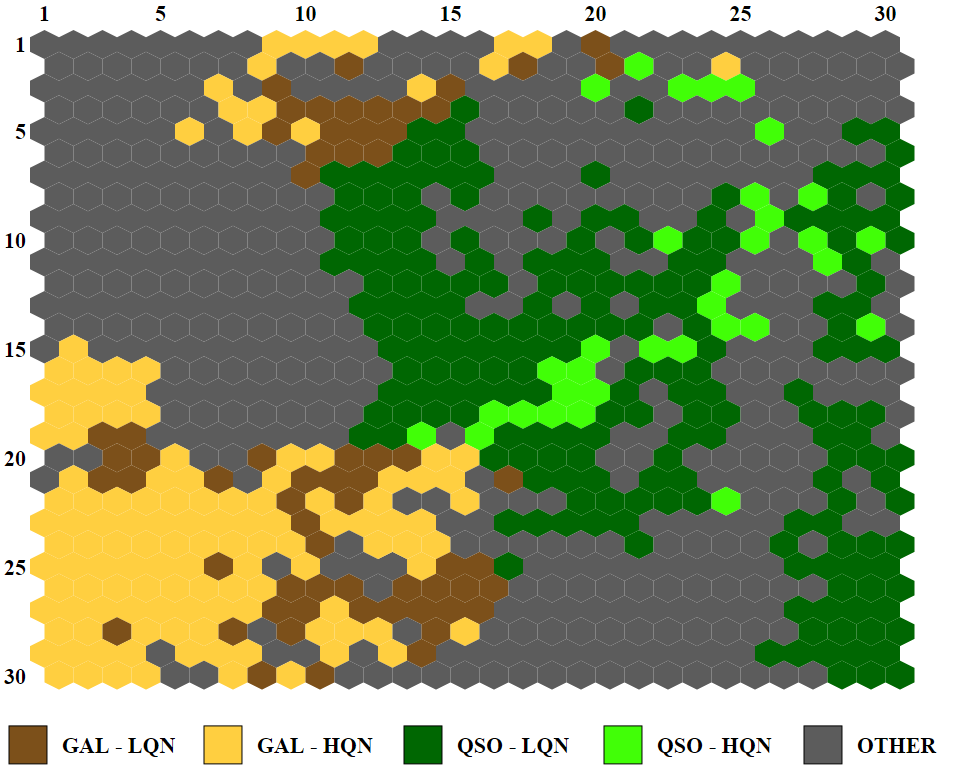}
\caption{OA class labels for extragalactic objects. HQN = high quality neuron (quality 0--3), LQN = low quality neuron (quality 4--6). 
}
\label{fig:oa_labels_extragalactic}
\end{center}
\end{figure}

\begin{figure}[t] 
\begin{center}
\includegraphics[width=0.4\textwidth]{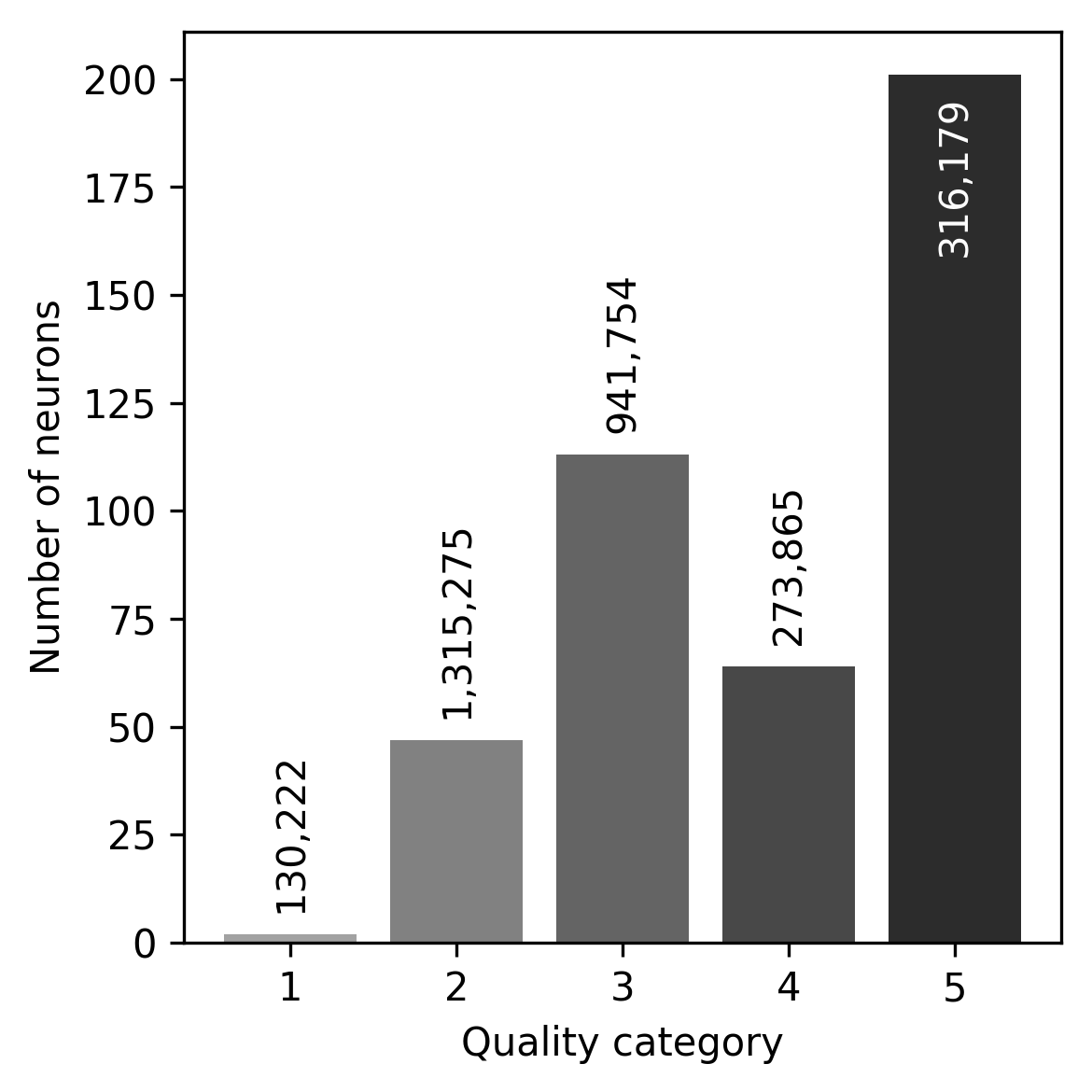}
\caption{Distribution of OA neurons labelled as extragalactic for each quality category. The number on each bar gives the number of sources in the {\tt qso\_candidates} or {\tt galaxy\_candidates} tables. No extragalactic objects appear in any of the best (0) or worst (6) quality neurons.}
\label{fig:oa_quality_categories}
\end{center}
\end{figure}

The unsupervised algorithm OA was used to analyse the sources with lower DSC class probabilities (Sect.~\ref{sec:modules:oa}).
Here we focus on those neurons that were assigned to an extragalactic class label ({\tt QSO} or {\tt GAL}). These are shown in Fig.~\ref{fig:oa_labels_extragalactic} for two different subsets: high quality neurons (HQN), that represent quality categories 0 to 3, and low quality neurons (LQN), that represent categories 4 to 6. 
We further limit our analyses to those sources that appear in the integrated tables.
Figure~\ref{fig:oa_quality_categories} shows the number of neurons and objects assigned to each quality category. Approximately 80\% of the sources are assigned to a high quality neuron.

\begin{table*}[t]
\centering
\caption{Contingency table for OA classifications.
Each entry gives the percentage of objects classified by DSC (using {\tt classlabel\_dsc}), or processed by QSOC or UGC, that are assigned to OA high-quality neurons (HQN) or low-quality neurons (LQN), for sources in the {\tt qso\_candidates} and {\tt galaxy\_candidates} tables.
\label{tab:oa_confusion_matrix}
}
\footnotesize
\begin{tabular}{llrrrrrrr}
	\cline{3-8}
	\multicolumn{2}{c}{} & \multicolumn{6}{c}{\text{OA}}\\
    \cline{3-8}
	\multicolumn{2}{c|}{} & \multicolumn{2}{c|}{\text{QSO}} & \multicolumn{2}{c|}{\text{GAL}} & \multicolumn{2}{c|}{\text{other}}\\
	\cline{3-8}
	\multicolumn{2}{c}{} & \text{HQN} & \text{LQN} & \text{HQN} & \text{LQN} & \text{HQN} & \text{LQN} & \multicolumn{1}{c}{\text{Total}}\\
	\hline
	\multirow{3}{*}{\text{DSC}} & \text{quasar} & $20$\% & $15$\% & $1$\% & $1$\% & $45$\% & $18$\% & $2\,158\,916$\\
	& \text{galaxy} & $7$\% & $2$\% & $77$\% & $6$\% & $3$\% & $5$\% & $851\,127$\\
	& \text{other} & $7$\% & $8$\% & $53$\% & $1$\% & $24$\% & $7$\% & $1\,993\,592$\\
	\hline
	\multirow{2}{*}{\text{QSOC}} & \text{quasar} & $19$\% & $15$\% & $2$\% & $1$\% & $45$\% & $17$\% & $3\,069\,458$\\
	& \text{other} & $3$\% & $1$\% & $87$\% & $3$\% & $4$\% & $2$\% & $1\,934\,177$\\
	\hline
	\multirow{2}{*}{\text{UGC}} & \text{galaxy} & $3$\% & $1$\% & $83$\% & $6$\% & $2$\% & $5$\% & $199\,093$\\
	& \text{other} & $13$\% & $10$\% & $33$\% & $2$\% & $30$\% & $12$\% & $4\,804\,542$\\
	\hline
\end{tabular}
\end{table*}

All OA sources were processed by DSC, as well as QSOC or UGC depending on their DSC probabilities.
Table~\ref{tab:oa_confusion_matrix} is a contingency table showing the fraction of objects in common between these classifiers and the various OA neurons. 
For this we use {\tt classlabel\_dsc}.
QSOC and UGC do not classify sources; for this purpose we just look at the ones they provide redshifts for. 
Among the galaxies identified by DSC, 83\% of them were also found to be galaxies by the OA module, of which 77\% landed in a high quality neuron and 6\% in a low quality one.
The coincidence increases for UGC, with 89\% of its galaxies found in a galaxy neuron, of which 83\% have high quality. The coincidence for the quasars is substantially lower, around 35\% for both DSC and QSOC, with no substantial difference between high and low quality neurons. We also see that a large fraction of those objects that were not classified as a quasar or galaxy by DSC, or that were not analysed by QSOC, are classified as galaxies by OA: 54\% and 90\%, respectively, where most of them belong to a high quality neuron. 

OA processes sources that tend to be faint with noisy \bprp\ spectra, some of which OA had to modify (e.g.\ remove negative fluxes) so that it could process them.  Table~\ref{tab:oa_confusion_matrix} suggests that the OA classification complements the results from the other modules. OA coincides with DSC and UGC when identifying galaxies in particular, and identifies objects rejected by those modules that may be real galaxies. OA could also potentially help to identify extragalactic candidates that are not in the {\tt qso\_candidates} or {\tt galaxy\_candidates} tables.

\section{External comparison}\label{sec:external_comparison}

\subsection{WISE and proper motions}\label{sec:wise_and_pms}

\begin{figure}[t]
\begin{center}
\includegraphics[width=0.40\textwidth,angle=0]{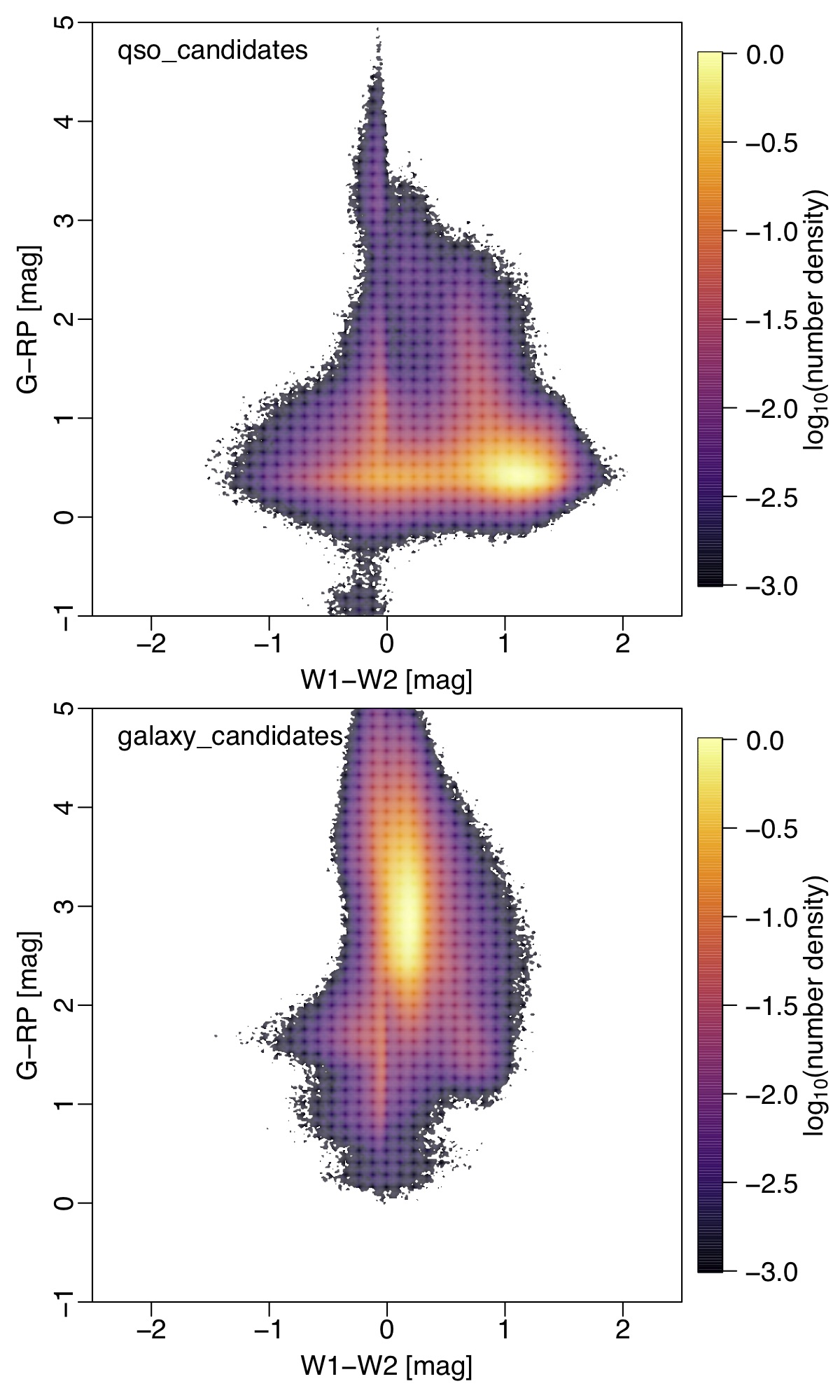}
\caption{\gaia-catWISE colour--colour diagrams. Top: All sources in the {\tt qso\_candidates} table. Bottom: All sources in the {\tt galaxy\_candidates} table. In both cases regions around the LMC and SMC have been excluded.
The colour scale shows the density of sources on a log scale relative to the peak density (densities 1000 times lower than the peak are not shown).
\label{fig:dr3int5_bothtables_ccd_grp_w12}
}
\end{center}
\end{figure}

\begin{figure}[t]
\begin{center}
\includegraphics[width=0.40\textwidth,angle=0]{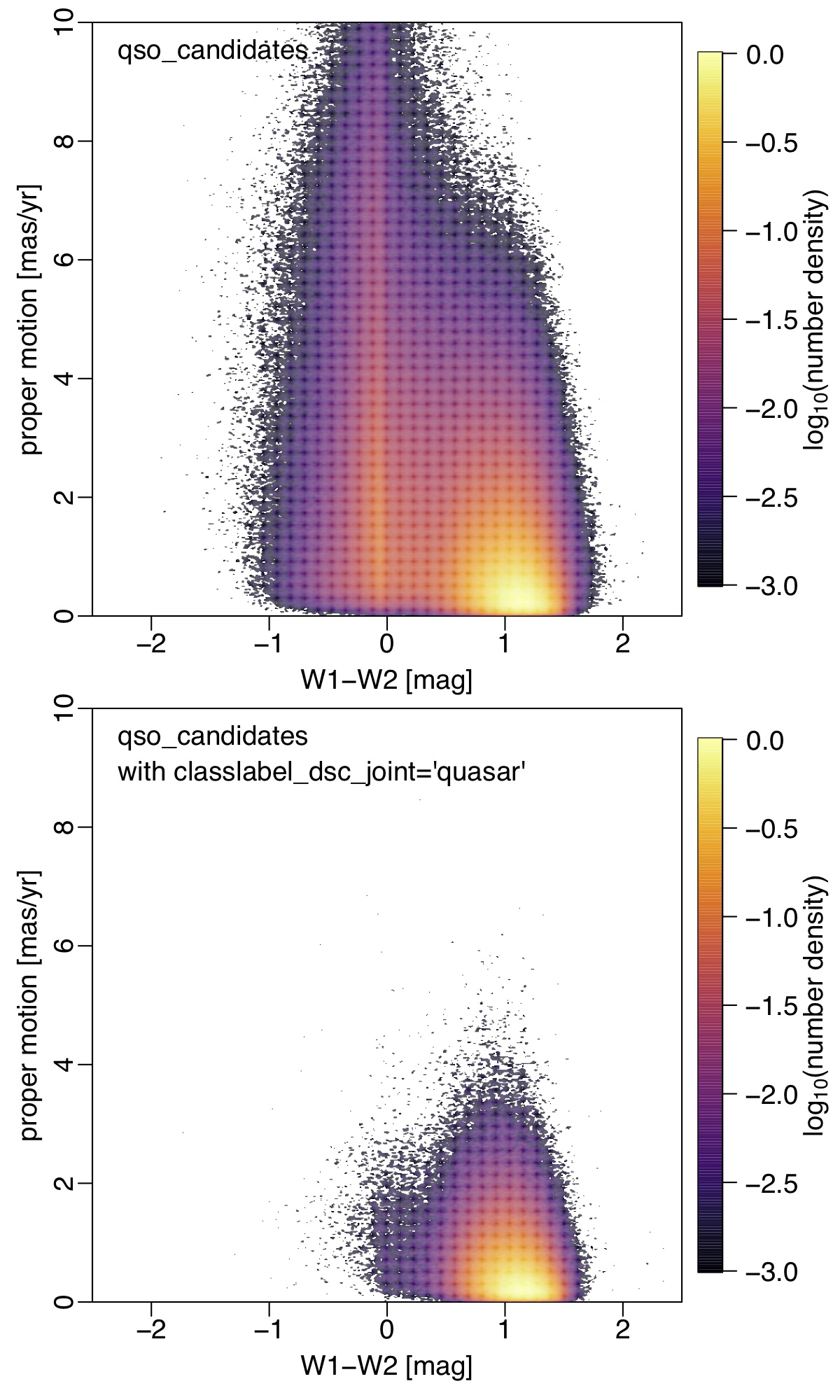}
\caption{\gaia\ proper motion vs catWISE W1$-$W2 colour for all sources in the {\tt qso\_candidates} table (top)
and the subset with {\tt classlabel\_dsc\_joint\,=\,quasar} (bottom).
Regions around the LMC and SMC are excluded.
The colour scale shows the density of sources on a log scale relative to the peak density (densities 1000 times lower than the peak are not shown). 
\label{fig:dr3int5_qsotable_densities_pm_w12}
}
\end{center}
\end{figure}

To investigate the infrared colours of the sources in the integrated tables, we cross-matched them to 
the catWISE2020 catalogue \citep[including the 2021 catalogue updates]{2021ApJS..253....8M} using a 
1\arcsec matching radius. 
We found 4.31 million sources (65\%)
matches in the {\tt qso\_candidates} table, and 4.59 million (95\%)
in the {\tt galaxy\_candidates} table.
Excluding the regions around the LMC and SMC (defined in appendix~\ref{sec:adql_queries}) left 2.99 million matches 
in the {\tt qso\_candidates} table and 4.46 million 
in the {\tt galaxy\_candidates} table.
Figure~\ref{fig:dr3int5_bothtables_ccd_grp_w12} shows the distribution of these sources in a \gaia-catWISE colour--colour diagram.
We see that most galaxy candidates have W1$-$W2 colours between 0.0 and 0.5\,mag.
This agrees with the range identified by \cite{2012ApJ...753...30S} for galaxies without an active nucleus and redshifts below 0.6.
The quasar candidates, in contrast, show two overdensities in the catWISE colour. We explore this further by looking at the quasar candidates in the proper motion space, as shown in Fig.~\ref{fig:dr3int5_qsotable_densities_pm_w12}. The upper panel is for all sources with 5p or 6p solutions (2.87 million sources). We see that the bluer clump at around W1$-$W2\,$\simeq$\,0\,mag shows the full range of proper motions. Recall that non-zero proper motions of true quasars are spurious, either due to noise or to time-variable source structure. Nonetheless, the larger proper motions in the bluer clump compared to the redder clump is indicative of contamination by stars (and some galaxies), and the W1$-$W2 colour would seem to confirm that. Indeed, the lower panel of
Fig.~\ref{fig:dr3int5_qsotable_densities_pm_w12} is for the purer subset defined by {\tt classlabel\_dsc\_joint\,=\,quasar} (0.50 million sources), and this retains just the redder sources with proper motions that are more consistent with zero (plus noise).

\subsection{Quasars}\label{sec:external_comparison:quasars}

Here we look in more detail at the properties of known quasars in \gaia.
For this purpose we cross-matched quasars from SDSS-DR14 \citep{2018A&A...613A..51P}
that have a visually confirmed redshifts ({\tt source\_z\,=\,VI} or {\tt source\_z\,=\,DR7Q})
to all \gaia sources (those in {\tt gaia\_source})
using a 1\arcsec \ matching radius. Such a match is nominally identical to the quasars selected for training DSC (see \citealt{DR3-DPACP-158}). However, we further limited this set to those with complete data in all photometric bands, at least five observations in BP and RP, complete astrometry (i.e.\ 5p or 6p solutions), and with $G<20.75$\,mag.
This gave 232\,794 sources covering the redshift range 0.038--5.305. 

A quasar in SDSS-DR14 is defined according to spectroscopic criteria. Specifically, they are sources with: (a) either at least one broad emission line with a full width at half maximum larger than 500\,\kms, or interesting or complex absorption features; and (b) sufficiently large intrinsic luminosity ($M_{\rm i} [z = 2] < -20.5$).
Since only one broad emission line is required, some objects may otherwise be classified as type 2 AGNs (those with predominantly narrow emission lines). The second part of the first condition aims to include Broad Absorption Line (BAL) quasars. This definition is free of morphological criteria. 
 

\begin{figure} 
\begin{center}
\includegraphics[width=0.5\textwidth,angle=0]{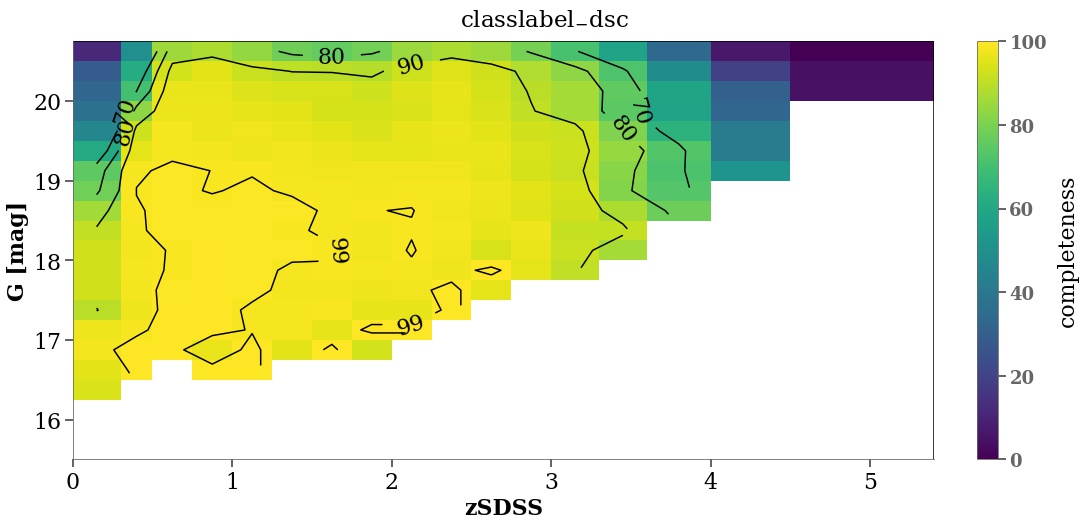}
\includegraphics[width=0.5\textwidth,angle=0]{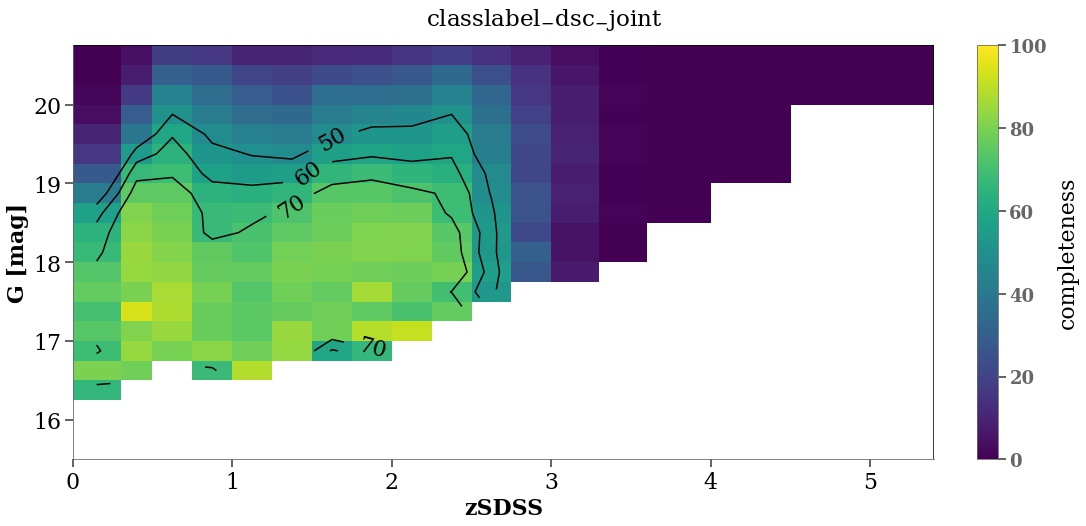}
\caption{Completeness of {\tt classlabel\_dsc\,=\,quasar} (top) and {\tt classlabel\_dsc\_joint\,=\,quasar} (bottom) with respect to the SDSS-DR14Q-\gdr3 cross-match as a function of \gmag\ and SDSS redshift. Empty bins (white) have fewer than 25 sources.
\label{fig:dr14q_dsc_joint_q_Gz_compl}} 
\end{center} 
\end{figure}

The sample defined above is similar to the superset from which the DSC-Allosmod training set was drawn. However, DSC did not force classifications on them, so we can use it to assess DSC's completeness as a function of magnitude and redshift (further assessments can be found in \citealt{LL:CBJ-094}).
This is shown in Fig.~\ref{fig:dr14q_dsc_joint_q_Gz_compl}, using the two class labels from DSC. The dependence on redshift is expected because of its weak correlation with \bpminrp, which increases the confusion with stars at high redshifts and with galaxies at low redshifts. 
Nonetheless, the completeness 
is above 80\% for redshifts between 0.3 and 3.6 and \gmag$\leq$\,20.25\,mag.
The lower completeness at fainter magnitudes is also expected, because lower quality data are more likely to be classified by DSC as the majority class of stars, according to the global prior (Sect.~\ref{sec:modules:dsc}), especially for the more conservative 
{\tt classlabel\_dsc\_joint} label. This also explains why the overall completeness is much lower for this label, although it is still 
above 60\% from $z=0$ to $z=2.5$ for \gmag$\leq$\,19.25\,mag. 
The overall completeness of {\tt classlabel\_dsc} is 215\,721/232\,794=93\% and of {\tt classlabel\_dsc\_joint} is  97\,995/232\,794=42\%.
However, given the non-uniform selection function of SDSS for obtaining spectra, we should be careful not to over-interpret this specific assessment of the DSC's completeness.


\begin{figure}
\begin{center}
\includegraphics[width=0.5\textwidth,angle=0]{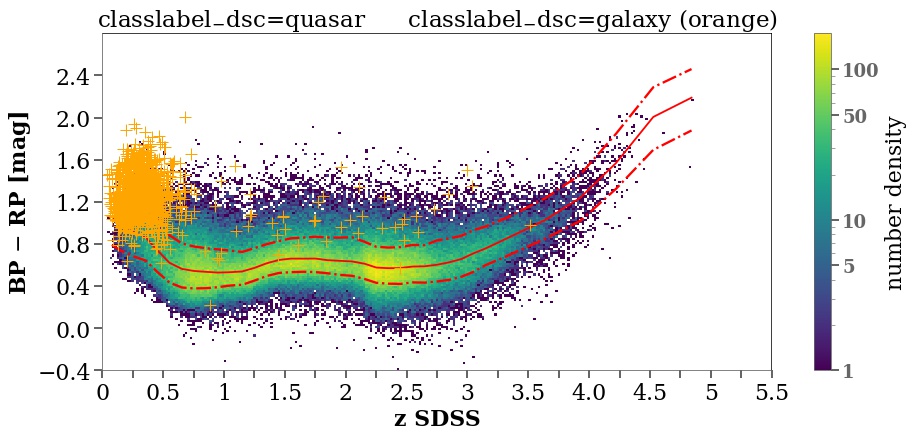}
\includegraphics[width=0.5\textwidth,angle=0]{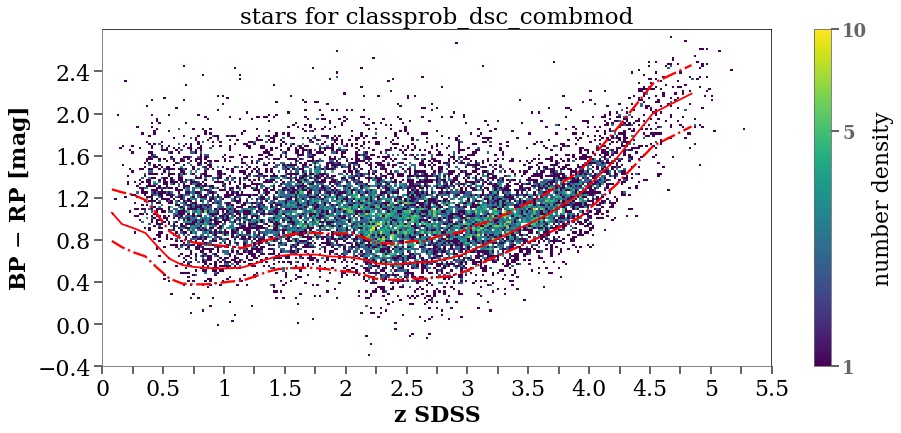}
\includegraphics[width=0.5\textwidth,angle=0]{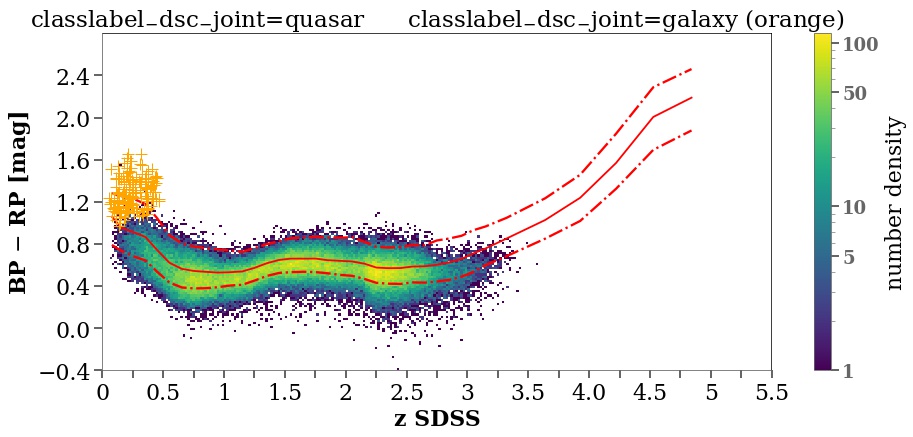}
\caption{\bpminrp\, vs redshift for the subsets  {\tt classlabel\_dsc} (top) and {\tt classlabel\_dsc\_joint} (bottom) of the SDSS-DR14Q-\gdr3 cross-match. The orange points are {\tt classlabel\_dsc\,=\,galaxy} (top) and {\tt classlabel\_dsc\_joint\,=\,galaxy} (bottom). 
The middle panel shows 
the quasars with {\tt classprob\_dsc\_combmod}$>$ 0.5 for any of the three stellar classes used in DSC-Combmod.
The red curves show the median \bpminrp\, and the  16\% and 84\% quantiles for all sources in the cross-match, regardless of the DSC class (so are the same in all panels).}  \label{fig:dr14q_dsc_joint_bprp_z}
\end{center} 
\end{figure}

The  \bpminrp\ vs redshift relation for the sources classified as quasars by {\tt classlabel\_dsc} and {\tt classlabel\_dsc\_joint} is shown in Fig.~\ref{fig:dr14q_dsc_joint_bprp_z}  (top and bottom panels). The pattern of undulations is expected, and is due to the quasar emission lines moving across the bands with redshift. The tail to the red for $z>3.5$ corresponds to the Ly$\alpha$ forest entering and then filling the BP band.  
We see that some low redshift objects classified by SDSS as quasars are classified as galaxies by DSC.
These quasars likely have a higher contribution of the host galaxy to the total emission, making them redder. 
The quasars with {\tt classlabel\_dsc\_joint\,=\,quasar} follow neatly the bluest part of the colour-$z$ relation, avoiding the regions with \bpminrp above the median (compare top and bottom panels). This result and Fig.~\ref{fig:dr14q_dsc_joint_q_Gz_compl} indicate that this class selects mainly bright quasars, and from these only the bluest ones. {\tt classlabel\_dsc\,=\,quasar} complements the {\tt classlabel\_dsc\_joint\,=\,quasar} class by covering the quasars that are redder due to their intrinsic emission, Galactic extinction, or local absorption (as in BALs, for example). The middle panel of Fig.~\ref{fig:dr14q_dsc_joint_bprp_z} shows that the incompleteness of {\tt classlabel\_dsc} at high-$z$ is due to the misclassification of many of these quasars as stars (see also Fig.~\ref{fig:dr14q_dsc_joint_q_Gz_compl}). Similarly, this plot shows that the quasars in the envelope of the reddest colours over the range $z=$\,0.5--3.5 are also classified as stars.

\begin{figure} 
\begin{center}
\includegraphics[width=0.4\textwidth,angle=0]{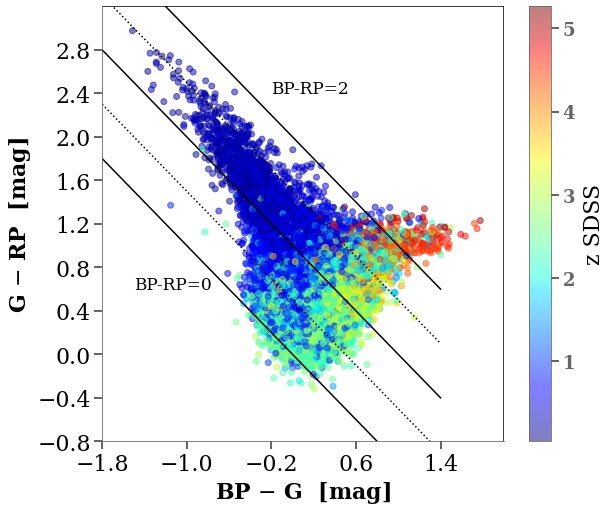} 
\includegraphics[width=0.4\textwidth,angle=0]{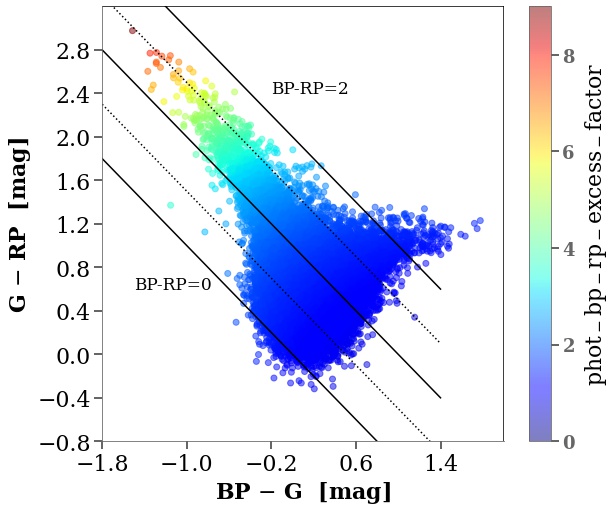}
\caption{Colour--colour diagram for the set in Fig. \ref{fig:dr14q_dsc_joint_bprp_z} colour-coded by redshift (top) and \gaia \texttt{phot\_bp\_rp\_excess\_factor} (bottom). The straight lines correspond to fixed \bpminrp colour, with values 2, 1.5, 1, 0.5 and 0. 
\label{fig:dr14q_CCD}}
\end{center} 
\end{figure}

Figure~\ref{fig:dr14q_CCD} shows the colour--colour diagram for the sample colour-coded by SDSS redshift (top panel) and by \gaia \linktoparam{sec_dm_main_source_catalogue/ssec_dm_gaia_source.html\#gaia_source-phot_bp_rp_excess_factor}{phot\_bp\_rp\_excess\_factor} (bottom panel). 
In the upper panel we see a clear trend of colour with redshift, with low redshifts located in the upper left and redshift increasing as we descend, but with the highest redshifts on the far right. The bottom panel shows that the region of low-$z$ sources corresponds to sources with larger \texttt{phot\_bp\_rp\_excess\_factor}, which indicates that the combined BP and RP bands contain more flux than the \gmag band. This region overlaps with the location of the galaxies (see plots of the DSC source densities in~\citealt{DR3-DPACP-158}). This suggests that the excess in \gbp and \grp with respect to \gmag is due to the 
wider photometric windows of \gbp and \grp compared to \gmag, which 
allows detection of the quasar's host galaxy emission over a wider region than in the \gmag band. This, together with the red \bpminrp colours, indicates a shift from quasars dominated by the nucleus to quasars with an important contribution of  the host galaxy. 
This transition can be also appreciated in the colour--colour diagram of the purer sub-samples from the {\tt qso\_candidates} and {\tt galaxy\_candidates} tables (Fig.~\ref{fig:dr3int5_bothtables_purersubset_ccd_cmd}), followed by the transition to the general galaxy population.

\subsection{Galaxies}

\begin{table*}[t]
\centering
\caption{SDSS spectral classes for objects in the {\tt galaxy\_candidates} table. The first row gives the number of sources found in SDSS for the whole table (first column) and for
each module (subsequent columns). The following rows give the percentage of sources of each SDSS class among these. The GALAXY row can be thought of as a measure of the purity against sources with SDSS spectral classifications, which by design is dominated by extragalactic sources, so is an overestimate of the purity of a sample selected at random from \gaia.
\label{tab:sdss_gal}
}
\begin{tabular}{ccccccccc}
\hline 
SDSS & {\tt galaxy\_} & {\tt classlabel\_} & {\tt classlabel\_} & Vari- & OA & UGC & Surface  & \\ 
& {\tt candidates} & {\tt dsc} & {\tt dsc\_joint} & Classification & & & brightness &\\
\hline 
TOTAL & 534\,154  & 477\,939 & 48\,460 & 360\,217 &105\,296 & 248\,196  & 96\,918 & \# \\ 
GALAXY  & 98.0   &97.9   & 95.1   &99.7 &95.6  & 97.9 & 99.7& \%  \\ 
QSO & 1.6   &1.7 & 4.9  &0.3 & 3.9  & 2.0  & 0.3 & \% \\ 
STAR & 0.4   &0.4 & 0.1  & 0.0 & 0.5  & 0.0  & 0.0 & \% \\ 
\hline 
\end{tabular}
\end{table*} 

We compared the content of the {\tt galaxy\_candidates} table with  the spectral classes in SDSS DR16.
We cross-matched the catalogues 
with a 1\,arcsec radius and removed duplicated sources or ones where {\tt zWarning} was not equal to zero.
Of the 4\,842\,342 sources in the {\tt galaxy\_candidates} table, we found 534\,154 matches in SDSS. 98.0\% of these are classified by SDSS as GALAXY, 1.6\% as QSO, and 0.4\% as STAR. Table~\ref{tab:sdss_gal} shows these percentages for each module that contributed to the
{\tt galaxy\_candidates} table. For example, 48\,460 sources have 
{\tt classlabel\_dsc\_joint\,=\,galaxy} in the matched set, and 95.1\% of these have SDSS class GALAXY.
This percentage is a measure of the purity of {\tt classlabel\_dsc\_joint} but only against those sources that have SDSS spectral classifications.
It is considerably higher than the purity of this DSC class label reported in Sect.~\ref{sec:modules:dsc} (and detailed in \citealt{DR3-DPACP-158}), even though this purity estimate was also based on SDSS. The reason is that this earlier purity estimate was computed for a set of sources selected at random from \gaia: It includes the significant contamination from non-galaxies, which outnumber galaxies by a factor of about one thousand in \gaia. The higher figure reported in Table~\ref{tab:sdss_gal}, in contrast, is just for those sources that have SDSS spectral classifications. By design of SDSS this includes proportionally very few stars and so very few potential contaminants. The numbers in Table~\ref{tab:sdss_gal} are a therefore a significant overestimate of the true purity and should be treated with caution. 

\begin{figure}
\centering
\includegraphics[width=0.5\textwidth,angle=0]{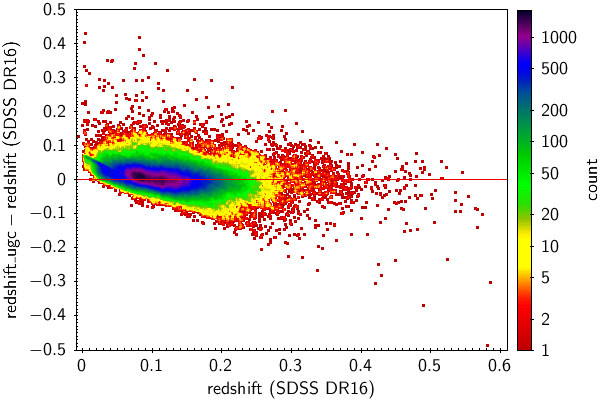}
\caption{Difference between {\tt redshift\_ugc} and SDSS DR16 redshift, as a function of the latter, for the 248\,356 sources in common with SDSS spectral class GALAXY.}
\label{fig:ugc_sdss_redshift}
\end{figure}

Of the 1\,367\,153 sources in the {\tt galaxy\_candidates} table that have redshifts provided by UGC, 248\,356 match to sources in the SDSS-DR16 {\tt specObj} table that are classified as GALAXY. \new{The small discrepancy with the number in Table~\ref{tab:sdss_gal} is due to slightly different cross-match criteria.} Figure~\ref{fig:ugc_sdss_redshift} shows the difference 
between {\tt redshift\_ugc} and the SDSS-DR16 redshift as a function of the latter. The average of this difference is 0.06 with a standard deviation of 0.054 (which reduces to 0.029 when the 67 sources with redshifts above 0.6 are excluded).  Generally the agreement is good, although UGC seems to systematically overestimate very low redshifts. 

\section{Composite quasar spectrum}\label{sec:composite}


Composite quasar spectra have many uses. First and foremost they are used as a reference in cross-correlations of individual spectra in order to classify these and determine their redshifts. Composite spectra are also used to identify faint spectral features that would otherwise be undetectable, to calibrate absolute magnitudes through the $k$-correction, as well as to construct colour--colour relations for identifying and characterising quasars based on photometry. Here we construct composite spectra to unveil the capability offered by the \gaia \bprp spectrophotometers to characterise quasars. 

\begin{figure*}
\centering
\includegraphics[width=\textwidth]{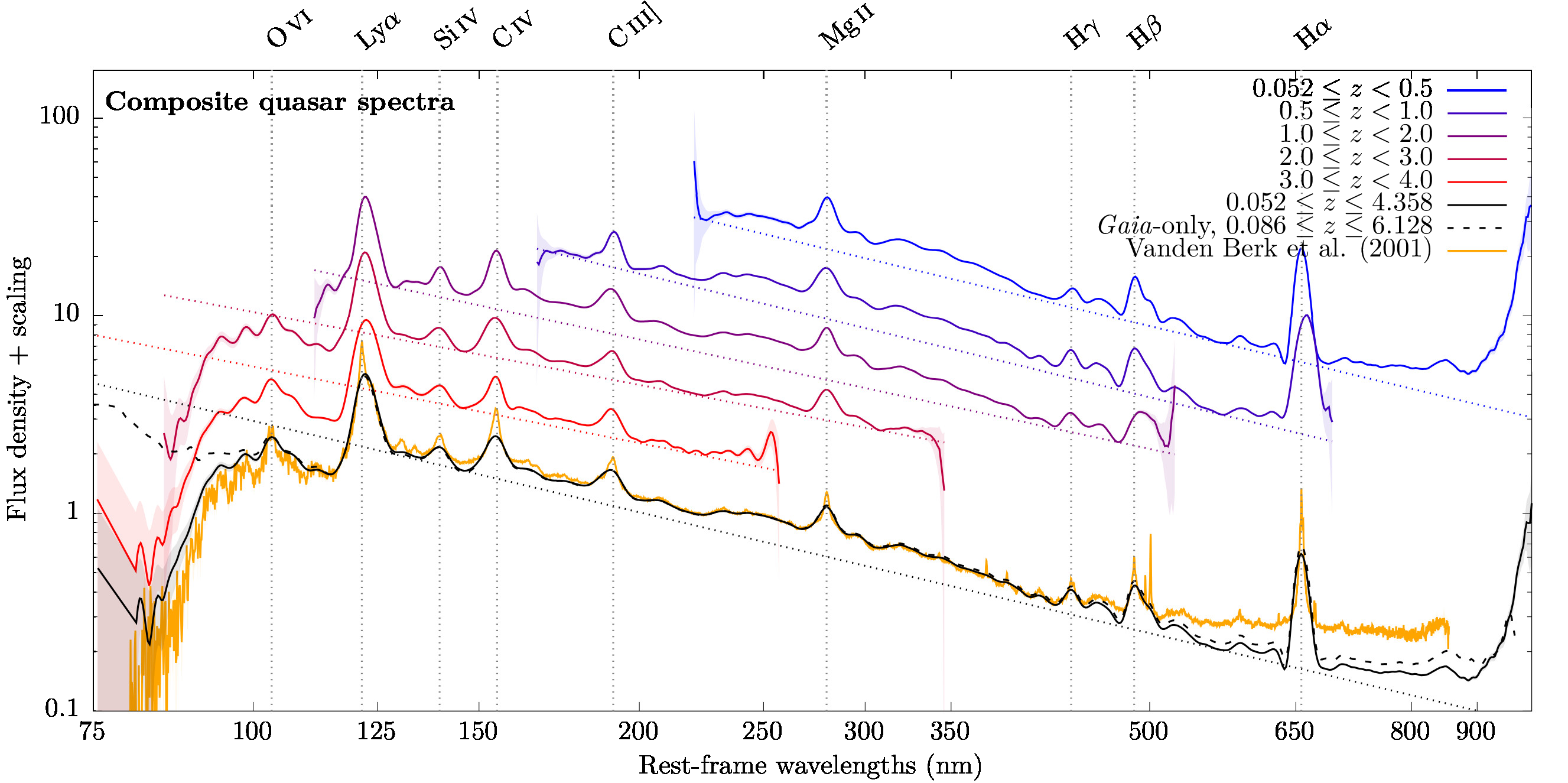}
\caption{Composite quasar spectra. The thick solid lines show composites made from the 42\,944 \bprp spectra with spectroscopically-confirmed redshifts from the Milliquas 7.2 quasar catalogue of \cite[i.e. {\tt type\,=\,Q} in Milliquas]{2021Flesch}. The different colours are for different redshift ranges.
The thick dotted black line shows the composite made from 111\,563 \bprp spectra with reliable QSOC redshift estimates (identified using the query given in appendix~\ref{sec:sample_qso_composite_gaia_only}). The diagonal dotted line under each spectrum shows the quasar continuum, as described in Sect.~\ref{sec:composite} and defined in Table~\ref{tab:qso_composite_parameters}. Vertical dotted lines indicate common quasar emission lines. 
For comparison purposes we also show the median SDSS composite spectrum of \cite{2001AJ....122..549V} (orange line).
The flux densities are tabulated in Table \ref{tab:qso_composite_fluxes}.
\label{fig:qso_composite}
}
\end{figure*}

\begin{table}
    \caption{Composite quasar spectra shown in Figure \ref{fig:qso_composite}. The columns are the arbitrarily scaled flux densities, $F$,
    and associated uncertainties, $F_{\rm err}$, computed at rest-frame wavelengths $\lambda$. The values below are for the Milliquas composite covering the redshift range $0.052$ to $4.358$. A full electronic version of this table is available with the online version of this article, along with the quasar composite spectra for the narrower redshift ranges shown in Figure \ref{fig:qso_composite} as well as the \gaia-only composite.}
    \label{tab:qso_composite_fluxes}
    \centering
    \begin{tabular}{rrr}
    \hline
    $\lambda$ (nm) & $F$ & $F_{\rm err}$ \\
    \hline \hline
    75.669 & 0.00296 & 0.00292 \\
    81.074 & 0.00157 & 0.00133 \\
    81.318 & 0.00189 & 0.00128 \\
    81.562 & 0.00208 & 0.00124 \\
    81.807 & 0.00207 & 0.00119 \\
    {\dots} & {\dots} & {\dots} \\
    980.799 & 0.00457 & 0.00095 \\
    983.746 & 0.00504 & 0.00142 \\
    986.701 & 0.00499 & 0.00188 \\
    989.666 & 0.00500 & 0.00222 \\
    992.639 & 0.00630 & 0.00297 \\
    \hline
    \end{tabular}
\end{table}

In order to build composite \bprp\ spectra we use the quasar sample described in Sect.~\ref{sssec:bp_rp_spectra_quasars}, which is based on the Milliquas 7.2 quasar catalogue of \cite{2021Flesch}. Our sample comprises 42\,944 sources for which we use the Milliquas redshifts. We rely on an external catalogue instead of the \gdr{3} internal classifications coupled with QSOC redshift determinations,
because the various instruments that contributed to the Milliquas catalogue have  better spectral resolutions and sensitivities than \gaia. 
We nonetheless also compute a composite \gaia-only spectrum based on 111\,563 sources coming from the \gaia\ classifications and input lists together with the redshifts from QSOC. The exact sample used for this is defined in appendix~\ref{sec:sample_qso_composite_gaia_only}. The method by which we compute composite spectra is described in appendix~\ref{sec:composite_qso_algo}.
The method relies on a single parameter, the logarithmic wavelength sampling of the composite spectra, chosen here to be $\log S = 0.003$ (i.e.\ $S \approx 1.003$) as a compromise between S/N and execution time. This sampling also applies to transformation matrices --  $\boldsymbol{\mathsf{M_i}}$ in Eq.~\ref{eq:qso_composite_chi2} -- that cover the observed wavelength region 309.5--1100.5\,nm.

The resulting composite spectrum inferred using the Milliquas sample is shown in Fig.~\ref{fig:qso_composite} as the solid black line and the values listed in \ref{tab:qso_composite_fluxes}. It covers a rest-frame wavelength range from $75.67$\,nm to $992.64$\,nm. The \gaia-only composite spectrum is shown as the dotted line and covers the wavelength range $47.08$\,nm to $962.83$\,nm. The composite spectra are trimmed in order to discard wavelength regions having flux density S/N less than one. After a multiplicative re-scaling of the flux densities so that their continua align, we found absolute differences between these two composites -- relative to the Ly$\alpha$ flux density in the Milliquas composite spectrum -- of less than $4$\% over the rest-frame wavelength region $100$--$900$\,nm, but up to 60\% for regions bluewards of this range. The cause of these larger deviations is either contamination by sources with erroneous redshift estimates, as a consequence of the low purity of QSOC in this very high redshift region (see \citealt{DR3-DPACP-158}), or border effects in the externally calibrated \bprp spectra that we describe below.
Figure~\ref{fig:qso_composite} also shows, for comparison, a median composite spectrum from SDSS \citep{2001AJ....122..549V} that covers a rest-frame wavelength range similar to that of our full composite spectrum.

\begin{figure}[t]
    \centering
    \includegraphics[width=0.49\textwidth]{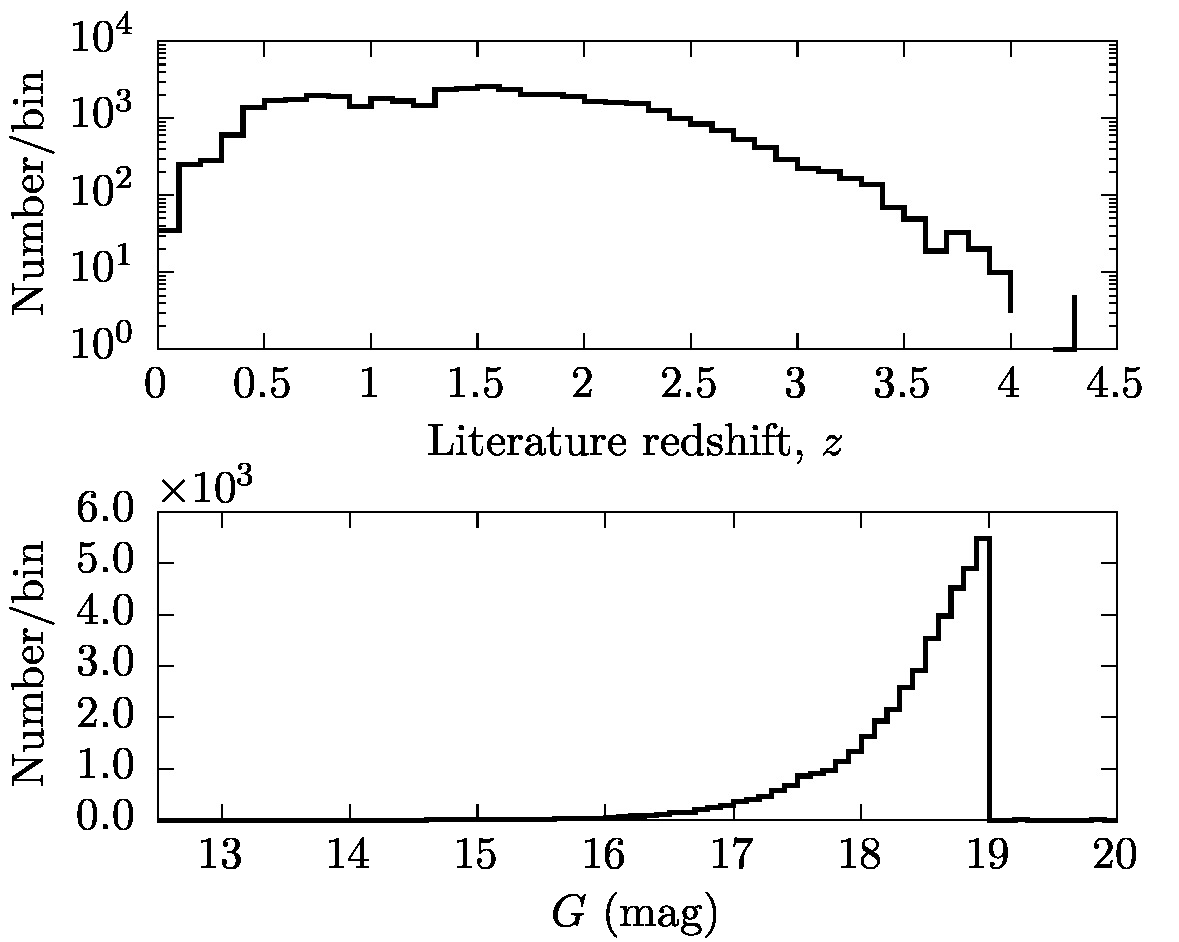}
    \caption{Distribution of the literature redshift from the Milliquas 7.2 catalogue \citep{2021Flesch} and \gaia $G$-band magnitudes for the 42\,944 sources used in the computation of the composite quasar spectra in Fig.~\ref{fig:qso_composite}.
    \label{fig:qso_composite_G_z}
    }
\end{figure}

Figure \ref{fig:qso_composite_G_z} shows the redshift and magnitude distributions of the sources used to build the
Milliquas-based
composite spectrum.  While the redshift distribution is as expected, with imprints of the selection and observational bias of each survey composing the Milliquas catalogue, the sharp drop at $G>19$\,mag is due to the filtering of the set of \bprp spectra published in \gdr{3}. The 17 sources with $G \geq 19$\,mag are present because: 11 are associated with a best-matching node of the OA module (see Sect.~\ref{sec:modules:oa}), four are used as external calibrators by CU5, and two are white dwarf candidates used in \cite{DR3-DPACP-93}.

\begin{table*}[t]
\centering
\footnotesize
\caption{Physical quantities derived from the composite spectra of quasars from the Milliquas catalogue shown in Fig.~\ref{fig:qso_composite}. The reduced chi-square, $\chi_\nu^2$, is obtained from Eq.~\ref{eq:qso_composite_chi2}. The frequency continuum slope, $\alpha_{\nu}$, is computed from the wavelength continuum slope, $\alpha_{\lambda} = - \alpha_{\nu} - 2$, where the continuum is modelled through a power law of the form $C_{\lambda} \propto \lambda^{\alpha_{\lambda}}$. In Fig.~\ref{fig:qso_composite}, the continuum is plotted as the lowest line joining the two most widely separated points in the range $121$--$600$\,nm without crossing the spectrum in this range. The emission line location, $\lambda_{\rm obs}$, and maximal line flux density, $F$, are retrieved from the fit of a quadratic polynomial in the vicinity of the laboratory wavelength, $\lambda_{\rm lab}$, after a local continuum has been subtracted that was computed in the same way as for $\alpha_{\lambda}$. The full-width at half maximum (FWHM) is retrieved from these continuum-subtracted emission lines using a linear interpolation of the spectral flux densities. All uncertainties are calculated, to first order, using the formal uncertainties on the composite spectra obtained from Eq.~\ref{eq:qso_composite_chi2_Cm}. 
}
\begin{tabular}{lclrrr} 
\hline
Composite spectrum & Continuum slope, $\alpha_\nu$ & Line+$\lambda_{\rm lab}$ (nm) & $\lambda_{\rm obs}$ (nm) & $F/F_{\rm ref}^a$ & FWHM (nm) \\
\hline
\hline
$0.052 \leq z < 0.5$ & $-0.4506 \pm 0.1453$  & \ion{Mg}{ii}$\lambda 279.875$ & $280.064 \pm 0.234$ & $1.00000 \pm 0.01036$ & $8.308 \pm 0.214$ \\
$2\,571$ sources & & H$\gamma$$\lambda 434.168$ & $435.221 \pm 0.588$ & $0.13927 \pm 0.00369$ & $10.432 \pm 0.586$ \\
$\chi_\nu^2 = 20.8$ & & H$\beta$$\lambda 486.268$ & $486.936 \pm 0.081$ & $0.46937 \pm 0.00233$ & $12.874 \pm 0.131$ \\
& & H$\alpha$$\lambda 656.461$ & $656.240 \pm 0.010$ & $1.27766 \pm 0.00095$ & $17.971 \pm 0.035$ \\
\hline
$0.5 \leq z < 1$ & $-0.4291 \pm 0.0570$ & \ion{C}{iii}]$\lambda 190.873$ & $191.144 \pm 0.134$ & $1.80831 \pm 0.03931$ & $5.583 \pm 0.262$ \\
$8\,810$ sources & & \ion{Mg}{ii}$\lambda 279.875$ & $279.644 \pm 0.084$ & $1.00000 \pm 0.00238$ & $10.613 \pm 0.120$ \\
$\chi_\nu^2 = 7.0$ & & H$\gamma$$\lambda 434.168$ & $433.932 \pm 0.048$ & $0.25984 \pm 0.00099$ & $10.355 \pm 0.123$ \\
& & H$\beta$$\lambda 486.268$ & $486.825 \pm 0.042$ & $0.53950 \pm 0.00085$ & $18.263 \pm 0.082$ \\
& & H$\alpha$$\lambda 656.461$ & $661.947 \pm 0.100$ & $1.36875 \pm 0.00735$ & $23.063 \pm 0.684$ \\
\hline
$1 \leq z < 2$ & $-0.6115 \pm 0.0227$ & Ly$\alpha$$\lambda 121.567$ & $122.264 \pm 0.029$ & $1.00000 \pm 0.00826$ & $4.411 \pm 0.097$ \\
$20\,755$ sources & & \ion{Si}{iv}$\lambda 139.676$ & $139.925 \pm 0.051$ & $0.14881 \pm 0.00292$ & $3.637 \pm 0.111$ \\
$\chi_\nu^2 = 4.9$ & & \ion{C}{iv}$\lambda 154.906$ & $154.636 \pm 0.050$ & $0.29494 \pm 0.00154$ & $4.477 \pm 0.043$ \\
& & \ion{C}{iii}]$\lambda 190.873$ & $190.124 \pm 0.093$ & $0.13300 \pm 0.00039$ & $7.729 \pm 0.068$ \\
& & \ion{Mg}{ii}$\lambda 279.875$ & $280.056 \pm 0.053$ & $0.08408 \pm 0.00026$ & $8.026 \pm 0.047$ \\
& & H$\gamma$$\lambda 434.168$ & $433.450 \pm 0.101$ & $0.01752 \pm 0.00013$ & $11.683 \pm 0.292$ \\
& & H$\beta$$\lambda 486.268$ & $490.168 \pm 5.268$ & $0.03491 \pm 0.00070$ & $23.559 \pm 2.499$ \\
\hline
$2 \leq z < 3$ & $-0.7752 \pm 0.0113$ & \ion{O}{vi}$\lambda 103.383$ & $103.589 \pm 0.152$ & $0.18338 \pm 0.00686$ & $3.279 \pm 0.342$ \\
$9\,870$ sources & & Ly$\alpha$$\lambda 121.567$ & $122.156 \pm 0.022$ & $1.00000 \pm 0.00160$ & $4.908 \pm 0.024$ \\
$\chi_\nu^2 = 5.8$& & \ion{Si}{iv}$\lambda 139.676$ & $139.738 \pm 0.086$ & $0.11467 \pm 0.00068$ & $4.927 \pm 0.129$ \\
& & \ion{C}{iv}$\lambda 154.906$ & $154.465 \pm 0.049$ & $0.19586 \pm 0.00041$ & $5.854 \pm 0.114$ \\
& & \ion{C}{iii}]$\lambda 190.873$ & $190.589 \pm 0.047$ & $0.11133 \pm 0.00107$ & $6.742 \pm 0.164$ \\
& & \ion{Mg}{ii}$\lambda 279.875$ & $279.941 \pm 0.091$ & $0.07435 \pm 0.00019$ & $9.471 \pm 0.078$ \\
\hline
$3 \leq z < 4$ & $-0.7155 \pm 0.0548$ & \ion{O}{vi}$\lambda 103.383$ & $103.381 \pm 0.112$ & $0.21926 \pm 0.00277$ & $5.202 \pm 0.935$ \\
$929$ sources & & Ly$\alpha$$\lambda 121.567$ & $122.398 \pm 0.086$ & $1.00000 \pm 0.00489$ & $5.591 \pm 0.083$ \\
$\chi_\nu^2=6.0$ & & \ion{Si}{iv}$\lambda 139.676$ & $139.980 \pm 0.232$ & $0.12360 \pm 0.00895$ & $4.588 \pm 0.486$ \\
& & \ion{C}{iv}$\lambda 154.906$ & $154.581 \pm 0.151$ & $0.24250 \pm 0.00306$ & $3.928 \pm 0.107$ \\
& & \ion{C}{iii}]$\lambda 190.873$ & $190.161 \pm 0.118$ & $0.13617 \pm 0.00097$ & $6.488 \pm 0.094$ \\
\hline
$0.052 \leq z \leq 4.358$ & $-0.4639 \pm 0.0057$ & \ion{O}{vi}$\lambda 103.383$ & $103.500 \pm 0.103$ & $0.19285 \pm 0.00584$ & $3.784 \pm 0.517$ \\
$42\,944$ sources & & Ly$\alpha$$\lambda 121.567$ & $122.202 \pm 0.037$ & $1.00000 \pm 0.00259$ & $5.088 \pm 0.036$ \\
$\chi_\nu^2=6.6$ & & \ion{Si}{iv}$\lambda 139.676$ & $139.838 \pm 0.141$ & $0.11467 \pm 0.00146$ & $4.672 \pm 0.139$ \\
& & \ion{C}{iv}$\lambda 154.906$ & $154.542 \pm 0.052$ & $0.21615 \pm 0.00132$ & $5.383 \pm 0.066$ \\
& & \ion{C}{iii}]$\lambda 190.873$ & $190.232 \pm 0.056$ & $0.11681 \pm 0.00072$ & $7.938 \pm 0.097$ \\
& & \ion{Mg}{ii}$\lambda 279.875$ & $280.006 \pm 0.034$ & $0.08202 \pm 0.00025$ & $8.891 \pm 0.047$ \\
& & H$\gamma$$\lambda 434.168$ & $434.164 \pm 0.059$ & $0.02165 \pm 0.00016$ & $10.706 \pm 0.129$ \\
& & H$\beta$$\lambda 486.268$ & $486.846 \pm 0.109$ & $0.04867 \pm 0.00011$ & $16.473 \pm 0.179$ \\
& & H$\alpha$$\lambda 656.461$ & $656.257 \pm 0.014$ & $0.13924 \pm 0.00011$ & $18.026 \pm 0.032$ \\
\hline
\multicolumn{6}{l}{$^a$ If Ly$\alpha$ is covered by the composite spectrum, it is used as a reference flux density, $F_{\rm ref}$, otherwise \ion{Mg}{ii} is used.}
\end{tabular}
\label{tab:qso_composite_parameters}
\end{table*}

In addition to the full composite spectrum computed over the whole redshift range from $0.052$ to $4.358$, we also computed composite spectra over several narrower redshift ranges, chosen such that their logarithmic rest-frame wavelength coverage are approximately of equal size: $0.052 \leq z < 0.5$, $0.5 \leq z < 1$, $1 \leq z < 2$, $2 \leq z < 3$ and $3 \leq z \leq 4.358$. Composite spectra associated with each redshift range are shown in Fig.~\ref{fig:qso_composite}.
The number of sources used in each redshift range, as well as the resulting reduced chi-square derived from Eq.~\ref{eq:qso_composite_chi2}, are provided in Table~\ref{tab:qso_composite_parameters} along with the frequency ($\nu$) continuum slope, $\alpha_\nu$ and some information on the strongest observed emission lines: the rest-frame wavelengths, the relative flux density at their peak compared to either Ly$\alpha$ or \ion{Mg}{ii}, and the full width at half maximum (FWHM). Unsurprisingly, all reduced chi-squares are larger than unity: Each composite spectrum evidently does not completely model the intrinsic variance seen in the observations. However, the moderate values that are observed are indicative of reasonable fits to the observations, which is corroborated by visual inspection and by the good agreement with the median composite spectrum of \cite{2001AJ....122..549V}. The full composite spectrum (with $\chi_\nu^2=6.6$) also explains 
99.2\% of the observed variance in the entire set of spectra, although most of this variance comes from the spectral continuum. The maximum S/N ranges from 224 per $\log S$ interval in the composite spectrum with $3 \leq z < 4$, to 645 per $\log S$ interval in the composite spectrum associated with the $0.5 \leq z < 1$ redshift range.

The highest S/N of 610 per $\log S$ interval of the full composite spectrum allows us to identify many more emission lines than are otherwise visible in the \bprp spectra of individual sources. Whereas only the Ly$\alpha$, \ion{C}{iv}, \ion{C}{iii}], \ion{Mg}{ii}, H$\beta$, and H$\alpha$ emission lines are commonly seen in the observed \bprp spectra of quasars, a visual inspection of the full composite spectrum reveals 22 emission lines, which are listed in Table~\ref{tab:qsoc_composite_emission_lines}. Despite the low resolution of \bprp spectra (Sect.~\ref{sec:bp_rp_spectra}), all common quasar emission lines were retrieved during this inspection procedure, in addition to some weak or rarely-seen emission lines. Emission lines from the wavelength region covering the Ly$\alpha$ forest are similarly recovered in an unambiguous way. We achieve good agreement with laboratory wavelength positions, with a maximum absolute difference of $| \Delta \lambda | = | \lambda_{\rm lab} - \lambda_{\rm obs} | =  0.951$\,nm for the \ion{C}{iii}] emission line, where we consider only the nearest emission line in case these are blended. The apparent blueshift of the \ion{C}{iii}] emission line resides in its asymmetry, which is due to the presence of the \ion{Si}{iii}]$\,\lambda\,189.203$\,nm in its neighbourhood. The same rationale applies to the shift of the Ly$\alpha$ and H$\beta$ emission lines ($\Delta \lambda = -0.683$\,nm and $-0.433$\,nm, respectively) due to the presence of the \ion{N}{v}$\,\lambda\,124.014$\,nm and [\ion{O}{iii}] doublets respectively.

Our composite spectra are highly coherent with one another, with  little variation depending on the redshift ranges that were used. After a multiplicative re-scaling of the flux densities to align the continua, we found differences relative to the Ly$\alpha$ flux density of less than $6$\% when compared to the full composite spectrum over several rest-frame wavelength regions (230--900\,nm for the $0.052 \leq z < 0.5$ composite spectrum; 180--680 nm for the $0.5 \leq z < 1.0$ composite spectrum; 115--500 nm for the $1 \leq z < 2$ composite spectrum; 85--320 nm for the $2 \leq z < 3$ composite spectrum and 85--240 nm for the $3 \leq z < 4$ composite spectrum). Some border effects that were ignored in the previous comparison
can still be seen in all composite spectra. These are due to the low S/N as well as systematic errors at the borders of the externally calibrated \bprp spectra of quasars that we used to build the composite spectra.
Such an effect is particularly noticeable in Fig. \ref{fig:qso_composite} redwards of $900$\,nm, where a sharp rise is visible that is also seen in about 90\% of the individual spectra of quasars we used. This artefact is consequently taken as a genuine signal by our method. 

The differences noticed when comparing the full composite spectrum to the SDSS one of \cite{2001AJ....122..549V} redwards of the H$\beta$ emission line are presumably due to different angular scales of the \gaia \bprp footprint and the SDSS fibres, which are $2.1$\arcsec (across scan direction; \citealt{2021A&A...652A..86C}) and $3$\arcsec respectively. This results in the contamination of the composite spectrum by the light from the host galaxy being more suppressed in the \gaia observations.

\begin{table}[]
\centering
\footnotesize
\caption{Quasar emission lines found in the Milliquas-based composite spectrum and covering the redshift range $0.052 \leq z \leq 4.35$ (Fig.~\ref{fig:qso_composite}). Each emission line is visually inspected before a quadratic polynomial is fit in the vicinity of its apparent peak using five samples of flux density. The maximum of the quadratic curve provides the observed rest-frame wavelength position, $\lambda_{\rm obs}$, of the line and its maximum flux density compared to Ly$\alpha$, $F/F_{{\rm Ly}\alpha}$ (where the number of significant digits is provided in parenthesis). Because of the intricacies inherent in the fit of a local continuum to faint, broad, and/or blended emission lines, such a continuum was not subtracted from the flux densities reported here.
This explains the differences between this table and the values found in Table~\ref{tab:qso_composite_parameters}.}
\begin{tabular}{lcc}
\hline
Emission line & $\lambda_{\rm obs}$ & $F/F_{{\rm Ly}\alpha}$ \\
+$\lambda_{\rm lab}$ (nm) & (nm) & \\
\hline \hline
{Ly$\epsilon$$\,\lambda\,93.780$} & $94.356 \pm 1.327$ & $0.344(2)$ \\
{Ly$\delta$$\,\lambda\,94.974$} & $\cdots$ & $\cdots$ \\
{\ion{C}{iii}$\,\lambda\,97.702$} & $98.632 \pm 0.237$ & $0.393(2)$ \\
{\ion{N}{iii}$\,\lambda\,99.069$} & $\cdots$ & $\cdots$ \\
{Ly$\beta$$\,\lambda\,102.572$} & $103.423 \pm 0.139$ & $0.479(2)$ \\
{\ion{O}{vi}$\,\lambda\,103.383$} & $\cdots$ & $\cdots$ \\
{\ion{Fe}{ii}$\,\lambda\,111.208$}$^a$ & $112.098 \pm 0.500$ & $0.328(2)$ \\
{Ly$\alpha$$\,\lambda\,121.567$} & $122.250 \pm 0.033$ & $1.000(2)$ \\
{\ion{O}{i}$\,\lambda\,130.435$} & $130.495 \pm 0.545$ & $0.402(2)$ \\
{\ion{Si}{ii}$\,\lambda\,130.682$} & $\cdots$ & $\cdots$ \\
{\ion{Si}{iv}$\,\lambda\,139.676$} & $139.568 \pm 0.098$ & $0.427(2)$ \\
{\ion{O}{iv}]$\,\lambda\,140.206$} & $\cdots$ & $\cdots$ \\
{\ion{C}{iv}$\,\lambda\,154.906$} & $154.469 \pm 0.061$ & $0.485(3)$ \\
{\ion{C}{iii}$\,\lambda\,190.873$} & $189.922 \pm 0.102$ & $0.327(3)$ \\
{\ion{Fe}{III}$\,\lambda\,205.271$}$^a$ & $205.600 \pm 0.555$ & $0.230(3)$ \\
{\ion{C}{ii}]$\,\lambda\,232.644$} & $232.390 \pm 0.350$ & $0.202(3)$ \\
{[\ion{Ne}{iv}]$\,\lambda\,242.383$} & $242.617 \pm 0.324$ & $0.198(3)$ \\
{\ion{Mg}{ii}$\,\lambda\,279.875$} & $279.888 \pm 0.039$ & $0.212(3)$ \\
{\ion{He}{i}$\,\lambda\,318.867$} & $318.612 \pm 0.295$ & $0.136(3)$ \\
{H$\gamma$$\,\lambda\,434.168$} & $434.052 \pm 0.067$ & $0.081(3)$ \\
{\ion{Fe}{ii}$\,\lambda\,453.780$}$^a$ & $454.281 \pm 0.569$ & $0.070(4)$ \\
{H$\beta$$\,\lambda\,486.268$} & $486.701 \pm 0.118$ & $0.085(4)$ \\
{[\ion{O}{iii}]$\,\lambda\,496.030$} & $\cdots$ & $\cdots$ \\
{[\ion{O}{iii}]$\,\lambda\,500.824$} & $\cdots$ & $\cdots$ \\
{[\ion{N}{i}]$\,\lambda\,520.053$} & $522.270 \pm 0.615$ & $0.054(4)$ \\
{\ion{He}{i}$\,\lambda\,587.729$} & $588.441 \pm 0.411$ & $0.044(4)$ \\
{H$\alpha$$\,\lambda\,656.461$} & $656.283 \pm 0.014$ & $0.123(4)$ \\
{\ion{He}{i}$\,\lambda\,706.720$} & $706.666 \pm 0.945$ & $0.034(4)$ \\
{\ion{O}{i}$\,\lambda\,844.868$} & $850.091 \pm 0.413$ & $0.033(4)$ \\
{\ion{Ca}{ii}$\,\lambda\,850.036$} & $\cdots$ & $\cdots$ \\
\hline
\multicolumn{3}{l}{$^a$ Broad feature composed of Fe multiplets.}
\end{tabular}
\label{tab:qsoc_composite_emission_lines}
\end{table}

Figure~\ref{fig:qso_composite} and Table~\ref{tab:qso_composite_parameters} reveal differences in the continuum slopes, $\alpha_{\nu}$, over the various redshift ranges. These are mostly a result of the different rest-frame wavelength coverage of each composite spectrum. Indeed, the presence of broad Fe multiplets and the Balmer continuum in the wavelength region $200$--$550$\,nm -- the so-called $300$\,nm bump -- complicates the continuum fitting in this range due to the lack of pure continuum. Consequently the values of $\alpha_{\nu}$ decrease once we consider quasars with redshifts $z > 1$. The value we found on the full composite spectrum, $\alpha_\nu = -0.464 \pm 0.005$, agrees reasonably well with literature values. The median composite spectrum from \cite{2001AJ....122..549V} shown in Fig.~\ref{fig:qso_composite}, for example, has $\alpha_\nu = -0.46$.

We inspected the measurements associated with the dominant quasar emission lines given in Table~\ref{tab:qsoc_composite_emission_lines}. We did not find any meaningful correlation between their values and the redshift range that was used for computing each composite spectrum.

\section{How to select purer sub-samples}\label{sec:howto_purer_samples}

The {\tt qso\_candidates} and {\tt galaxy\_candidates} tables collate together results for most extragalactic candidates in \gdr{3} from a number of modules, as described in Sect.~\ref{sec:integrated_tables}. Overall these tables aim for high completeness, rather than high purity. We have done this intentionally to allow the user to select their own sub-samples using the quality indicators and class probabilities. Here we describe how to select purer sub-samples from these tables.

\begin{table}[t]
\caption{ADQL query to select the purer quasar sub-sample.
\label{tab:qso_purer_subset}
}
\hrule
{\small
\begin{verbatim}
SELECT * FROM gaiadr3.qso_candidates
WHERE (gaia_crf_source='true' OR 
       host_galaxy_flag<6 OR
       classlabel_dsc_joint='quasar' OR
       vari_best_class_name='AGN')
\end{verbatim}
}
\hrule
\end{table}

\begin{table}[t]
\caption{ADQL query to select the purer galaxy sub-sample.
\label{tab:galaxy_purer_subset}
}
\hrule
{\small
\begin{verbatim}
SELECT * FROM gaiadr3.galaxy_candidates
WHERE (radius_sersic IS NOT NULL OR
       classlabel_dsc_joint='galaxy' OR
       vari_best_class_name='GALAXY')
\end{verbatim}
}
\hrule
\end{table}

For the quasars, EO and CRF3 used input lists of quasars identified by other surveys, so their samples are believed to be quite pure, above 90\%. \new{From the EO sample we exclude those with close neighbours ({\tt host\_galaxy\_flag\,=\,6})}.
For DSC, the joint subset has a purity of 62\%, increasing to 79\% when the Galactic plane ($|b|<11.54\deg$) is avoided~\citep{DR3-DPACP-158}.
The Vari classifier results for AGN, which already exclude the Galactic plane, has been assessed to have a purity of over 90\% \citep{DR3-DPACP-165}. We therefore recommend the query in Table~\ref{tab:qso_purer_subset} to select a purer sub-sample of quasars. 
This returns 1\,942\,825 sources, which is 29\% of the original table.
\new{Using the same approach at the end of section~\ref{sec:integrated_tables}, and assuming a 96\% purity for the non-DSC modules, we estimate the overall purity of this sub-sample to be 95\%.}
\newer{Of these, 1.7 million have published redshifts from QSOC.}

We use similar criteria to define a purer galaxy sub-sample, except that here there is no contribution from CRF3. Again we take all of the sources from EO (provided by an input list), the purer subset of DSC (64\% pure; 82\% outside the Galactic plane), and all of Vari. This query, in Table~\ref{tab:galaxy_purer_subset}, returns 2\,891\,132 sources, 60\% of the original table. 
We estimate the purity of this sub-sample to be \new{94\%}.
\newer{Of these, 1.1 million have published redshifts from UGC.}

There are 14\,471 sources in common between these two purer sub-samples, and their union contains 4.8 million sources.

\begin{figure*}
\begin{center}
\includegraphics[width=0.49\textwidth,angle=0]{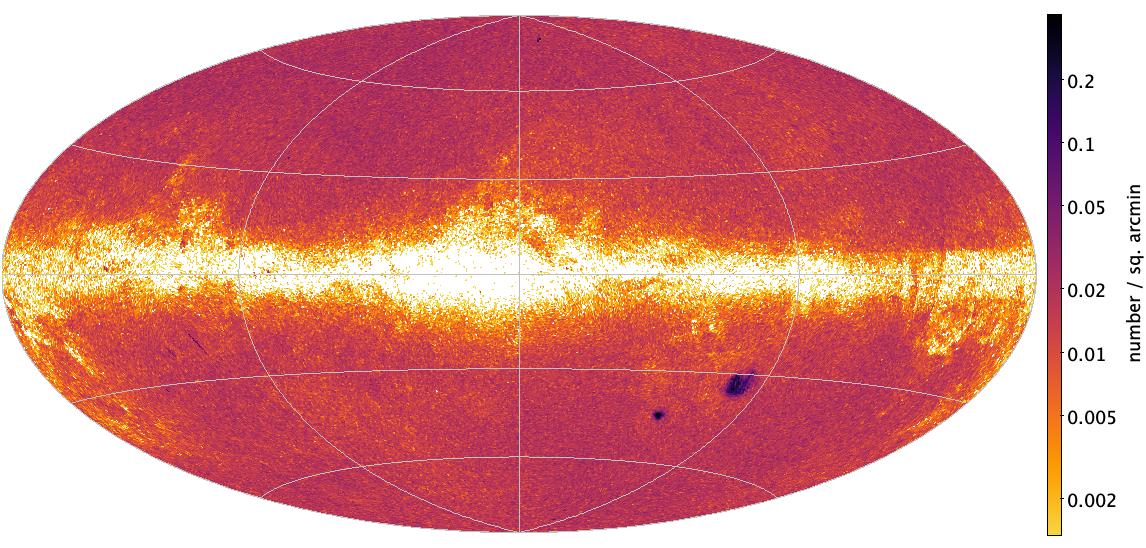}
\includegraphics[width=0.49\textwidth,angle=0]{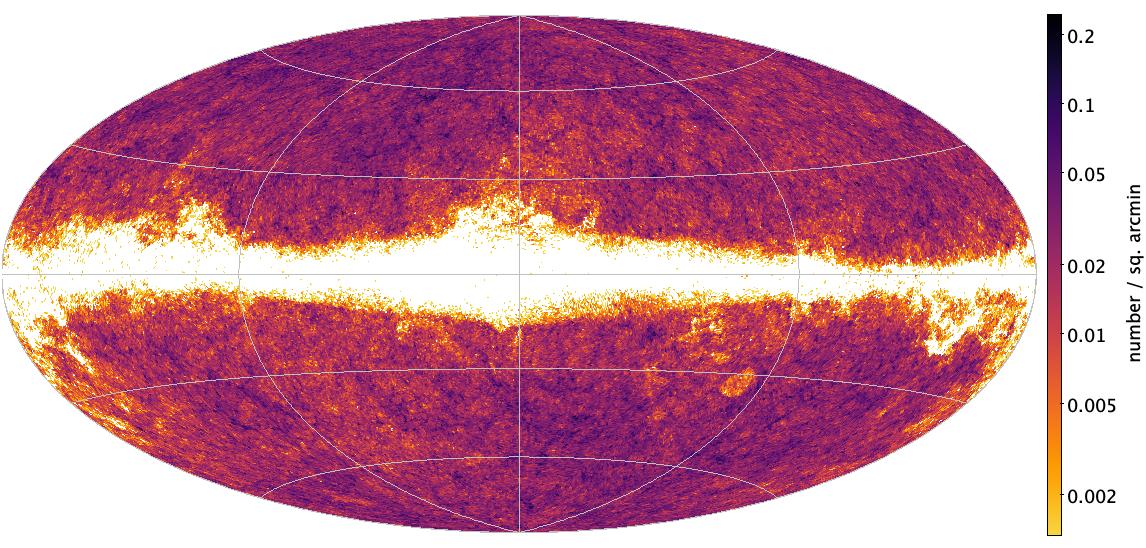}
\caption{Galactic sky distribution of all the purer sub-sample of sources in the
{\tt qso\_candidates} table (left) and {\tt galaxy\_candidates} table (right).
The plot is shown at HEALpixel level 7 (0.210 sq.\ deg.) in Hammer--Aitoff projection.
The colour scale, which is logarithmic, covers the full range for each panel, so is different for each panel.
Compare to  Fig.~\ref{fig:qso_galaxy_skyplots} for the full tables.
\label{fig:qso_galaxy_purersubset_skyplots}
}
\end{center}
\end{figure*}

The sky distributions for these purer sub-samples are shown in Fig.~\ref{fig:qso_galaxy_purersubset_skyplots} and can be compared with those for the full tables in Fig.~\ref{fig:qso_galaxy_skyplots}. We immediately see how the purer sub-samples have lower densities in the Galactic plane. There are still artefacts from the \gaia\ scanning law, which is an indication of the less than perfect purity. We also still see overdensities at the LMC and SMC. This comes mostly from DSC, because unlike the other modules DSC did not do any sky-position-dependent filtering. Such sources can be easily removed by the user, as shown in appendix~\ref{sec:sample_purer_dsc}.

\begin{figure}
    \centering
    \includegraphics[width=0.49\textwidth]{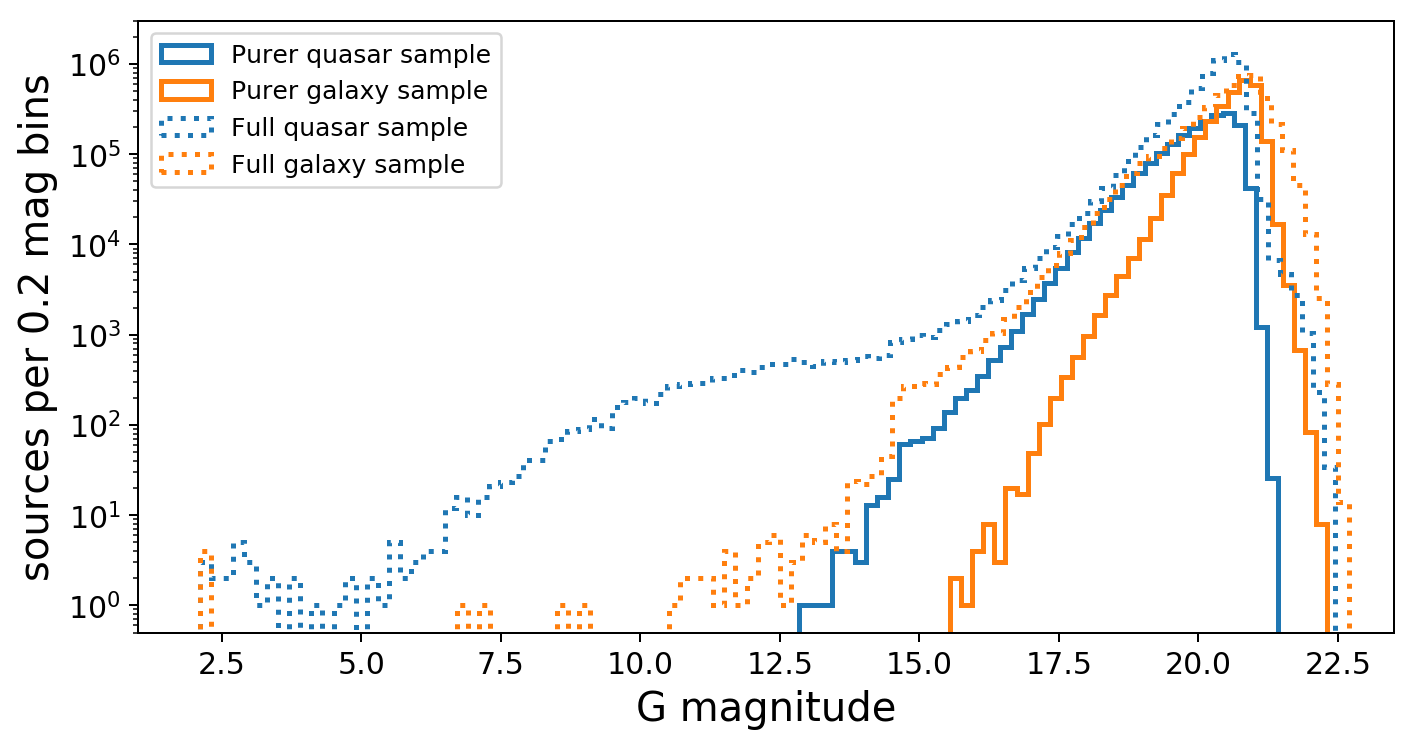}
    \caption{G-band magnitude distribution of the purer sub-sample of objects in the {\tt qso\_candidates} (blue) and {\tt galaxy\_candidates} (orange) table on a logarithmic scale. The dotted lines show the distributions for the full tables.
    \label{fig:qso_galaxy_purersubset_gmagdist}
    }
\end{figure}

The magnitude distributions of the purer sub-samples are shown in Fig.~\ref{fig:qso_galaxy_purersubset_gmagdist}. Compared to the distribution for the full set (dotted lines), we see that the purer sub-sample has excluded the brightest sources (the presence of which appears exaggerated, however, due to the logarithmic number scale). The faintest quasars have also been removed.

\begin{figure}
\begin{center}
\includegraphics[width=0.40\textwidth,angle=0]{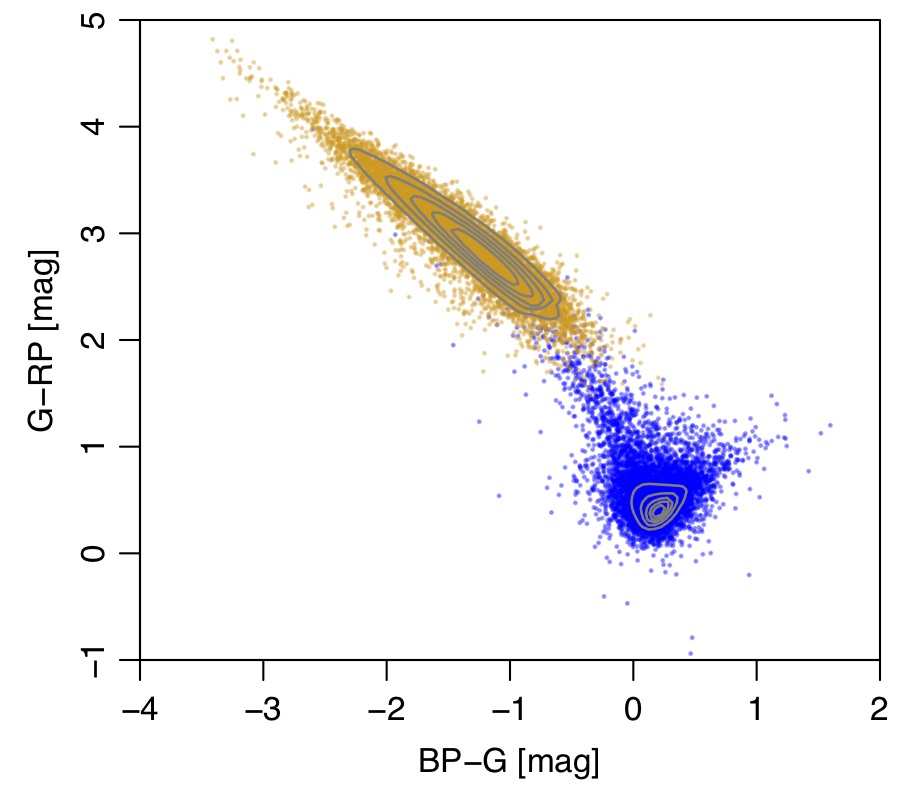}
\includegraphics[width=0.40\textwidth,angle=0]{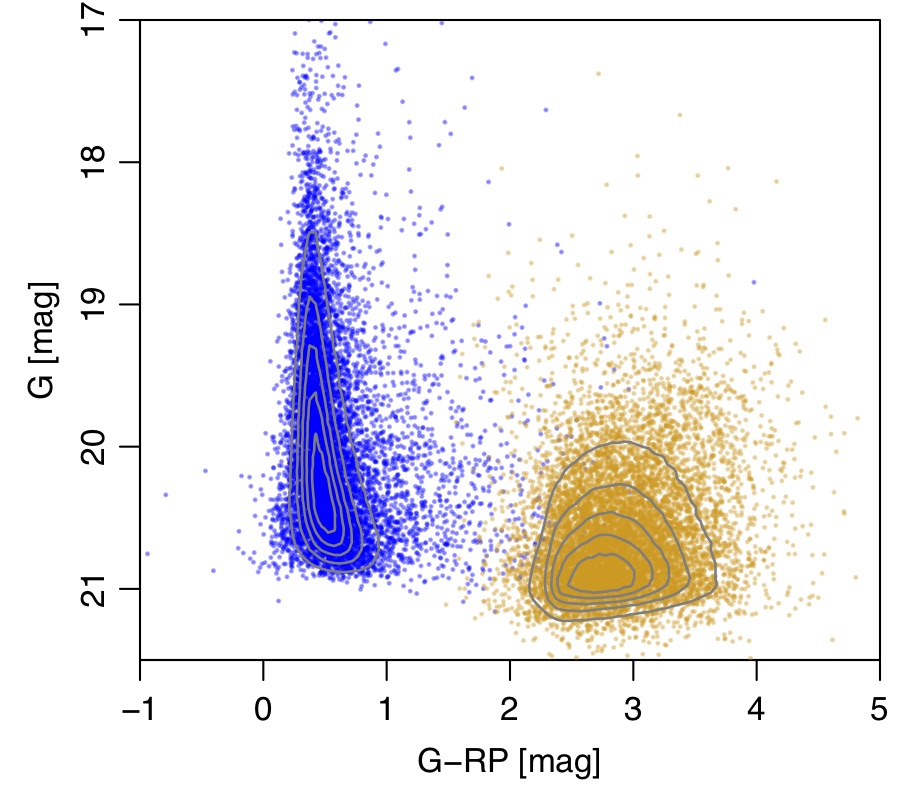}
\caption{Colour--colour diagram (top) and colour--magnitude diagram (bottom) for the purer sub-sample of sources in the {\tt qso\_candidates} table (blue) and {\tt galaxy\_candidates} table (orange). The contours show density on a linear scale. The points are a random selection of 10\,000 sources for each class.
Compare to  Fig.~\ref{fig:dr3int5_bothtables_ccd_cmd} for the full tables.  
\label{fig:dr3int5_bothtables_purersubset_ccd_cmd}
}
\end{center}
\end{figure}

Colour--magnitude and colour--colour diagrams of the purer sub-samples are shown in 
Fig.~\ref{fig:dr3int5_bothtables_purersubset_ccd_cmd} and can be compared with the same diagrams for the full tables in 
Fig.~\ref{fig:dr3int5_bothtables_ccd_cmd}. This shows that the purer sub-samples have a tighter colour distribution, and remove many of the fainter galaxies.

There are other ways to obtain purer sub-samples of the integrated tables. One could, for example, select on a higher probability threshold of the DSC probabilities (the probabilities for all three DSC classifiers are listed in the {\tt astrophysical\_parameters} table). An example is shown in appendix~\ref{sec:sample_purer_dsc}. The variation of purity with threshold is explored in \cite{LL:CBJ-094}. 
The joint flag used in the purer sub-samples (Tables~\ref{tab:qso_purer_subset} and Tables~\ref{tab:galaxy_purer_subset}) corresponds to a threshold of 0.5.
One could also use a higher threshold on the {\tt vari\_best\_class\_score} and {\tt vari\_agn\_membership\_score} rankings from the Vari-Classification and Vari-AGN modules
listed in the {\tt qso\_candidates} table.

In Sect.~\ref{sec:astrometric_analysis}, we identified a purer sub-sample of the {\tt qso\_candidates} table 
via an analysis of astrometric distributions
that are consistent with a uniform sky distribution of infinitely distant objects.
The sources in this astrometric selection are indicated by the boolean flag {\tt astrometric\_selection\_flag} in the {\tt qso\_candidates} table.  Although this certainly excludes genuine quasars, it should be a reasonably pure list. Of the 1\,897\,754
sources in the astrometric sample, 1\,801\,255
are in common with the purer quasar sub-sample defined in Table~\ref{tab:qso_purer_subset}, and the union of these two sets contains 2\,039\,324
sources.
An equivalent flag is not available for the {\tt galaxy\_candidates} table, for reasons discussed in Sect.~\ref{sec:astrometric_analysis}.

\section{Conclusions}\label{sec:conclusions}

We have described the data products released in \gdr{3} for 11.3 million candidate quasars and galaxies. This set arises from both a classification using the \gaia\ data and from an analysis of sources identified by external surveys cross-matched to \gaia. The information on these sources is presented in the {\tt qso\_candidates} and {\tt galaxy\_candidates} integrated tables. Further information, also on additional lower probability candidates, is provided in several other tables (see Sect.~\ref{sec:tables}). Our integrated tables are completeness driven, so many sources in them will not be true extragalactic objects. We therefore also provide a purer sub-sample of 4.8 million quasars and galaxies (see Sect.~\ref{sec:howto_purer_samples}).

We foresee a number of use cases for our results, including:
aiding confirmation of candidates found in other surveys;
identifying unusual or rare objects;
providing targets for spectroscopic follow-up; 
providing input data for more focused classifications or characterisations.
It is our hope and expectation that the community can build on and improve our results also by combining them with data from other surveys. 

As with previous data releases, \gdr{3} is an intermediate data release, this time based on 34 months of mission data. The next data release will be based on 66 months of data, with correspondingly higher S/N and lower systematic errors. We plan to use those data, along with improvements in our algorithms, to update both our classifications and our characterisations of extragalactic objects.

\section*{Acknowledgements}
This work presents results from the European Space Agency (ESA) space
mission \gaia. \gaia\ data are processed by the Data Processing and
Analysis Consortium (DPAC). Funding for the DPAC is provided by
national institutions, in particular the institutions participating in
the \gaia\ MultiLateral Agreement (MLA). The \gaia\ mission website is
\url{https://www.cosmos.esa.int/gaia} and the archive website is
\url{https://archives.esac.esa.int/gaia}. Further acknowledgments are
given in appendix~\ref{sec:acknowl}.


\bibliographystyle{aa}
\bibliography{bibliography,dpac}
\appendix

\section{Acknowledgements\label{sec:acknowl}}
\addcontentsline{toc}{chapter}{Acknowledgements}

The \gaia\ mission and data processing have been financially supported by, in alphabetical order by country:
\begin{itemize}
\item the Algerian Centre de Recherche en Astronomie, Astrophysique et G\'{e}ophysique of Bouzareah Observatory;
\item the Austrian Fonds zur F\"{o}rderung der wissenschaftlichen Forschung (FWF) Hertha Firnberg Programme through grants T359, P20046, and P23737;
\item the BELgian federal Science Policy Office (BELSPO) through various PROgramme de D\'{e}veloppement d'Exp\'{e}riences scientifiques (PRODEX) grants and the Polish Academy of Sciences - Fonds Wetenschappelijk Onderzoek through grant VS.091.16N, and the Fonds de la Recherche Scientifique (FNRS), and the Research Council of Katholieke Universiteit (KU) Leuven through grant C16/18/005 (Pushing AsteRoseismology to the next level with TESS, GaiA, and the Sloan DIgital Sky SurvEy -- PARADISE);  
\item the Brazil-France exchange programmes Funda\c{c}\~{a}o de Amparo \`{a} Pesquisa do Estado de S\~{a}o Paulo (FAPESP) and Coordena\c{c}\~{a}o de Aperfeicoamento de Pessoal de N\'{\i}vel Superior (CAPES) - Comit\'{e} Fran\c{c}ais d'Evaluation de la Coop\'{e}ration Universitaire et Scientifique avec le Br\'{e}sil (COFECUB);
\item the Chilean Agencia Nacional de Investigaci\'{o}n y Desarrollo (ANID) through Fondo Nacional de Desarrollo Cient\'{\i}fico y Tecnol\'{o}gico (FONDECYT) Regular Project 1210992 (L.~Chemin);
\item the National Natural Science Foundation of China (NSFC) through grants 11573054, 11703065, and 12173069, the China Scholarship Council through grant 201806040200, and the Natural Science Foundation of Shanghai through grant 21ZR1474100;  
\item the Tenure Track Pilot Programme of the Croatian Science Foundation and the \'{E}cole Polytechnique F\'{e}d\'{e}rale de Lausanne and the project TTP-2018-07-1171 `Mining the Variable Sky', with the funds of the Croatian-Swiss Research Programme;
\item the Czech-Republic Ministry of Education, Youth, and Sports through grant LG 15010 and INTER-EXCELLENCE grant LTAUSA18093, and the Czech Space Office through ESA PECS contract 98058;
\item the Danish Ministry of Science;
\item the Estonian Ministry of Education and Research through grant IUT40-1;
\item the European Commission’s Sixth Framework Programme through the European Leadership in Space Astrometry (\href{https://www.cosmos.esa.int/web/gaia/elsa-rtn-programme}{ELSA}) Marie Curie Research Training Network (MRTN-CT-2006-033481), through Marie Curie project PIOF-GA-2009-255267 (Space AsteroSeismology \& RR Lyrae stars, SAS-RRL), and through a Marie Curie Transfer-of-Knowledge (ToK) fellowship (MTKD-CT-2004-014188); the European Commission's Seventh Framework Programme through grant FP7-606740 (FP7-SPACE-2013-1) for the \gaia\ European Network for Improved data User Services (\href{https://gaia.ub.edu/twiki/do/view/GENIUS/}{GENIUS}) and through grant 264895 for the \gaia\ Research for European Astronomy Training (\href{https://www.cosmos.esa.int/web/gaia/great-programme}{GREAT-ITN}) network;
\item the European Cooperation in Science and Technology (COST) through COST Action CA18104 `Revealing the Milky Way with \gaia (MW-Gaia)';
\item the European Research Council (ERC) through grants 320360, 647208, and 834148 and through the European Union’s Horizon 2020 research and innovation and excellent science programmes through Marie Sk{\l}odowska-Curie grant 745617 (Our Galaxy at full HD -- Gal-HD) and 895174 (The build-up and fate of self-gravitating systems in the Universe) as well as grants 687378 (Small Bodies: Near and Far), 682115 (Using the Magellanic Clouds to Understand the Interaction of Galaxies), 695099 (A sub-percent distance scale from binaries and Cepheids -- CepBin), 716155 (Structured ACCREtion Disks -- SACCRED), 951549 (Sub-percent calibration of the extragalactic distance scale in the era of big surveys -- UniverScale), and 101004214 (Innovative Scientific Data Exploration and Exploitation Applications for Space Sciences -- EXPLORE);
\item the European Science Foundation (ESF), in the framework of the \gaia\ Research for European Astronomy Training Research Network Programme (\href{https://www.cosmos.esa.int/web/gaia/great-programme}{GREAT-ESF});
\item the European Space Agency (ESA) in the framework of the \gaia\ project, through the Plan for European Cooperating States (PECS) programme through contracts C98090 and 4000106398/12/NL/KML for Hungary, through contract 4000115263/15/NL/IB for Germany, and through PROgramme de D\'{e}veloppement d'Exp\'{e}riences scientifiques (PRODEX) grant 4000127986 for Slovenia;  
\item the Academy of Finland through grants 299543, 307157, 325805, 328654, 336546, and 345115 and the Magnus Ehrnrooth Foundation;
\item the French Centre National d’\'{E}tudes Spatiales (CNES), the Agence Nationale de la Recherche (ANR) through grant ANR-10-IDEX-0001-02 for the `Investissements d'avenir' programme, through grant ANR-15-CE31-0007 for project `Modelling the Milky Way in the \gaia era’ (MOD4Gaia), through grant ANR-14-CE33-0014-01 for project `The Milky Way disc formation in the \gaia era’ (ARCHEOGAL), through grant ANR-15-CE31-0012-01 for project `Unlocking the potential of Cepheids as primary distance calibrators’ (UnlockCepheids), through grant ANR-19-CE31-0017 for project `Secular evolution of galxies' (SEGAL), and through grant ANR-18-CE31-0006 for project `Galactic Dark Matter' (GaDaMa), the Centre National de la Recherche Scientifique (CNRS) and its SNO \gaia of the Institut des Sciences de l’Univers (INSU), its Programmes Nationaux: Cosmologie et Galaxies (PNCG), Gravitation R\'{e}f\'{e}rences Astronomie M\'{e}trologie (PNGRAM), Plan\'{e}tologie (PNP), Physique et Chimie du Milieu Interstellaire (PCMI), and Physique Stellaire (PNPS), the `Action F\'{e}d\'{e}ratrice \gaia' of the Observatoire de Paris, the R\'{e}gion de Franche-Comt\'{e}, the Institut National Polytechnique (INP) and the Institut National de Physique nucl\'{e}aire et de Physique des Particules (IN2P3) co-funded by CNES;
\item the German Aerospace Agency (Deutsches Zentrum f\"{u}r Luft- und Raumfahrt e.V., DLR) through grants 50QG0501, 50QG0601, 50QG0602, 50QG0701, 50QG0901, 50QG1001, 50QG1101, 50\-QG1401, 50QG1402, 50QG1403, 50QG1404, 50QG1904, 50QG2101, 50QG2102, and 50QG2202, and the Centre for Information Services and High Performance Computing (ZIH) at the Technische Universit\"{a}t Dresden for generous allocations of computer time;
\item the Hungarian Academy of Sciences through the Lend\"{u}let Programme grants LP2014-17 and LP2018-7 and the Hungarian National Research, Development, and Innovation Office (NKFIH) through grant KKP-137523 (`SeismoLab');
\item the Science Foundation Ireland (SFI) through a Royal Society - SFI University Research Fellowship (M.~Fraser);
\item the Israel Ministry of Science and Technology through grant 3-18143 and the Tel Aviv University Center for Artificial Intelligence and Data Science (TAD) through a grant;
\item the Agenzia Spaziale Italiana (ASI) through contracts I/037/08/0, I/058/10/0, 2014-025-R.0, 2014-025-R.1.2015, and 2018-24-HH.0 to the Italian Istituto Nazionale di Astrofisica (INAF), contract 2014-049-R.0/1/2 to INAF for the Space Science Data Centre (SSDC, formerly known as the ASI Science Data Center, ASDC), contracts I/008/10/0, 2013/030/I.0, 2013-030-I.0.1-2015, and 2016-17-I.0 to the Aerospace Logistics Technology Engineering Company (ALTEC S.p.A.), INAF, and the Italian Ministry of Education, University, and Research (Ministero dell'Istruzione, dell'Universit\`{a} e della Ricerca) through the Premiale project `MIning The Cosmos Big Data and Innovative Italian Technology for Frontier Astrophysics and Cosmology' (MITiC);
\item the Netherlands Organisation for Scientific Research (NWO) through grant NWO-M-614.061.414, through a VICI grant (A.~Helmi), and through a Spinoza prize (A.~Helmi), and the Netherlands Research School for Astronomy (NOVA);
\item the Polish National Science Centre through HARMONIA grant 2018/30/M/ST9/00311 and DAINA grant 2017/27/L/ST9/03221 and the Ministry of Science and Higher Education (MNiSW) through grant DIR/WK/2018/12;
\item the Portuguese Funda\c{c}\~{a}o para a Ci\^{e}ncia e a Tecnologia (FCT) through national funds, grants SFRH/\-BD/128840/2017 and PTDC/FIS-AST/30389/2017, and work contract DL 57/2016/CP1364/CT0006, the Fundo Europeu de Desenvolvimento Regional (FEDER) through grant POCI-01-0145-FEDER-030389 and its Programa Operacional Competitividade e Internacionaliza\c{c}\~{a}o (COMPETE2020) through grants UIDB/04434/2020 and UIDP/04434/2020, and the Strategic Programme UIDB/\-00099/2020 for the Centro de Astrof\'{\i}sica e Gravita\c{c}\~{a}o (CENTRA);  
\item the Slovenian Research Agency through grant P1-0188;
\item the Spanish Ministry of Economy (MINECO/FEDER, UE), the Spanish Ministry of Science and Innovation (MICIN), the Spanish Ministry of Education, Culture, and Sports, and the Spanish Government through grants BES-2016-078499, BES-2017-083126, BES-C-2017-0085, ESP2016-80079-C2-1-R, ESP2016-80079-C2-2-R, FPU16/03827, PDC2021-121059-C22, RTI2018-095076-B-C22, and TIN2015-65316-P (`Computaci\'{o}n de Altas Prestaciones VII'), the Juan de la Cierva Incorporaci\'{o}n Programme (FJCI-2015-2671 and IJC2019-04862-I for F.~Anders), the Severo Ochoa Centre of Excellence Programme (SEV2015-0493), and MICIN/AEI/10.13039/501100011033 (and the European Union through European Regional Development Fund `A way of making Europe') through grant RTI2018-095076-B-C21, the Institute of Cosmos Sciences University of Barcelona (ICCUB, Unidad de Excelencia `Mar\'{\i}a de Maeztu’) through grant CEX2019-000918-M, the University of Barcelona's official doctoral programme for the development of an R+D+i project through an Ajuts de Personal Investigador en Formaci\'{o} (APIF) grant, the Spanish Virtual Observatory through project AyA2017-84089, the Galician Regional Government, Xunta de Galicia, through grants ED431B-2021/36, ED481A-2019/155, and ED481A-2021/296, the Centro de Investigaci\'{o}n en Tecnolog\'{\i}as de la Informaci\'{o}n y las Comunicaciones (CITIC), funded by the Xunta de Galicia and the European Union (European Regional Development Fund -- Galicia 2014-2020 Programme), through grant ED431G-2019/01, the Red Espa\~{n}ola de Supercomputaci\'{o}n (RES) computer resources at MareNostrum, the Barcelona Supercomputing Centre - Centro Nacional de Supercomputaci\'{o}n (BSC-CNS) through activities AECT-2017-2-0002, AECT-2017-3-0006, AECT-2018-1-0017, AECT-2018-2-0013, AECT-2018-3-0011, AECT-2019-1-0010, AECT-2019-2-0014, AECT-2019-3-0003, AECT-2020-1-0004, and DATA-2020-1-0010, the Departament d'Innovaci\'{o}, Universitats i Empresa de la Generalitat de Catalunya through grant 2014-SGR-1051 for project `Models de Programaci\'{o} i Entorns d'Execuci\'{o} Parallels' (MPEXPAR), and Ramon y Cajal Fellowship RYC2018-025968-I funded by MICIN/AEI/10.13039/501100011033 and the European Science Foundation (`Investing in your future');
\item the Swedish National Space Agency (SNSA/Rymdstyrelsen);
\item the Swiss State Secretariat for Education, Research, and Innovation through the Swiss Activit\'{e}s Nationales Compl\'{e}mentaires and the Swiss National Science Foundation through an Eccellenza Professorial Fellowship (award PCEFP2\_194638 for R.~Anderson);
\item the United Kingdom Particle Physics and Astronomy Research Council (PPARC), the United Kingdom Science and Technology Facilities Council (STFC), and the United Kingdom Space Agency (UKSA) through the following grants to the University of Bristol, the University of Cambridge, the University of Edinburgh, the University of Leicester, the Mullard Space Sciences Laboratory of University College London, and the United Kingdom Rutherford Appleton Laboratory (RAL): PP/D006511/1, PP/D006546/1, PP/D006570/1, ST/I000852/1, ST/J005045/1, ST/K00056X/1, ST/\-K000209/1, ST/K000756/1, ST/L006561/1, ST/N000595/1, ST/N000641/1, ST/N000978/1, ST/\-N001117/1, ST/S000089/1, ST/S000976/1, ST/S000984/1, ST/S001123/1, ST/S001948/1, ST/\-S001980/1, ST/S002103/1, ST/V000969/1, ST/W002469/1, ST/W002493/1, ST/W002671/1, ST/W002809/1, and EP/V520342/1.
\end{itemize}
We made use of the following tools in the preparation of this paper:
\citep[SIMBAD,][]{2000AAS..143....9W} and VizieR \citep{2000A&AS..143...23O} operated at (\href{http://cds.u-strasbg.fr/}{CDS}) Strasbourg; 
NASA \href{http://adsabs.harvard.edu/abstract_service.html}{ADS}; 
\href{http://www.starlink.ac.uk/topcat/}{TOPCAT} \citep{2005ASPC..347...29T};
Matplotlib \citep{Hunter:2007};
IPython \citep{PER-GRA:2007};  
Astropy, a community-developed core Python package for Astronomy \citep{2018AJ....156..123A};
\href{https://cran.r-project.org}{R} \citep{RManual}.

Funding for SDSS-III has been provided by the Alfred P. Sloan Foundation, the Participating Institutions, the National Science Foundation, and the U.S. Department of Energy Office of Science. The SDSS-III web site is http://www.sdss3.org/. 
SDSS-III is managed by the Astrophysical Research Consortium for the Participating Institutions of the SDSS-III Collaboration including the University of Arizona, the Brazilian Participation Group, Brookhaven National Laboratory, Carnegie Mellon University, University of Florida, the French Participation Group, the German Participation Group, Harvard University, the Instituto de Astrofisica de Canarias, the Michigan State/Notre Dame/JINA Participation Group, Johns Hopkins University, Lawrence Berkeley National Laboratory, Max Planck Institute for Astrophysics, Max Planck Institute for Extraterrestrial Physics, New Mexico State University, New York University, Ohio State University, Pennsylvania State University, University of Portsmouth, Princeton University, the Spanish Participation Group, University of Tokyo, University of Utah, Vanderbilt University, University of Virginia, University of Washington, and Yale University.

Funding for the Sloan Digital Sky Survey IV has been provided by the Alfred P. Sloan Foundation, the U.S.  Department of Energy Office of Science, and the Participating Institutions.  SDSS-IV acknowledges support and resources from the Center for High Performance Computing at the University of Utah. The SDSS website is www.sdss.org.  SDSS-IV is managed by the Astrophysical Research Consortium for the Participating Institutions of the SDSS Collaboration including the Brazilian Participation Group, the Carnegie Institution for Science, Carnegie Mellon University, Center for Astrophysics | Harvard \& Smithsonian, the Chilean Participation Group, the French Participation Group, Instituto de Astrof\'isica de Canarias, The Johns Hopkins University, Kavli Institute for the Physics and Mathematics of the Universe (IPMU) / University of Tokyo, the Korean Participation Group, Lawrence Berkeley National Laboratory, Leibniz Institut f\"ur Astrophysik Potsdam (AIP), Max-Planck-Institut f\"ur Astronomie (MPIA Heidelberg), Max-Planck-Institut f\"ur Astrophysik (MPA Garching), Max-Planck-Institut f\"ur Extraterrestrische Physik (MPE), National Astronomical Observatories of China, New Mexico State University, New York University, University of Notre Dame, Observat\'ario Nacional / MCTI, The Ohio State University, Pennsylvania State University, Shanghai Astronomical Observatory, United Kingdom Participation Group, Universidad Nacional Aut\'onoma de M\'exico, University of Arizona, University of Colorado Boulder, University of Oxford, University of Portsmouth, University of Utah, University of Virginia, University of Washington, University of Wisconsin, Vanderbilt University, and Yale University.

\section{ADQL queries}\label{sec:adql_queries}

\subsection{Purer sub-samples}

The queries that provide the recommended purer sub-samples
of quasars and galaxies are shown in Tables~\ref{tab:qso_purer_subset}
and~\ref{tab:galaxy_purer_subset}. These sub-samples are discussed in Sect.~\ref{sec:howto_purer_samples}.

\subsection{Sources used for the construction of a \gaia-only composite spectrum of quasars}\label{sec:sample_qso_composite_gaia_only}

The following query returns the {\tt source\_id} of the set of sources used to construct the \gaia-only composite spectra of quasars. It selects 111\,563
sources from the purer quasar sub-sample (Table~\ref{tab:qso_purer_subset}) that have both a reliable redshift estimate from QSOC \citep[\linktoparam{sec_dm_extra--galactic_tables/ssec_dm_qso_candidates.html\#qso_candidates-flags_qsoc}{flags\_qsoc}\,=\,0 or \linktoparam{sec_dm_extra--galactic_tables/ssec_dm_qso_candidates.html\#qso_candidates-flags_qsoc}{flags\_qsoc}\,=\,16, as explained in][]{DR3-DPACP-158} and \bprp coefficients that are published in \gdr{3}.

{\small
\begin{verbatim}
SELECT source_id 
FROM gaiadr3.qso_candidates
JOIN gaiadr3.xp_summary USING (source_id)
WHERE (classlabel_dsc_joint='quasar'
       OR vari_best_class_name='AGN'
       OR host_galaxy_detected='true'
       OR gaia_crf_source='true') 
  AND (flags_qsoc = 0 OR flags_qsoc = 16)
\end{verbatim}
}

\subsection{Using DSC probability thresholds}\label{sec:sample_purer_dsc}

The following query uses thresholds on DSC-Specmod and DSC-Allosmod, which are in {\tt astrophysical\_parameters} table, to select sources from the {\tt qso\_candidates} table.
This example uses thresholds of 0.9 and
returns 371\,708 sources. This compares to
547\,201 returned when using thresholds of 0.5 (which is equivalent to, but slower than, selecting on {\tt classlabel\_dsc\_joint\,=\,quasar} supplied in the {\tt qso\_candidates} table itself).
An analogous query can be used for the {\tt galaxy\_candidates} table.
{\small
\begin{verbatim}
SELECT source_id
FROM gaiadr3.qso_candidates
JOIN gaiadr3.astrophysical_parameters USING
     (source_id)
WHERE classprob_dsc_specmod_quasar>0.9 AND
      classprob_dsc_allosmod_quasar>0.9 
\end{verbatim}
}

If we want to exclude generous regions around the LMC and SMC from the DSC purer subset, we can use the following query (change the initial line to select the fields you want).
It removes 22\,705 sources in the LMC and 7456 in the SMC, to leave 517\,040 sources.
{\small
\begin{verbatim}
SELECT source_id
FROM (SELECT *
	  FROM gaiadr3.qso_candidates
      WHERE classlabel_dsc_joint='quasar') AS temp
JOIN gaiadr3.gaia_source USING (source_id)
WHERE 1!=CONTAINS(
  POINT('ICRS', 81.3, -68.7),
  CIRCLE('ICRS', ra, dec, 9)) AND
  1!=CONTAINS(
  POINT('ICRS', 16.0, -72.8),
  CIRCLE('ICRS', ra, dec, 6))
\end{verbatim}
}

Combining the two queries above to use higher thresholds on the DSC probabilities and to exclude the LMC and SMC, we get the following query, which returns 366\,574 sources.
{\small
\begin{verbatim}
SELECT source_id
FROM (SELECT *
	  FROM gaiadr3.qso_candidates
	  JOIN gaiadr3.astrophysical_parameters USING
	  (source_id)
      WHERE classprob_dsc_specmod_quasar>0.9 AND
      classprob_dsc_allosmod_quasar>0.9) AS temp
JOIN gaiadr3.gaia_source USING (source_id)
WHERE 1!=CONTAINS(
  POINT('ICRS', 81.3, -68.7),
  CIRCLE('ICRS', ra, dec, 9)) AND
  1!=CONTAINS(
  POINT('ICRS', 16.0, -72.8),
  CIRCLE('ICRS', ra, dec, 6))
\end{verbatim}
}

\section{Computation of the composite spectra of quasars}
\label{sec:composite_qso_algo}

The computation of a composite spectrum of quasars from individual spectra may appear to be a straightforward task, but it turns out not to be for a number of reasons.
\begin{enumerate}
    \item Quasars cover a large range of redshifts, equivalently a large range of rest-frame wavelengths, whereas each spectrum covers only a limited fraction of these rest-frame wavelengths.
    \item Spectra have very different apparent luminosities, either because some are intrinsically brighter or fainter, or because of their difference in redshift, or because they are gravitationally lensed. Accordingly, each spectrum contributing to the composite spectrum must be scaled in order to reduce the dispersion of flux in rest-frame wavelength. 
    \item Spectra often have correlated noise in their fluxes that should be taken into account.
    \item Quasars can have different continuum slopes (or any other background signal) that we may want to subtract in order to produce a pure emission line composite spectrum.
    \item Spectra used to build the composite spectrum may have different resolutions, sampling and line spread function, as in \bprp spectra, that should be first homogenized so as to model the sole signal of interest.
\end{enumerate}

With all of these arguments in mind, we developed a new method for computing a composite \bprp spectrum based on maximum likelihood estimation through the minimization of\footnote{In our notation, $\left\|\vec{x}\right\|_{\mat{W}}^2 = \left\|\mat{W} \vec{x}\right\|^2 = \vec{x}^T \mat{W}^T \mat{W} \vec{x}$.}
\begin{equation}
    \chi^2 = \sum_{i=1}^N \; \left\| \; \vec{x}_i - \mat{M_i} \; \left[ \mat{P} \vec{f}_i + \vec{m} s_i \right] \; \right\|_{\mat{W_i}}^2,
\label{eq:qso_composite_chi2}
\end{equation}
where
\begin{itemize}
    \item $N$ is the number of observations (spectra).
    
    \item $\vec{x}_i$ is the $i$-th observation vector, here taken as a concatenation of the BP and RP spectral coefficients, which are coefficients associated with a linear spline basis functions that represent the \bprp spectra \cite{2021A&A...652A..86C}.
    
    \item $\mat{W_i}$ is the weight matrix associated with $\vec{x}_i$. If $\mat{L_i}$ is the Cholesky decomposition of the covariance matrix associated with $\vec{x}_i$, $\mat{C_i} = \mat{L_i} \mat{L_i}^T$, then $\mat{W_i} = \mat{L_i}^{-1}$ such that $\mat{C_i}^{-1} = \mat{W_i}^T \mat{W_i}$.
    
    \item $\vec{m}$ is the composite spectrum we are inferring. We additionally infer the scaling factors, $s_i$, one associated with each observation $i$.
    
    \item $\mat{P}$ is a matrix composed of a set of basis functions used to model a background signal to be subtracted from the spectra. The linear coefficients associated with $\mat{P}$ for the $i$-th observation are given by the column vector $\vec{f}_i$. An example use of $\mat{P}$ would be to model the quasar continua as a low order polynomial (whose coefficients are computed in $\vec{f}_i$) and to subtract these continua from the spectra in Eq.~\ref{eq:qso_composite_chi2}. This produces a pure emission line composite spectrum in $\vec{m}$. The matrix $\mat{P}$ could be a set of vectors resulting from a previous minimization of Eq.~\ref{eq:qso_composite_chi2}, namely $\mat{P} = \mat{P_t} = \begin{pmatrix} \mat{P_{t-1}} & \vec{m}_t\end{pmatrix}$ where $\vec{m}_t$ minimizes $\chi_t^2 = \sum_i \; \left\| \; \vec{x}_i - \mat{M_i} \; \left[ \mat{P_{t-1}} \vec{f}_i^{(t)} + \vec{m}_t s_i^{(t)} \right] \; \right\|_{\mat{W_i}}^2$. This method can be seen as a weighted principal component analysis, in the sense that $\mat{P_t}$ are the minimal set of $t$ components that minimize $\chi_t^2$. Although mentioned here for completeness, we decided not to subtract the quasar continua in the present study, so we set $\mat{P} = 0$ and $\vec{f}_i = 0$.
    
    \item $\mat{M_i}$ is a transformation matrix, associated with $\vec{x}_i$, that projects $\mat{P}$ and $\vec{m}$ into the space of $\vec{x}_i$. In the present application, the goal of $\mat{M_i}$ is twofold. First it isolates the rest-frame wavelength regions from $\mat{P}$ and $\vec{m}$ that correspond to the observed wavelength region in $\vec{x}_i$ (the source redshift must therefore be taken into account). Second, the shifted and resampled spectrum is converted into BP/RP spectral coefficients through the use of the GaiaXPy simulator. See the documentation on \href{https://ioadpc.ast.cam.ac.uk/examplepydocs/gaiaxpy.simulator.html}{\texttt{simulate\_continuous}} for more information on the calibration procedure. 
\end{itemize}

In the minimization of Eq.~\ref{eq:qso_composite_chi2}, $\vec{m}$ and $\vec{a}_i = \begin{pmatrix}\vec{f}_i \\ s_i\end{pmatrix}$ are free to vary. However, for a given value of $\vec{m}$, we can differentiate Eq.~ \ref{eq:qso_composite_chi2} with respect to $\vec{a}_i$ and set the resulting gradient to zero to get\footnote{Direct inversion of the normal equations is known to be numerically unstable and should be avoided.}
\begin{equation}
\vec{a}_i = \left( \mat{T_i}^T \mat{W_i}^T \mat{W_i} \mat{T_i}\right)^{-1} \mat{T_i}^T \mat{W_i}^T \mat{W_i} \vec{x}_i \hspace{0.5cm}\text{where}\hspace{0.5cm} \mat{T_i} = \mat{M_i} \begin{pmatrix}\mat{P} & \vec{m} \end{pmatrix}.
\label{eq:qso_composite_chi2_ai}
\end{equation}
Substituting this last equation into Eq.~\ref{eq:qso_composite_chi2} makes it depend only on $\vec{m}$, such that any (global) optimization algorithm can be used with a number of unknowns given by the number of fluxes in $\vec{m}$. In the present study, we use an expectation-maximization algorithm with momentum in batch mode. The steps of the expectation-maximization algorithm are: (i) compute $\vec{a}_i$ using Eq.~\ref{eq:qso_composite_chi2_ai}; (ii) fit $\vec{m}$ to all $\vec{x}_i$ given the previously computed $\vec{a}_i$,
\begin{equation}
\vec{m} = \left( \sum_{i=1}^N s_i^2 \; \mat{M_i}^T \mat{W_i}^T \mat{W_i} \mat{M_i} \right)^{-1} \left( \sum_{i=1}^N s_i \; \mat{M_i}^T \mat{W_i}^T \mat{W_i} \left[ \vec{x}_i - \mat{M_i} \mat{P} \vec{f}_i \right] \right).
\end{equation}
\new{Steps (i) and (ii) are then iterated until the reduced chi-square improves by no more that 0.001 for 16 consecutive iterations.}

To first order, the covariance matrix associated with the formal uncertainties of the computed composite spectrum can be approximated through the asymptotic normality property of the maximum likelihood estimator as
\begin{equation}
\mat{\Sigma_{m}} \approx \left( \; \sum_{i=1}^N \mat{J_i}^T \mat{W_i}^T \mat{W_i} \mat{J_i} \; \right)^{-1}
\label{eq:qso_composite_chi2_Cm}
\end{equation}
where
\begin{equation*}
\mat{J_i} = \frac{\partial \mat{T_i} \vec{a}_i}{\partial \vec{m}} = \mat{T_i} \frac{\partial \vec{a}_i}{\partial \vec{m}} + \mat{M_i} \; s_i
\end{equation*}
and
\begin{equation*}
\frac{\partial \vec{a}_i}{\partial \vec{m}} = \left( \mat{T_i}^T \mat{W_i}^T \mat{W_i} \mat{T_i} \right)^{-1} \begin{pmatrix}
-s_i \left[\mat{M_i} \mat{P}\right]^T \\
\left[ \vec{x}_i - \mat{M_i} \mat{P} \vec{f}_i \right]^T - 2 s_i \left[ \mat{M_i} \vec{m} \right]^T
\end{pmatrix} \mat{W_i}^T \mat{W_i} \mat{M_i}.
\end{equation*}
Equation \ref{eq:qso_composite_chi2_Cm} is only valid for large values of $N$. How large $N$ should be is problem dependent. We performed simulations using 65\,536 noisy realisations of a problem with $N=64$ and eight variables in $\vec{m}$, where all matrices, except $\mat{W_i}$, are uniformly distributed in $[-1,1]$ and $\mat{W_i}$ is an orthogonal transformation of matrices whose eigenvalues are uniformly drawn in $[0.001,1]$. This led to a maximum absolute error in the median
correlation coefficients of $0.007$.

The Octave/Matlab source code for minimizing Equation \ref{eq:qso_composite_chi2} and for computing the approximate covariance matrix from Equation \ref{eq:qso_composite_chi2_Cm} is available at \href{https://github.com/ldelchambre/gls_mean/}{https://github.com/ldelchambre/gls\_mean/}.

\end{document}